\documentclass[lettersize,journal,hyphens]{IEEEtran}

\usepackage{amsmath,amsfonts}
\usepackage{algorithm,color}
\usepackage{algpseudocode}
\usepackage{array}
\usepackage{longtable}
\usepackage{booktabs}
\usepackage{multirow}
\usepackage{makecell}
\usepackage{threeparttable}
\usepackage{afterpage}
\usepackage{graphicx}
\usepackage{xcolor}
\usepackage[caption=false,font=footnotesize]{subfig}
\usepackage{cite}
\usepackage[colorlinks=true,breaklinks=true]{hyperref}
\usepackage{textcomp}
\usepackage{ragged2e}
\usepackage{pifont}
\usepackage{stfloats}
\usepackage{cuted}
\usepackage{capt-of}
\usepackage[normalem]{ulem}
\usepackage{footnote}
\makesavenoteenv{table}
\usepackage{url}
\usepackage{verbatim}
\usepackage{csquotes}
\usepackage{tikz}
\usetikzlibrary{mindmap,shapes,backgrounds}
\usetikzlibrary{positioning,calc,shapes,fit}
\usetikzlibrary{shapes.geometric,arrows,positioning,fit,backgrounds,calc,shadows,decorations.pathreplacing}
\usepackage[edges]{forest}

\newcommand{\cmark}{\ding{51}}
\newcommand{\xmark}{\ding{55}}
\newcommand{\pmark}{\textasciitilde}
\definecolor{revblue}{RGB}{0,0,200}
\newcommand{\rev}[1]{#1}
\newcolumntype{P}[1]{>{\centering\arraybackslash}p{#1}}
\newcolumntype{L}[1]{>{\raggedright\arraybackslash}p{#1}}

\DeclareUnicodeCharacter{2265}{\ensuremath{\geq}}

\MakeOuterQuote{"}
\begin{document}

\title{Neuro-Symbolic AI for Cybersecurity: State of the Art, Challenges, and Opportunities}

\author{{Safayat Bin Hakim, Muhammad Adil, \IEEEmembership{Senior Member,~IEEE}, Alvaro Velasquez, \IEEEmembership{Senior Member,~IEEE}, Shouhuai Xu, \IEEEmembership{Senior Member,~IEEE}, and Houbing Herbert Song, \IEEEmembership{Fellow,~IEEE}}
\thanks{Safayat Bin Hakim and Houbing H. Song are with the Department of Information Systems, University of Maryland, Baltimore County, Baltimore, MD 21250 USA (e-mail: shakim3@umbc.edu; songh@umbc.edu).
\par Muhammad Adil is with the Department of Computer Science and Engineering, University at Buffalo, Buffalo, NY 14260 USA (e-mail: muhammad.adil@ieee.org).
\par Alvaro Velasquez is with the Department of Computer Science, University of Colorado Boulder, Boulder, CO 80309 USA (e-mail: alvaro.velasquez@colorado.edu).
\par Shouhuai Xu is with the Laboratory for Cybersecurity Dynamics, Department of Computer Science, University of Colorado Colorado Springs, Colorado Springs, CO 80918 USA (e-mail: sxu@uccs.edu).
}}

\markboth{}%
{Hakim \MakeLowercase{\textit{et al.}}: Neuro-Symbolic AI for Cybersecurity: State of the Art, Challenges, and Opportunities}

\maketitle

\begin{abstract}
Cybersecurity demands both rapid pattern recognition and deliberative reasoning, yet purely neural or purely symbolic approaches each address only one side of this duality. Neuro-Symbolic (NeSy) AI bridges this gap by integrating learning and logic within a unified framework. This systematic review analyzes 103 publications across the neural-symbolic integration spectrum in cybersecurity through April 2026, organizing them via a three-tier taxonomy---deep integration, structured interaction, and contextual baselines---and a Grounding-Instructibility-Alignment (G-I-A) analytical lens. We find that multi-agent and structured-integration architectures across the surveyed spectrum substantially outperform single-agent approaches in complex scenarios, causal reasoning enables proactive defense beyond correlation-based detection, and knowledge-guided learning improves both data efficiency and explainability. These findings span intrusion detection, malware analysis, vulnerability discovery, and autonomous penetration testing, revealing that integration depth often correlates with capability gains across domains. A first-of-its-kind dual-use analysis further shows that autonomous offensive systems in the broader survey corpus are already achieving notable zero-day exploitation success at significantly reduced cost, fundamentally reshaping threat landscapes. However, critical barriers persist: evaluation standardization remains nascent, computational costs constrain deployment, and effective human-AI collaboration is underexplored. We distill these findings into a prioritized research roadmap emphasizing community-driven benchmarks, responsible development practices, and defensive alignment to guide the next generation of NeSy cybersecurity systems.
\end{abstract}


\section{Introduction}
\label{sec:introduction}

Traditional AI approaches in cybersecurity---pure neural networks lacking interpretability and pure symbolic systems lacking adaptability---exhibit three fundamental limitations: (1) inadequate conceptual grounding leading to brittleness against novel attacks \cite{sarhan2023from,guo2023review}, (2) limited instructibility impeding analyst-guided adaptation \cite{Pawlicki2022NeuCom}, (3) misalignment with true cybersecurity objectives \cite{Hitzler2020NeSySW}. Cybersecurity inherently demands both fast pattern recognition (detecting anomalies at network speed) and deliberative reasoning (tracing attack chains, verifying policy compliance)---a duality that parallels Kahneman's System~1/System~2 distinction and motivates Neuro-Symbolic (NeSy) AI as a principled integration of both capabilities \cite{kautz2022third,wang2024towards,bhuyan2024neuro,colelough2025neuro}. Yet this paradigm lacks systematic characterization.

The cybersecurity threat landscape undergoes constant transformation driven by the cyber attack-defense arms race \cite{ferrag2025generative,zhang2025llm}. Autonomous attacks achieve state-of-the-art zero-day exploitation with substantial cost reductions \cite{zhu2026teams}, while multi-agent systems like VulnBot demonstrate 30.3\% completion rates versus 9.09\% baselines \cite{kong2025vulnbot,zhu2026teams}, underscoring the need for principled frameworks that combine learning with reasoning \cite{tafreshian2024defensive,grini2025constrained,mitre_advml}.

\noindent\textbf{Research Gap and Contributions.} Despite emerging NeSy applications \cite{wang2024towards,bhuyan2024neuro,colelough2025neuro,velasquez2025neurosymbolic,shreha2026neuro}, no systematic analysis addresses: (1) unified evaluation frameworks, (2) dual-use offensive capabilities, (3) implementation barriers, (4) standardization roadmaps. Standardization gaps limit reproducible evaluation \cite{renkhoff2024survey}, computational complexity requires resource orchestration, and human-AI collaboration patterns determine adoption \cite{renkhoff2024survey,xiong2024converging}.

\begin{figure*}[!t]
 \centering
 \includegraphics[width=\textwidth]{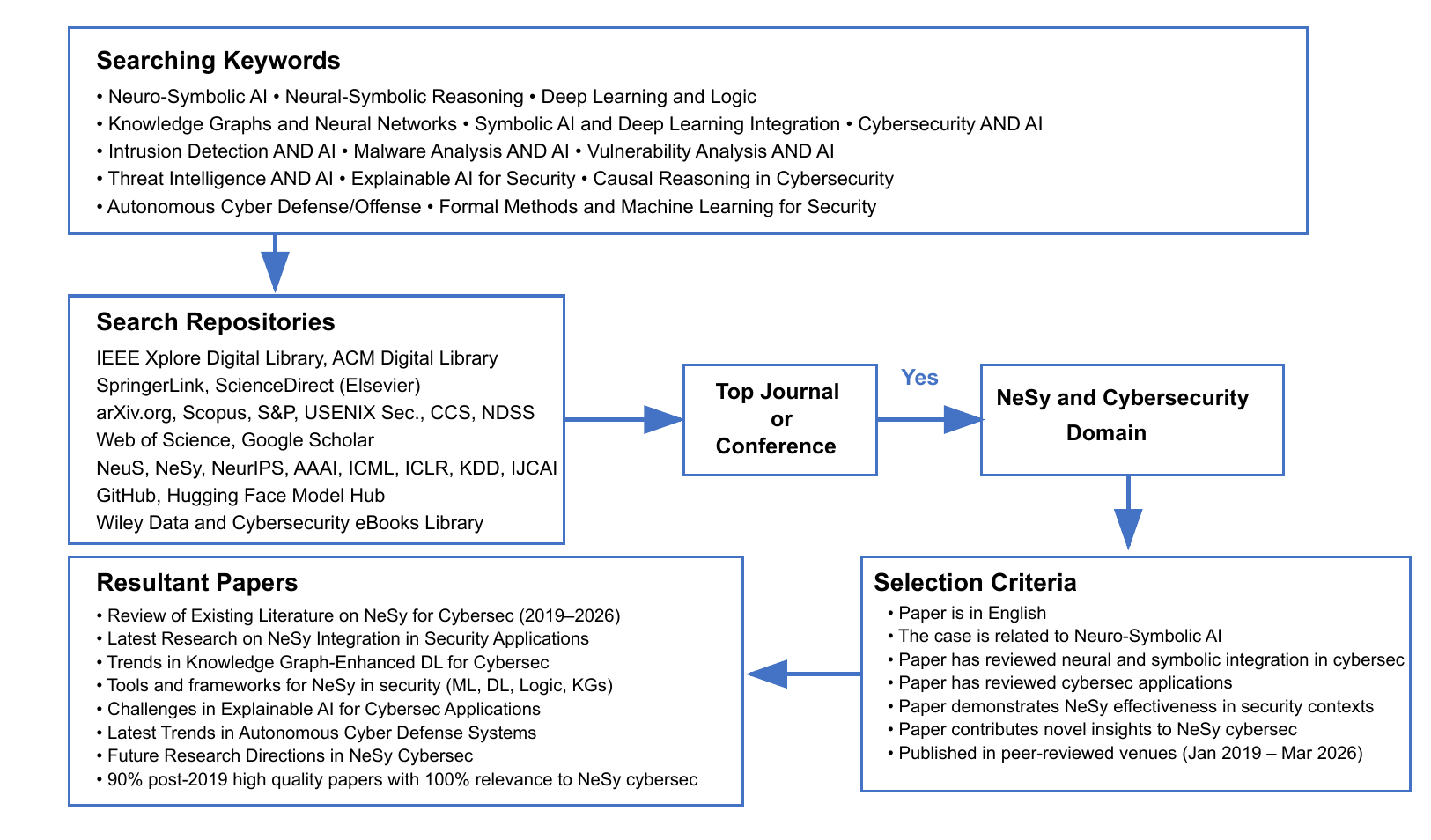}
 \caption{Systematic literature review for paper selection and screening criteria, illustrating the process used to identify and select the 103 publications analyzed in this survey.}
 \label{fig:methodology_flowchart}
\end{figure*}

We systematically analyze 103 publications spanning the neural-symbolic integration spectrum in cybersecurity (2019--2026), classifying each into three integration tiers: 22 deep NeSy systems with end-to-end integration, 55 structured NeSy systems with meaningful neural-symbolic interaction, and 26 contextual baselines included for comparative context. We introduce the Grounding-Instructibility-Alignment (G-I-A) framework as a structured analytical lens for NeSy cybersecurity assessment. Our methodology follows SPAR-4-SLR procedures \cite{paul2021spar4slr} with rigorous selection (Figure~\ref{fig:methodology_flowchart}), achieving 90\% post-2019 high-quality papers with 100\% relevance to NeSy cybersecurity. The resulting analysis makes several contributions. First, we provide a comprehensive SOTA analysis showing 20--50\% improvements in autonomous operations alongside advanced zero-day exploitation capabilities. Second, we present the first dual-use analysis examining both defensive innovations and autonomous offensive capabilities. Third, we systematically evaluate implementation challenges, instructibility mechanisms, and trust development factors. Finally, we identify critical evaluation gaps, propose NeSy-specific benchmarks and deployment frameworks, and outline prioritized research directions emphasizing grounding mechanisms, instructible collaboration, and responsible innovation governance.

\noindent\textbf{Research Questions.} \textbf{RQ1}: What levels of neural-symbolic integration have been achieved in cybersecurity, and what performance advantages does deeper integration provide? \textbf{RQ2}: What are the dual-use implications of NeSy offensive capabilities for defensive cybersecurity? \textbf{RQ3}: What technical and organizational barriers constrain operational deployment of NeSy cybersecurity systems? \textbf{RQ4}: What evaluation frameworks and benchmarks exist for NeSy cybersecurity, and where are the critical standardization gaps? \textbf{RQ5}: How do human-AI collaboration patterns and trust factors influence NeSy cybersecurity system effectiveness? \textbf{RQ6}: What are the most promising research directions for advancing NeSy cybersecurity capabilities while ensuring grounding, instructibility, and alignment with defensive objectives?

\subsection{Review Methodology}
\label{subsec:methodology}

To ensure comprehensive coverage, transparency, and reproducibility, we employ SPAR-4-SLR \cite{paul2021spar4slr,khan2023anthropomorphism}, a systematic procedure for computer science domains. The \textbf{Planning} phase established foundations through problem formulation, theoretical framework development anchored in hybrid intelligence systems, and protocol specification spanning January 1, 2019 to March 31, 2026 (last search: April 1, 2026).

The \textbf{Conducting} phase implemented systematic literature search across IEEE Xplore, ACM DL, SpringerLink, ScienceDirect, arXiv, Scopus, Web of Science, Google Scholar (supplementary citation backfilling only), plus specialized venues (NeSy, NeuS, S\&P, CCS, USENIX Security, NDSS, NeurIPS, ICML, AAAI, ICLR). Search strings combined NeSy terminology (``neuro-symbolic,'' ``neurosymbolic,'' ``neural-symbolic,'' ``hybrid AI,'' ``knowledge-guided learning'') with cybersecurity domains (``cybersecurity,'' ``network security,'' ``intrusion detection,'' ``malware analysis,'' ``vulnerability analysis''), supplemented with domain-specific terms (``knowledge graph,'' ``explainable AI,'' ``symbolic reasoning,'' ``logic tensor networks,'' ``causal reasoning in cybersecurity,'' ``MITRE ATT\&CK''). Records were deduplicated by DOI/title-year matching.

\begin{figure*}[!t]
 \centering
 \includegraphics[width=\textwidth]{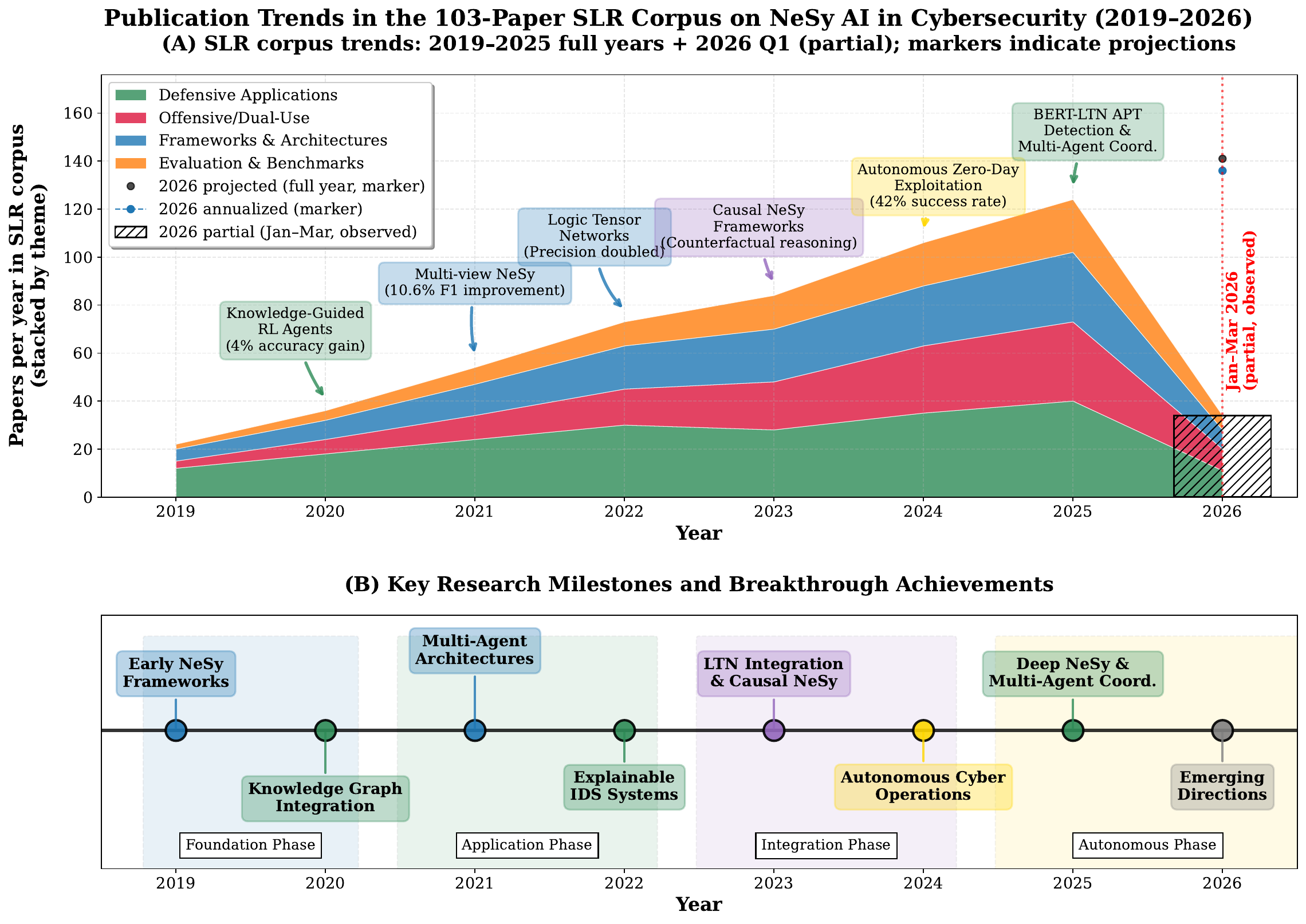}
 \caption{\small Publication trends in the 103-paper SLR corpus on NeSy AI in cybersecurity (2019--2026).
 (A) Per-year counts stacked by research theme (Defensive Applications, Offensive/Dual-Use, Frameworks \& Architectures, Evaluation \& Benchmarks).
 The 2025--2026 bars reflect partial-year observed data only (hatched); markers indicate simple full-year projections for context and are not used in the quantitative synthesis.
 (B) Timeline of notable research milestones, aligned with four developmental phases from foundational frameworks to autonomous cyber operations.}
 \label{fig:nesy_evolution_overview}
\end{figure*}

Our three-stage selection \cite{paul2021spar4slr} applied predefined criteria: \emph{Stage 1} screened 347 papers via title/abstract, identifying potential NeSy cybersecurity applications; deduplication removed 102 records, yielding 245 unique records. \emph{Stage 2} conducted comprehensive full-text review applying detailed inclusion/exclusion criteria, retaining 189 papers excluding works lacking sufficient neural-symbolic integration. \emph{Stage 3} performed systematic quality assessment for research rigor, methodological soundness, and practical significance, resulting in 103 surveyed publications plus approximately 93 support references for final analysis (Figure~\ref{fig:nesy_evolution_overview}).

\noindent\textbf{Inclusion/Exclusion Criteria.} Papers were included if they: (1) presented a system, method, or framework with identifiable neural \emph{and} symbolic components applied to a cybersecurity task, (2) were published in peer-reviewed venues or established preprint servers (arXiv with institutional affiliation) between January 2019 and March 2026, and (3) provided sufficient technical detail for classification. Papers were excluded if they: (a) addressed only general AI/ML without a cybersecurity application, (b) employed purely neural or purely symbolic methods without cross-paradigm interaction, or (c) were position papers, editorials, or workshop abstracts lacking empirical or architectural contributions. Pure neural and pure symbolic systems of particular relevance to the NeSy cybersecurity landscape were retained as contextual baselines (Type~C) to enable comparative analysis. Complete per-paper classification details, including integration tier assignments and component identification, are provided in Supplementary Table~S1. To support transparency of the curated survey artifacts, we also provide a public supplementary repository at \url{https://github.com/sbhakim/gia-nesy-cybersecurity-survey}.

To ensure reliability, two independent reviewers conducted screening with 20\% double-coding (n=49) achieving $\kappa=0.89$ for inclusion decisions and $\kappa=0.85$ for quality scores; disagreements resolved through structured discussion and expert consultation.

Data extraction employed structured forms capturing study metadata, NeSy integration strategies, cybersecurity applications, evaluation methodologies, performance metrics, deployment considerations, human factors, and limitations. We synthesized findings through a combination of qualitative thematic coding, quantitative performance aggregation, taxonomic classification, and gap analysis, organized around the six research questions.

\section{Theoretical Underpinnings and Advanced Integration Approaches}
\label{sec:foundations}

NeSy architectures unite fast, sub-symbolic pattern recognition (Kahneman's System~1) with deliberative symbolic reasoning (System~2) \cite{kautz2022third}, combining the adaptability of neural learning with the transparency and logical consistency of symbolic methods \cite{wang2024towards,marra2024statistical,bhuyan2024neuro,gaur2024building}. Pure neural systems excel at pattern recognition but lack conceptual grounding essential for security principles, leading to brittleness against novel threats \cite{neupane2022explainable,onchis2022neurosymbolic}. Symbolic knowledge injection has been shown to reduce data requirements and improve robustness to noise and perturbations \cite{agiollo2023symbolic,rafanelli2024empirical}, while symbolic systems alone provide logical consistency but struggle with noisy, real-world data.

\subsection{Neural-Symbolic Integration: Definitions and Scope}
\label{subsec:nesy_definitions}

The term ``neuro-symbolic'' spans a broad definitional spectrum in the AI literature. At the strict end, Marra et al.~\cite{marra2024statistical} characterize NeSy systems as those achieving end-to-end differentiable integration of neural learning and symbolic reasoning, with Ciatto et al.~\cite{ciatto2024symbolic} providing a complementary taxonomy of symbolic knowledge extraction and injection methods that operationalize this integration. At the broad end, Kautz's~\cite{kautz2022third} influential taxonomy---originally six categories, later expanded to eight pathways~\cite{nerode2024integrating}---encompasses approaches ranging from symbolic preprocessing of neural inputs through full neural-symbolic integration, including the emerging Neuro[Symbolic()] paradigm where LLMs invoke formal reasoning tools. This survey deliberately adopts a broad scope to capture the full landscape relevant to cybersecurity practitioners, while making integration depth explicit through a three-tier classification:

\begin{itemize}
 \item \textbf{Type~A --- Deep NeSy (22 papers):} Systems achieving joint optimization or deeply interleaved neural-symbolic training. Examples include Logic Tensor Network-based intrusion detection systems \cite{bizzarri2024synergistic,onchis2022neurosymbolic}, differentiable logic integration in GNNs \cite{zhou2024knowgraph}, and iterative neural-symbolic refinement loops \cite{li2025automated}. These satisfy even the strictest NeSy definitions \cite{marra2024statistical}.
 \item \textbf{Type~B --- Structured NeSy (55 papers):} Systems with meaningful interaction between identifiable neural and symbolic components, where each contributes distinct capabilities---corresponding to multiple pathways in Kautz's expanded taxonomy \cite{kautz2022third,nerode2024integrating}. This includes knowledge graph-guided neural learning \cite{piplai2023knowledge,zhang2025improving}, LLMs paired with formal verification or symbolic planning tools (Kautz's Neuro[Symbolic()] paradigm) \cite{grov2024neurosymbolic,nieponice2025aracne}, causal-neural integration \cite{jaimini2024causal}, and ontology-constrained processing \cite{nalluri2025nscti,belcastro2025enhancing}.
 \item \textbf{Type~C --- Contextual Baselines (26 papers):} Pure neural or pure statistical systems included for comparative context to evaluate NeSy advantages. These are explicitly acknowledged as non-NeSy and are labeled as baselines throughout the analysis.
\end{itemize}

This taxonomy makes the survey's scope transparent: 74.8\% of surveyed papers (77 of 103) involve genuine neural-symbolic integration at varying depths, while the remaining 25.2\% provide essential comparative context. The distinction allows readers to assess NeSy claims at the appropriate integration level rather than treating all surveyed systems as equivalent. Complete per-paper tier assignments with neural component, symbolic component, and domain identification are provided in Supplementary Table~S1.

Table~\ref{tab:notation} summarizes notations used throughout.

\begin{table}[!t]
\renewcommand\arraystretch{1.02}
\centering
\caption{Notation and Symbols Used Throughout This Work}
\label{tab:notation}
\scriptsize
\setlength{\tabcolsep}{1.5pt}
\begin{tabular}{|>{\centering\arraybackslash}m{3.05cm}|>{\raggedright\arraybackslash}m{4.95cm}|}
\hline
\textbf{Symbol} & \textbf{Description} \\
\hline \hline
\multicolumn{2}{|c|}{\textit{Core NeSy System Components}} \\
\hline
$\Phi_\theta$ & Neural component parameterized by $\theta$ \\
$\Psi_{\mathcal{K}}$ & Symbolic component operating over knowledge base $\mathcal{K}$ \\
$\mathcal{X}, \mathcal{Y}$ & Input and output spaces \\
$\Theta$ & Parameter space for neural components \\
$\mathcal{K}$ & Knowledge base containing symbolic knowledge \\
\hline
\multicolumn{2}{|c|}{\textit{G-I-A Framework}} \\
\hline
$\mathcal{G}(\theta, \mathcal{K})$ & Grounding quality measure \\
$\mathcal{I}(\theta, \mathcal{K}, \mathcal{H})$ & Instructibility effectiveness measure \\
$\mathcal{A}(\theta, \mathcal{K}, \mathcal{O})$ & Alignment coherence measure \\
$\mathcal{Z}$ & Set of cybersecurity concepts \\
$\mathcal{H}$ & Set of human feedback instances \\
$\mathcal{O}$ & Set of organizational cybersecurity objectives \\
$w_o$ & Weight for objective $o \in \mathcal{O}$ \\
$\lambda_G, \lambda_I, \lambda_A$ & G-I-A component weighting parameters \\
$\mathcal{L}_{\text{G-I-A}}$ & Integrated G-I-A optimization objective \\
$\mathcal{L}_N$ & Standard neural training loss \\
\hline
\multicolumn{2}{|c|}{\textit{Multi-Agent Systems}} \\
\hline
$\mathcal{S} = \{a_1, \ldots, a_k\}$ & Multi-agent system with $k$ agents \\
$\alpha_i$ & Specialization weight for agent $a_i$ \\
$\beta$ & Coordination effectiveness parameter \\
$P_{\text{multi}}$ & Multi-agent collaborative performance \\
$P_{\text{individual}}(a_i)$ & Individual performance of agent $a_i$ \\
$\text{Synergy}(a_i, a_j)$ & Cross-validation benefit between agents \\
$\omega_i$ & Agent weighting parameter \\
$\text{Agreement}(d_i, d_j)$ & Agreement measure between agent decisions \\
$\tau$ & Decision threshold parameter \\
\hline
\multicolumn{2}{|c|}{\textit{Causal Reasoning}} \\
\hline
$\mathcal{M} = (\mathcal{V}, \mathcal{E}, f)$ & Cybersecurity causal model \\
$\mathcal{V} = \{X_1, \ldots, X_n\}$ & Set of security events \\
$\mathcal{E}$ & Set of causal relationships \\
$w: \mathcal{E} \rightarrow [0,1]$ & Causal strength function \\
$\text{CF}(y, x, x')$ & Counterfactual analysis function \\
$\text{do}(\cdot)$ & Causal intervention operator \\
\hline
\multicolumn{2}{|c|}{\textit{Loss Functions and Optimization}} \\
\hline
$\mathcal{L}_{\text{total}}$ & Joint optimization objective \\
$\mathcal{L}_S$ & Symbolic reasoning consistency loss \\
$\mathcal{L}_{INT}$ & Integration effectiveness loss \\
$\mathcal{L}_A$ & Alignment penalty term (optional, see Sec. VI-B) \\
$\lambda_{\text{sym}}, \gamma, \lambda$ & Loss weighting and task decomposition parameters \\
$T_{N \rightarrow S}, T_{S \rightarrow N}$ & Knowledge transfer functions \\
$\text{CommOverhead}$ & Communication overhead cost \\
$\text{Cost}(f_i)$ & Task decomposition cost function \\
\hline
\multicolumn{2}{|c|}{\textit{Performance Metrics}} \\
\hline
$\text{Consistency}(\cdot,\cdot)$ & Alignment measure between components \\
$\text{Adaptation}(\Delta\theta_h,\,\Delta\mathcal{K}_h)$ & System responsiveness measure \\
$\text{Objective}(\Phi_\theta,\,\Psi_{\mathcal{K}},\,o)$ & Objective consistency measure \\
$C$ & Confidence/classification score \\
$\mathcal{R}$ & Symbolic rule set from MITRE ATT\&CK and domain expertise \\
$\mathcal{D}$ & Agent decision set $\{d_1, \ldots, d_k\}$ \\
\hline
\end{tabular}
\end{table}

\subsection{Grounding-Instructibility-Alignment (G-I-A) Framework}

\begin{figure}[!htbp]
\centering
\includegraphics[width=\columnwidth]{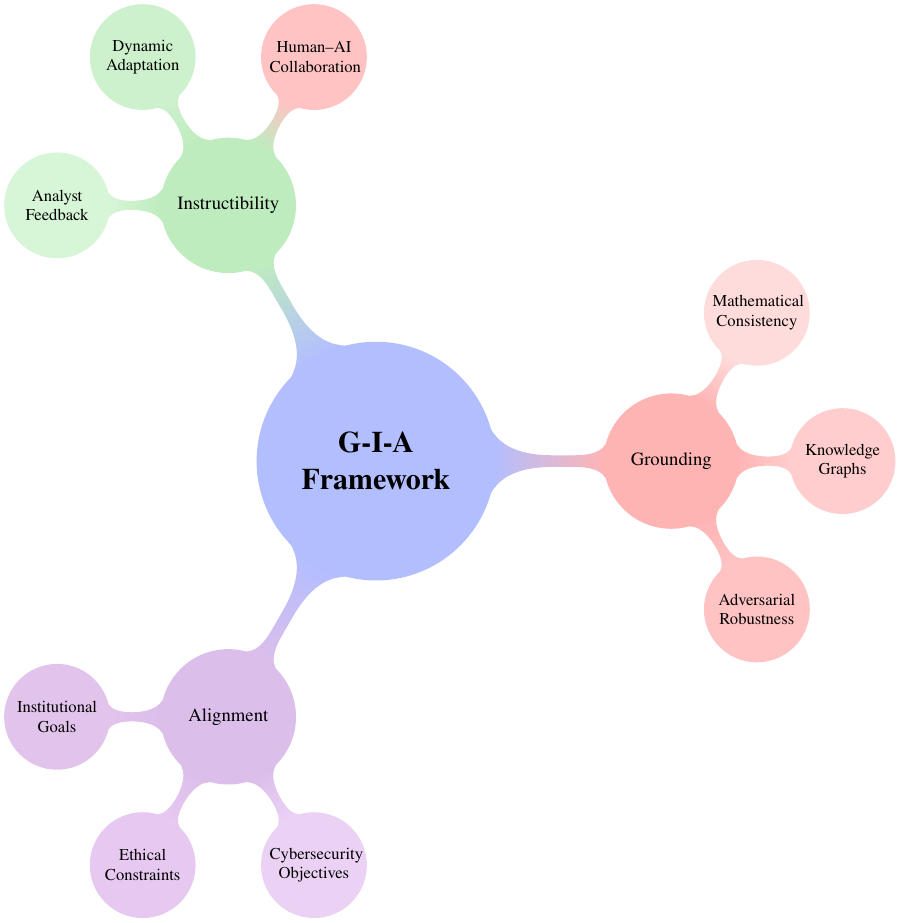}
\caption{The G-I-A Framework for assessing NeSy cybersecurity systems.}
\label{fig:gia_framework_diagram}
\end{figure}

At the deep-integration end of the spectrum, NeSy approaches are characterized by neural and symbolic components that synergistically enhance each other through continuous bidirectional information exchange, distinguishing them from simple pipeline or ensemble combinations that lack end-to-end optimization \cite{colelough2025neuro,xiong2024converging,marra2024statistical}. We formalize core G-I-A requirements through unified mathematical frameworks enabling systematic optimization and evaluation (Figure~\ref{fig:gia_framework_diagram}).

\textit{Grounding Quality} measures system ability to establish meaningful connections between outputs and cybersecurity concepts:
\begin{equation}
\small
\mathcal{G}(\theta, \mathcal{K}) = \frac{1}{|\mathcal{Z}|} \sum_{c \in \mathcal{Z}} \text{Consistency}(\Phi_\theta(x_c), \Psi_{\mathcal{K}}(x_c, c)),
\label{eq:grounding_quality}
\end{equation}
where $\mathcal{Z}$ represents cybersecurity concepts, $x_c$ denotes input examples related to concept $c$, and Consistency measures alignment between neural predictions and symbolic reasoning.

\textit{Instructibility} quantifies system responsiveness to analyst feedback:
\begin{equation}
\small
\mathcal{I}(\theta, \mathcal{K}, \mathcal{H}) = \mathbb{E}_{h \in \mathcal{H}} \left[ \text{Adaptation}(\Delta\theta_h, \Delta\mathcal{K}_h) \right],
\label{eq:instructibility}
\end{equation}
where $\mathcal{H}$ represents feedback instances, and Adaptation quantifies effective guidance incorporation.

\textit{Alignment} ensures consistency with cybersecurity objectives:
\begin{equation}
\small
\mathcal{A}(\theta, \mathcal{K}, \mathcal{O}) = \sum_{o \in \mathcal{O}} w_o \cdot \text{Objective}(\Phi_\theta, \Psi_{\mathcal{K}}, o),
\label{eq:alignment}
\end{equation}
where $\mathcal{O}$ represents organizational objectives, $w_o$ denotes importance weights, and Objective measures how well the integrated system serves each specific goal.

The integrated optimization objective balances all requirements:
\begin{equation}
\small
\mathcal{L}_{\text{G-I-A}}(\theta, \mathcal{K}) = \mathcal{L}_N - \lambda_G \mathcal{G}(\theta, \mathcal{K}) - \lambda_I \mathcal{I}(\theta, \mathcal{K}, \mathcal{H}) - \lambda_A \mathcal{A}(\theta, \mathcal{K}, \mathcal{O}).
\label{eq:gia_objective}
\end{equation}

This minimizes standard neural training loss $\mathcal{L}_N$ while maximizing the three G-I-A components, with weighting parameters $\lambda_G$, $\lambda_I$, $\lambda_A$ allowing practitioners to emphasize different aspects based on deployment requirements. These formulations provide precise conceptual definitions for each G-I-A dimension, establishing what each measures and enabling structured qualitative comparison across NeSy cybersecurity systems. Table~\ref{tab:gia_validation} illustrates their application to representative systems; we present G-I-A as an analytical framework for examining integration quality, not as a computable scoring algorithm. The table is intended to support comparative discussion of representative cases rather than to claim a validated benchmark or inter-rater scoring protocol.

Table~\ref{tab:gia_validation} illustrates G-I-A assessment on representative systems. KnowGraph \cite{zhou2024knowgraph} achieves high grounding (4.2/5) through weighted first-order logic integration, moderate instructibility (3.1/5) via rule modification, and strong alignment (3.8/5) through defensive focus, corresponding with superior out-of-distribution performance (91.2\% AUC). Multi-agent systems in the broader survey corpus \cite{kong2025vulnbot,singh2024hierarchical} demonstrate consistently high alignment scores through defensive specialization, corresponding with substantial performance gains including VulnBot's 30.3\% versus 9.09\% completion rates. Across these illustrative cases, systems with stronger G-I-A profiles tend to exhibit better robustness and operational alignment, though we note that these assessments reflect author judgment on a small sample and should be interpreted as qualitative observations rather than statistical findings.

\begin{table}[!htbp]
\renewcommand\arraystretch{1.15}
\centering
\caption{G-I-A Framework Illustration: System Assessment and Performance Association}
\label{tab:gia_validation}
\scriptsize
\begin{threeparttable}
\begin{tabular}{|l|c|c|c|l|}
\hline
\textbf{System} & \textbf{G} & \textbf{I} & \textbf{A} & \textbf{Key Performance Metric} \\
\hline \hline
KnowGraph \cite{zhou2024knowgraph} & 4.2 & 3.1 & 3.8 & 91.2\% Inductive AUC \\
HPTSA$^\dagger$ \cite{zhu2026teams} & 3.5 & 2.8 & 2.1 & 42\% Zero-Day Success \\
VulnBot \cite{kong2025vulnbot} & 3.8 & 3.5 & 4.1 & 30.3\% Completion Rate \\
H-MARL Defense$^\dagger$ \cite{singh2024hierarchical} & 3.7 & 3.8 & 4.0 & 61\% Recovery Precision \\
LTN-IDS \cite{grov2024neurosymbolic} & 4.0 & 3.4 & 3.9 & 21.3\% XSS Precision \\
IoT NeSy \cite{kalutharage2025neurosymbolic} & 4.1 & 3.6 & 4.2 & 97\% Accuracy + ATT\&CK \\
\hline
\end{tabular}
\begin{tablenotes}
\item \textit{Note:} G = Grounding, I = Instructibility, A = Alignment (all scores /5). AUC = area under receiver operating characteristic curve; LTN = logic tensor network; IDS = intrusion detection system; XSS = cross-site scripting. Scores were assigned by the authors based on published system descriptions and documented capabilities; we present these as indicative assessments to illustrate G-I-A dimensions, not as definitive quantitative evaluations.
\item[$\dagger$] Type~C contextual baseline (non-NeSy); included for comparative illustration of G-I-A dimensions.
\end{tablenotes}
\end{threeparttable}
\end{table}

Table~\ref{tab:gia_framework} operationalizes G-I-A integration through representation learning \cite{wang2024towards,ontiveros2025ground,li2023softened} and causal reasoning mechanisms \cite{jaimini2024causal,rawal2025causality} supporting cybersecurity objectives.

\begin{table*}[!htbp]
\renewcommand\arraystretch{1.3}
\centering
\caption{Grounding-Instructibility-Alignment (G-I-A) Framework Operationalization in NeSy Cybersecurity Systems}
\label{tab:gia_framework}
\resizebox{\linewidth}{!}{
\begin{tabular}{|c|c|c|c|c|}
\hline
\textbf{G-I-A Component} & \textbf{Mathematical Formulation} & \textbf{Cybersecurity Implementation} & \textbf{Operational Benefits} & \textbf{Example Applications} \\
\hline \hline
\textbf{Grounding} &
$\begin{aligned}
\mathcal{G}(\theta, \mathcal{K}) &= \frac{1}{|\mathcal{Z}|} \sum_{c \in \mathcal{Z}} \\
&\quad \text{Consistency}(\Phi_\theta(x_c), \Psi_{\mathcal{K}}(x_c, c))
\end{aligned}$ &
\makecell{Mapping outputs to\\cybersecurity concepts via\\knowledge graphs and\\domain ontologies} &
\makecell{Robust understanding of\\security principles;\\Resistance to adversarial\\attacks; Reliable\\generalization to novel threats} &
\makecell{KnowGraph (91.2\%\\inductive AUC), IoT IDS\\with 100\% ATT\&CK\\mapping} \\
\hline
\textbf{Instructibility} &
$\mathcal{I}(\theta, \mathcal{K}, \mathcal{H}) = 
\mathbb{E}_{h \in \mathcal{H}}\big[\text{Adaptation}(\Delta\theta_h, \Delta\mathcal{K}_h)\big]$ &
\makecell{Dynamic updates to\\knowledge base and\\parameters from\\analyst feedback} &
\makecell{Rapid adaptation to emerging\\threats; Enhanced human–AI\\collaboration; Continuous\\learning from expert guidance} &
\makecell{Multi-agent coordination\\with analyst feedback\\loops; Causal reasoning\\modification} \\
\hline
\textbf{Alignment} &
$\mathcal{A}(\theta, \mathcal{K}, \mathcal{O}) = 
\sum_{o \in \mathcal{O}} w_o \cdot \text{Objective}(\Phi_\theta, \Psi_{\mathcal{K}}, o)$ &
\makecell{Ensuring consistency with\\organizational objectives\\and ethical constraints\\via weighted optimization} &
\makecell{Ethical AI deployment;\\Policy compliance;\\Resource-efficient operation\\aligned with sustainability goals} &
\makecell{Defensive-biased\\development; 100×\\parameter reduction;\\Environmental\\sustainability focus} \\
\hline
\textbf{Integrated G-I-A} &
$\mathcal{L}_{\text{G-I-A}} = \mathcal{L}_N - \lambda_G \mathcal{G} - \lambda_I \mathcal{I} - \lambda_A \mathcal{A}$ &
\makecell{Joint optimization balancing\\all three components via\\weighted loss functions\\and coordinated training} &
\makecell{Superior performance\\(20–50\% improvement);\\Explainable decisions;\\Sustainable deployment;\\Responsible innovation} &
\makecell{Multi-agent pentesting\\with 30.3\% completion\\rates; Zero-day detection\\with ethical constraints} \\
\hline
\end{tabular}}
\end{table*}

\subsection{Advanced Integration Strategies}
\label{subsec:integration_strategies}

\begin{figure*}[!htbp]
 \centering
 \includegraphics[width=0.98\textwidth]{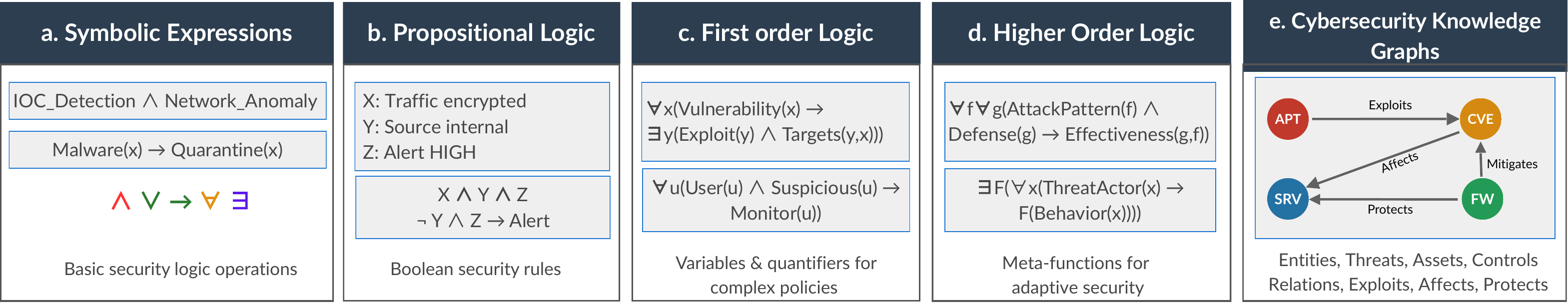}
 \caption{\small Advanced symbolic reasoning foundations enabling NeSy cybersecurity systems. 
 \textbf{(a)} Symbolic expressions with logical operators for precise security concept representation; example shows indicator of compromise (IOC) detection. 
 \textbf{(b)} Propositional logic enabling automated decision making with proper grounding. 
 \textbf{(c)} First-order logic with quantifiers supporting complex instructible policy frameworks. 
 \textbf{(d)} Higher-order logic enabling meta-analysis of attack patterns with adaptive defense alignment. 
 \textbf{(e)} Knowledge graphs with cybersecurity entities and relationships for systematic reasoning and analyst instruction.}
 \label{fig:symbolic_reasoning_foundations}
\end{figure*}

How neural and symbolic components are woven together varies considerably, and the choice of integration strategy shapes both what a system can do and where it can be deployed \cite{lu2024surveying,hagos2024neuro,tran2025neurosymbolic}. Recent work has broadened the design space to include verification-oriented frameworks, causal reasoning systems, and multi-agent coordination mechanisms. Figure~\ref{fig:symbolic_reasoning_foundations} illustrates the symbolic reasoning foundations that underpin these approaches.

\noindent\textbf{Mathematical Frameworks for Integration.} We formalize core strategies through unified mathematical frameworks enabling systematic analysis and optimization \cite{luo2024neurosymbolic,samaddar2025ood}. We define component loss functions for neural pattern recognition ($\mathcal{L}_N$), symbolic reasoning consistency ($\mathcal{L}_S$), and cross-modal integration effectiveness ($\mathcal{L}_{INT}$). The joint optimization objective balances:
\begin{equation}
\mathcal{L}_{\text{total}}(\theta, \mathcal{K}) = \mathcal{L}_N + \lambda_{\text{sym}} \mathcal{L}_S + \gamma \mathcal{L}_{INT},
\label{eq:joint_optimization}
\end{equation}
where knowledge transfer functions facilitate bidirectional information flows:
\begin{equation}
T_{N \rightarrow S}: \Phi_\theta \rightarrow \mathcal{K}_{\text{updated}}, \quad
T_{S \rightarrow N}: \mathcal{K} \rightarrow \theta_{\text{updated}},
\label{eq:knowledge_transfer}
\end{equation}
enabling instructible adaptation based on analyst feedback.

\subsubsection{Core Integration Paradigms}

\begin{figure*}[!t]
\centering
\includegraphics[width=0.80\textwidth]{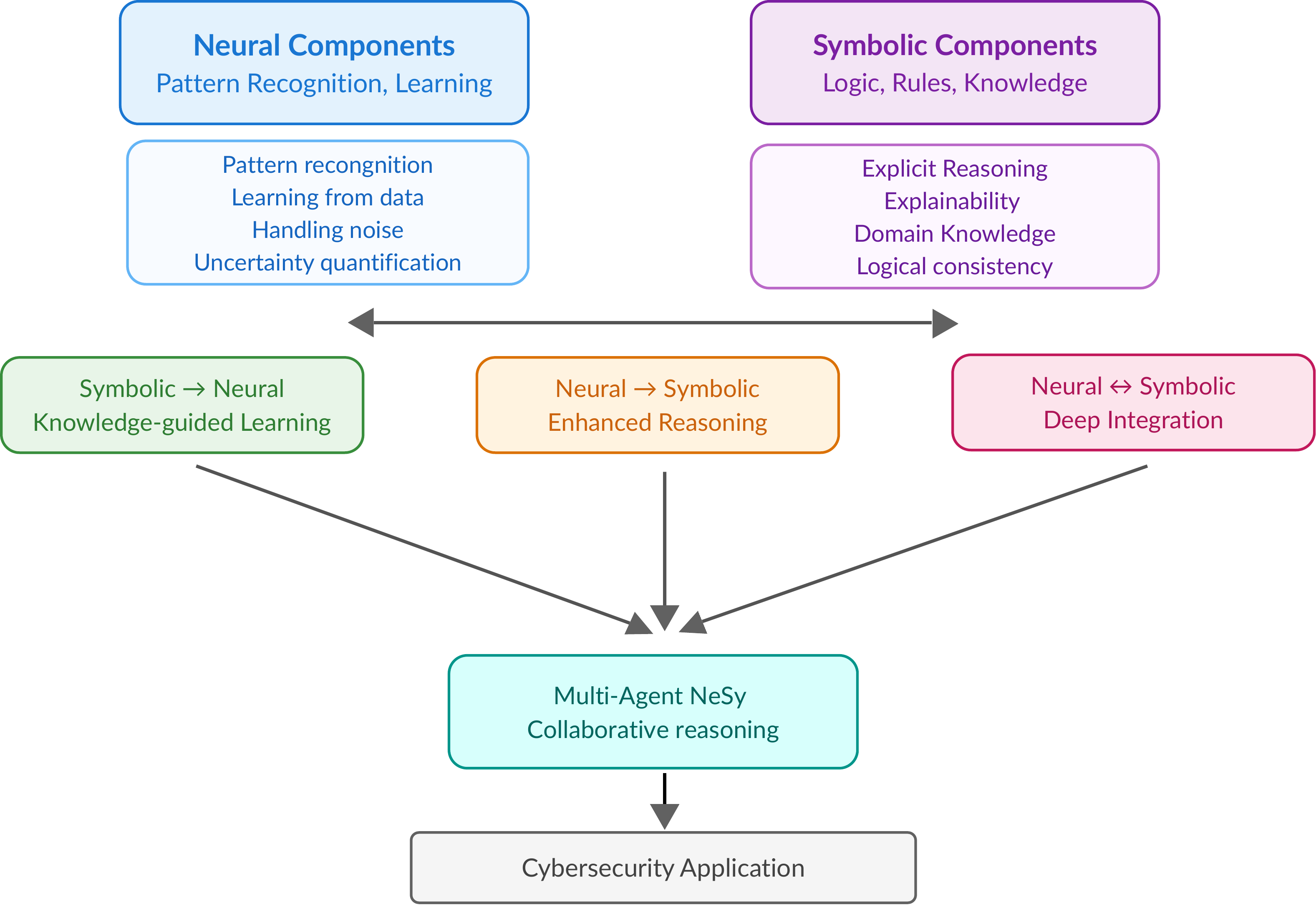}
\caption{\small SOTA NeSy integration strategies for cybersecurity applications, demonstrating bidirectional information flow, multi-agent architectures, and complementary capabilities supporting G-I-A objectives.}
\label{fig:nesy_integration}
\end{figure*}

NeSy integration follows three complementary strategies (Figure~\ref{fig:nesy_integration}). \textbf{Knowledge-guided learning} injects symbolic constraints---security policies, attack patterns, threat intelligence---into neural training objectives \cite{piplai2023knowledge,zhang2025improving}, achieving 8\% faster convergence and 4\% accuracy improvements \cite{piplai2020using}. Advanced approaches systematically integrate cybersecurity knowledge graphs capturing entities, relationships, and causal dependencies \cite{falcarin2024building,sikos2023cybersecurity,zhao2024survey}. Logic-Guided Neural Learning incorporates logical constraints directly into training \cite{bizzarri2024synergistic,onchis2022neurosymbolic}. Symbolic Prefix-Tuning uses knowledge graphs to generate dynamic prefixes injected into transformer layers \cite{hsu2023ampere}.

\textbf{Neural-enhanced reasoning} employs neural networks to scale symbolic components: knowledge graph completion identifies attack vectors \cite{sikos2023cybersecurity,liu2025graph}, neural-guided SMT solvers double verification performance \cite{blaauwbroek2024learning,piepenbrock2023neural,blaauwbroek2024graph2tac}, and rule discovery automates security policy generation \cite{fieblinger2024actionable,cheng2025crucialg}.

\textbf{Deep iterative integration} enables bidirectional refinement through Logic Tensor Networks \cite{bizzarri2024synergistic,bizzarri2024neuro} grounding logical terms in continuous vector spaces, and continual learning loops \cite{choi2025nesyc} where symbolic rules are dynamically reformulated based on new experiences, maintaining knowledge preservation during adaptation.

\subsubsection{Advanced Multi-Agent Architectures}

Multi-agent approaches consistently achieve superior performance through collaborative specialization \cite{kong2025vulnbot,zhu2026teams,singh2024hierarchical,hurten2024hierarchical}. VulnBot achieves 30.3\% completion rates versus 9.09\% for single-agent approaches, while teams demonstrate 53\% success rates on zero-day vulnerabilities \cite{kong2025vulnbot,zhu2026teams}. The collaborative performance model extends mathematical frameworks to multi-agent scenarios:
\begin{equation}
\small
P_{\text{multi}} = \sum_{i=1}^{k} \alpha_i \cdot P_{\text{individual}}(a_i) + \beta \cdot \sum_{i<j} \text{Synergy}(a_i, a_j)
\end{equation}

For optimal task decomposition, the optimization objective balances computational efficiency with coordination effectiveness:
\begin{equation}
\small
\min \sum_{i} \text{Cost}(f_i) + \lambda \cdot \text{CommOverhead}
\end{equation}
subject to completeness constraints ensuring threat coverage. These architectures employ hierarchical organization where coordination agents manage task distribution while specialized agents handle domain-specific analysis, enabling systematic decomposition of complex security problems.

\subsection{Advanced Frameworks and Technologies}
\label{subsec:frameworks}

Table~\ref{tab:frameworks} summarizes five critical NeSy frameworks. Logic Tensor Networks ground first-order logic in continuous spaces via differentiable Real Logic \cite{bizzarri2024neuro,bizzarri2024synergistic}, enabling systems maintaining logical consistency while supporting instructible adaptation for IDS \cite{onchis2022neurosymbolic,grov2024neurosymbolic}. Neural-guided SMT solvers achieve $2\times$ verification performance improvements \cite{blaauwbroek2024learning,piepenbrock2023neural,blaauwbroek2024graph2tac}, enabling scalable formal verification where systems like Z3 and cvc5 combine SAT solving with neural methods \cite{lu2023z3}. Knowledge Graph Neural Networks enhance architectures with structured cybersecurity knowledge \cite{falcarin2024building,zhang2025gnnsecurity,kurniawan2024cykg}, particularly effective for understanding complex attack campaigns \cite{chen2022aptkgl,alharbi2025enhancing}. Causal NeSy frameworks enable sophisticated reasoning about attack causality and counterfactual scenarios \cite{jaimini2024causal,rawal2025causality}, with applications including dynamic causal Bayesian optimization \cite{andrew2022developing}. LLM-symbolic integration represents cutting-edge development combining LLMs with symbolic reasoning \cite{clairouxtrepanier2024llm,zhang2025llm}, with applications including automated threat intelligence analysis, security policy formulation, and natural language interfaces. Recent frameworks like ARACNE demonstrate autonomous shell penetration testing through sophisticated LLM-symbolic integration \cite{nieponice2025aracne,deng2024pentestgpt}.

\begin{table*}[!t]
\centering
\renewcommand\arraystretch{1.02}
\caption{Advanced NeSy Framework Landscape for Cybersecurity}
\label{tab:frameworks}
\footnotesize
\begin{tabular}{|>{\centering\arraybackslash}p{1.8cm}|>{\centering\arraybackslash}p{2.9cm}|>{\centering\arraybackslash}p{3.0cm}|>{\centering\arraybackslash}p{2.9cm}|>{\centering\arraybackslash}p{2cm}|}
\hline
\textbf{Framework} & \textbf{Core Mechanism} & \textbf{Key Advantage} & 
\textbf{Cybersecurity Applications} & \textbf{References} \\
\hline
Logic Tensor Networks & Differentiable first-order logic grounding in continuous spaces & Maintains logical consistency during learning & IDS with policy constraints, malware classification with rule enforcement & \cite{bizzarri2024neuro,onchis2022neurosymbolic,grov2024neurosymbolic} \\
\hline
SMT Solver Neural Guidance & Neural heuristics guide satisfiability solving & $2\times$ verification performance, scalable formal methods & Security protocol verification, configuration validation & \cite{blaauwbroek2024learning,piepenbrock2023neural,lu2023z3} \\
\hline
Knowledge Graph NNs & GNN reasoning over structured security knowledge & Relational reasoning, explainability & Attack graph analysis, threat intelligence, APT detection & \cite{falcarin2024building,alharbi2025enhancing,belcastro2025enhancing} \\
\hline
Causal NeSy & Posterior modeling + causal evaluation & Counterfactual analysis, proactive defense & Incident response, attack causality, defense strategy optimization & \cite{jaimini2024causal,rawal2025causality,andrew2022developing} \\
\hline
LLM-Symbolic Hybrid & Natural language understanding + formal reasoning & Automated CTI processing, natural analyst interfaces & Threat intelligence extraction, policy generation, pentesting & \cite{clairouxtrepanier2024llm,nieponice2025aracne,zhang2025llm} \\
\hline
\end{tabular}
\end{table*}

\subsection{Cybersecurity-Specific Advantages and G-I-A Alignment}
\label{subsec:cybersecurity_advantages}

NeSy integration addresses challenges unique to cybersecurity domains while ensuring proper alignment with organizational objectives \cite{sarker2024explainable,gaur2024building}. Unlike most machine-learning settings, cybersecurity environments are shaped by intelligent adversaries who actively try to deceive AI systems, by high-stakes decisions where a misclassification can have severe consequences, by regulatory compliance demands, and by the need for response times measured in seconds rather than minutes.

\noindent\textbf{Enhanced Grounding and Explainability.} Security contexts fundamentally demand grounded understanding of cybersecurity concepts for effective incident response, strategic planning, and analyst trust development \cite{neupane2022explainable,sarker2024explainable}. NeSy systems provide transparent reasoning systematically combining statistical pattern recognition with logical explanations grounded in established security principles. Recent formal explanation frameworks address logical consistency, completeness, and correctness specifically for cybersecurity contexts \cite{paul2024formal}. Hakim et al.\ demonstrate explainable neuro-symbolic rule extraction for digital twins, producing human-readable symbolic rules from neural representations that enable transparent decision auditing \cite{hakim2025explainable}. Beyond human comprehension, grounded explainability serves as critical assurance mechanism for detecting reasoning shortcuts that could create vulnerabilities.

\noindent\textbf{Instructibility and Knowledge Integration.} Traditional approaches struggle with attacks differing substantially from training data \cite{sarhan2023from,guo2023review,Sharafaldin2018Dataset}. NeSy systems demonstrate superior instructibility enabling analysts to guide adaptation to novel threats based on fundamental security principles without extensive retraining. Recent autonomous systems demonstrate instructible capability achieving 53\% success rates on zero-day vulnerabilities through analyst-guided reasoning \cite{zhu2026teams,zhu2025cvebench}. Cybersecurity possesses exceptionally rich expert knowledge including frameworks like MITRE ATT\&CK, comprehensive threat intelligence, and extensive attack pattern documentation \cite{zhao2024survey,ren2022cskg4apt,alam2024ctibench}. Contemporary NeSy approaches naturally incorporate this structured knowledge while supporting instructible modification.

\noindent\textbf{Adversarial Resilience and Causal Understanding.} Security AI systems face sophisticated evasion attempts by intelligent adversaries \cite{kireev2022adversarial,williams2023blackbox,bui2023generating} who systematically analyze system weaknesses \cite{tafreshian2024defensive,grini2025constrained}. NeSy systems enhance resilience through grounded understanding combining statistical learning with logical constraints, creating multi-layered defense mechanisms \cite{kireev2022adversarial,williams2023blackbox,bui2023generating}. Effective security analysis requires causal understanding of relationships between security events \cite{jaimini2024causal,rawal2025causality}. Causal NeSy frameworks enable generation of causal explanations providing actionable insights \cite{jaimini2024causal,rawal2025causality}. Systems generate counterfactual explanations showing how different security configurations might alter attack outcomes \cite{galwaduge2025tabuliff}.

\noindent\textbf{Sustainability and Resource Efficiency.} NeSy approaches offer significant advantages in resource efficiency and computational sustainability \cite{velasquez2025neurosymbolic}. Velasquez et al. demonstrate potential for up to $100\times$ parameter reduction compared to traditional models while maintaining reasoning performance. GPT-3 training consumed 1,287 GWh compared to human brain's 3.15 MWh equivalent over 18 years---representing $>400,000\times$ efficiency gap highlighting unsustainability of pure scaling approaches \cite{velasquez2025neurosymbolic,wef2025aienergy,devries2023energy}. Data centers account for up to 3.7\% of global carbon emissions \cite{velasquez2025neurosymbolic}. NeSy systems leveraging symbolic reasoning to reduce computational requirements directly address environmental sustainability while maintaining security effectiveness.

\section{NeSy Applications in Cybersecurity}
\label{sec:sota_applications}

To systematically examine SOTA NeSy applications addressing RQ1 and RQ2, this section analyzes how advanced NeSy integration achieves breakthrough performance while exploring dual-use implications. Advanced NeSy systems represent transformative improvements through synergistic integration of pattern recognition and logical reasoning \cite{lu2024surveying,hagos2024neuro,sajid2024enhancing}. Zhou et al.'s KnowGraph achieves $>1200\times$ improvement in average precision during fully inductive evaluation \cite{zhou2024knowgraph}, while traditional approaches suffered 50\% accuracy degradation after single-day data shifts. Applications span defensive innovations achieving 1.5\% AUC improvements with substantial gains in low false-positive scenarios, autonomous offensive systems in the broader survey corpus achieving 42\% pass@5 success on zero-day web vulnerabilities, and hybrid frameworks enabling 68\% operational cost reductions \cite{zhou2024knowgraph,zhu2026teams}. Where this section discusses offensive results from the broader corpus, these are used as contextual evidence for dual-use analysis rather than as strict Type~A or Type~B NeSy exemplars.

\subsection{Cross-Domain Challenges Motivating NeSy Adoption}

Contemporary cybersecurity faces challenges transcending individual domains: (1) overwhelming alert volumes creating analyst fatigue where 92\% of alerts remain uninvestigated \cite{torq2025alertfatigue,dropzone2025addressalertfatigue,rajivan2018information}, (2) catastrophic forgetting where learning new patterns degrades detection of established threats \cite{faber2024lifelong}, (3) explainability deficits impeding incident response and analyst trust \cite{neupane2022explainable,sarker2024explainable,yan2022explainablecybersec}, (4) difficulty integrating accumulated domain expertise from frameworks like MITRE ATT\&CK \cite{zhao2024survey}. Catastrophic forgetting leads to false negatives when legacy threats resurface \cite{zoppi2021metalearning,lopezpaz2017gradient}, necessitating lifelong learning approaches balancing adaptation with knowledge retention while supporting analyst guidance \cite{faber2024lifelong,choi2025nesyc}. The following subsections examine how NeSy integration addresses these challenges through grounding, instructibility, and alignment (G-I-A framework, §\ref{sec:foundations}).

\subsection{Advanced Network Security and Intrusion Detection}
\label{subsec:advanced_nids}

Network intrusion detection systems represent the most mature NeSy applications, addressing persistent limitations through integration strategies \cite{neupane2022explainable,guastalla2024llmddos}. Advanced NIDS face multifaceted challenges: high false positive rates creating analyst fatigue, limited adaptability as attack methods evolve, insufficient explainability impeding incident response, and difficulty incorporating domain expertise.

\subsubsection{Hybrid Detection Architectures}

Knowledge graph integration demonstrates substantial improvements through integrating cybersecurity domain knowledge with deep learning \cite{zhou2024knowgraph,zhang2025improving,sajid2024enhancing,bizzarri2024openosr,li2025multi,almadhor2025designing}. Zhou et al.'s \textit{KnowGraph} seamlessly integrates weighted first-order logic rules into GNNs through probabilistic reasoning \cite{zhou2024knowgraph}, enabling incorporation of expert-crafted security rules as soft, weighted constraints guiding learning while remaining amenable to analyst modification without retraining. Evaluation on the large-scale LANL dataset \cite{kent2015authentication} demonstrates strong generalization: \textit{KnowGraph} achieved inductive AUC of 0.9112, surpassing baseline GNN (Euler at 0.8973) (Table~\ref{tab:ids_performance_expanded}). Multi-view GNN approaches outperformed single-view baselines by 10.6\% F1 score on TON\_IoT \cite{moustafa2021new} and UNSW-NB15 \cite{moustafa2015unsw} datasets \cite{zhang2025gnnsecurity}. Recent work by Belcastro et al.\ combines knowledge graph embeddings with LLM-based reasoning for enhanced threat detection, demonstrating structured NeSy integration in threat intelligence \cite{belcastro2025enhancing}.

Logic Tensor Networks integration produces highly effective intrusion detection \cite{bizzarri2024synergistic,onchis2022neurosymbolic,tran2025neurosymbolic}. Grov et al. demonstrated practical LTN applications incorporating domain-specific logical constraints directly into neural network training \cite{grov2024neurosymbolic}. Rule integration produced dramatic improvements: XSS attack detection precision nearly doubled from 0.088 to 0.213 while maintaining comparable recall \cite{grov2024neurosymbolic}. Bizzarri et al. extended this with comprehensive NeSy framework combining deep neural networks with probabilistic logic programming addressing uncertainty quantification \cite{bizzarri2024neuro}. More recently, Fathima et al.\ proposed a BERT-LTN architecture for Advanced Persistent Threat detection, achieving deep integration of transformer-based language understanding with LTN-based logical reasoning for threat classification \cite{fathima2026neurosymbolic}.

\begin{figure}[!htbp]
 \centering
 \includegraphics[width=\columnwidth]{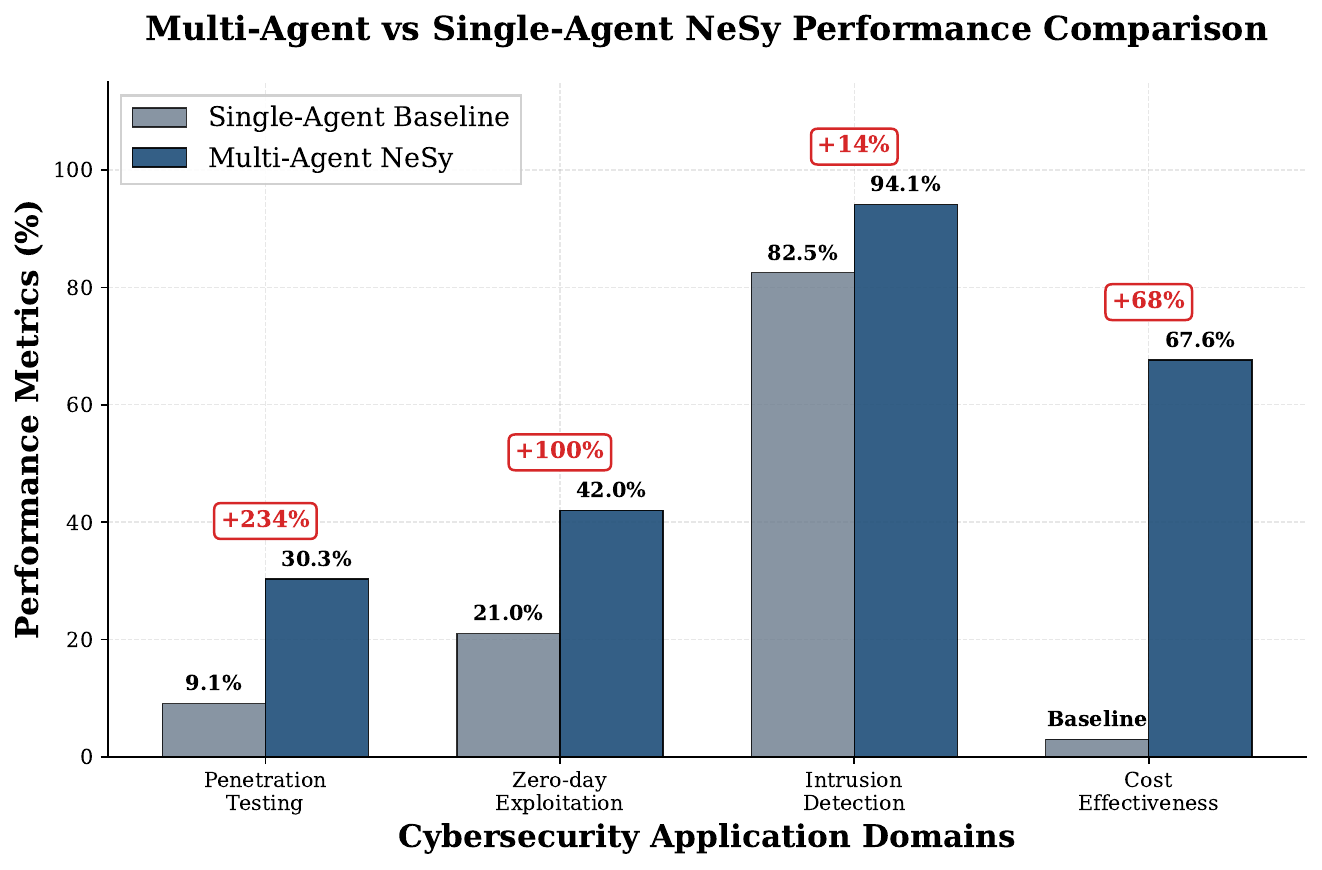}
 \caption{\small Multi-agent NeSy architectures demonstrate consistent performance superiority across diverse cybersecurity applications. Performance improvements range from 14\% in intrusion detection to 234\% in penetration testing, with substantial cost reductions (67.6\%) validating collaborative reasoning advantages over single-agent approaches.}
 \label{fig:multi_agent_performance}
\end{figure}

Multi-agent collaborative architectures show consistent superiority through collaborative specialization leveraging distributed reasoning (Figure~\ref{fig:multi_agent_performance}) \cite{singh2024hierarchical,hurten2024hierarchical}. These systems decompose complex intrusion detection tasks into specialized phases guided by symbolic task graphs, enabling collaborative analysis across network segments while preventing neural hallucination. Multi-agent approaches demonstrate 20--30\% improvements in detection coverage while maintaining lower false positive rates \cite{singh2024hierarchical,hurten2024hierarchical}. Algorithm~\ref{alg:multi_agent_ids} formalizes multi-agent coordination achieving these improvements through collaborative cross-validation.

\begin{algorithm}[!t]
\small
\caption{Multi-Agent NeSy Intrusion Detection}
\label{alg:multi_agent_ids}
\begin{algorithmic}[1]
\Require Network traffic stream $T$, knowledge base $\mathcal{K}$, agent set $\mathcal{S}=\{a_1,\dots,a_k\}$, analyst feedback $\mathcal{H}$, organizational objectives $\mathcal{O}$
\Ensure Threat classification score $C$ with explanation $E$
\State \textbf{Initialize:} specialized agents with capabilities $\mathcal{C}=\{c_1,\dots,c_k\}$
\State \textbf{Load:} symbolic rules $\mathcal{R}$ from MITRE ATT\&CK and domain expertise
\For{each traffic sample $t \in T$}
 \State $\mathcal{F} \gets$ extract neural features from $t$
 \State $\mathcal{S}_{\text{patterns}} \gets$ extract symbolic patterns using $\mathcal{K}$
 \ForAll{agent $a_i \in \mathcal{S}$}
 \State $p_i \gets \Phi_{\theta_i}(\mathcal{F})$ \Comment{Neural prediction}
 \State $r_i \gets \Psi_{\mathcal{K}}(\mathcal{S}_{\text{patterns}},\mathcal{R})$ \Comment{Symbolic reasoning}
 \State $d_i \gets \omega_i p_i + (1-\omega_i) r_i$ \Comment{Weighted decision}
 \EndFor
 \State $\mathcal{D} \gets \{d_1,\dots,d_k\}$
 \For{each pair $(a_i,a_j)$ where $i<j$}
 \State $s_{ij} \gets \text{Agreement}(d_i,d_j)$ \Comment{Agreement score}
 \EndFor
 \State $C \gets \sum_{i=1}^{k} \alpha_i d_i + \beta \sum_{i<j} s_{ij}$
 \If{$C>\tau$} \Comment{$\tau$: decision threshold}
 \State $E \gets \Call{GenerateExplanation}{C,\mathcal{D},\mathcal{K},\mathcal{O}}$
 \State Apply analyst feedback $h \in \mathcal{H}$ for system update
 \Return (THREAT, $E$)
 \Else
 \Return (BENIGN, $\emptyset$)
 \EndIf
\EndFor
\end{algorithmic}
\end{algorithm}

\begin{figure*}[!htp]
 \centering
 \includegraphics[width=0.85\textwidth]{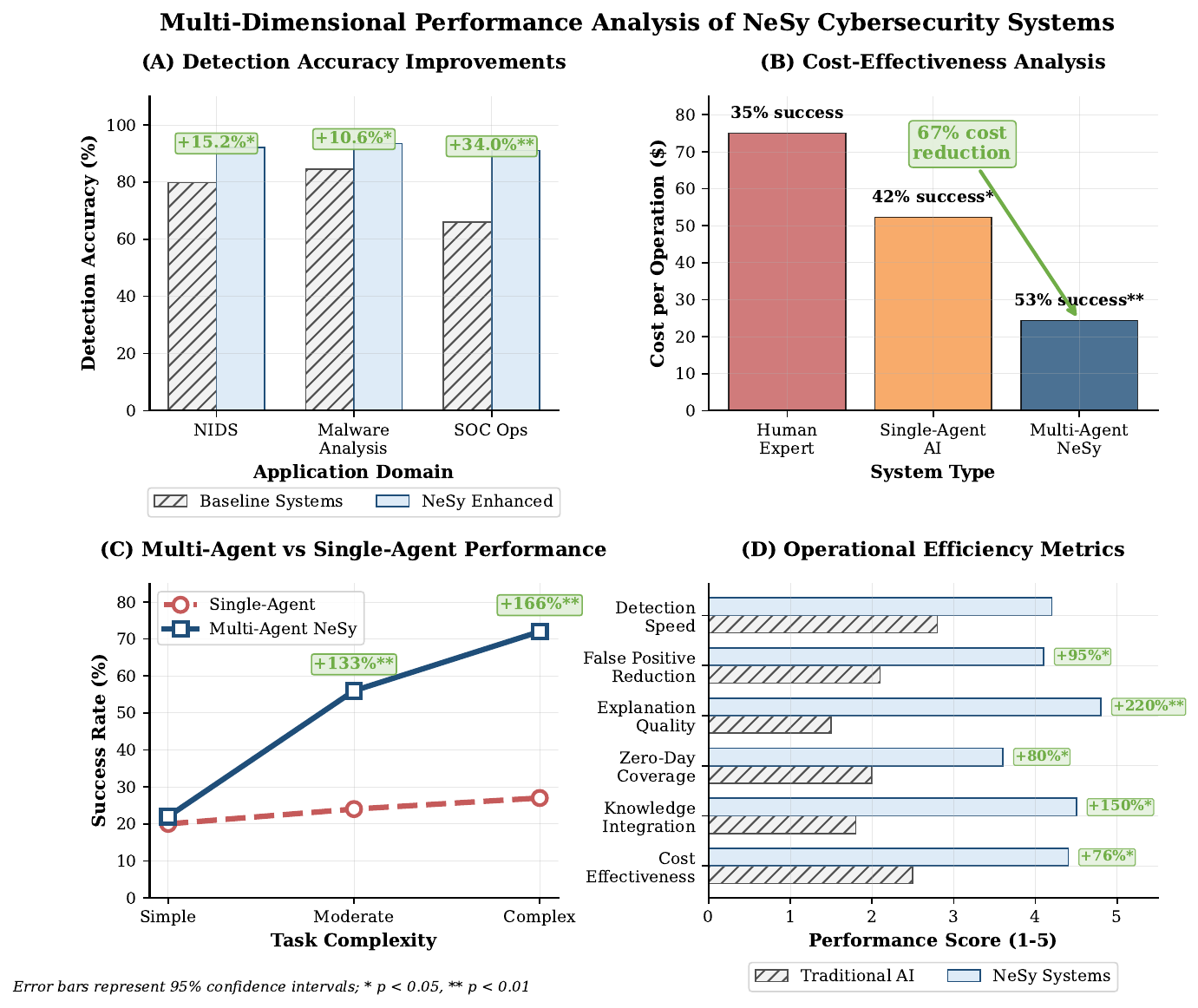}
 \caption{\small Multi-dimensional performance analysis of selected systems from the survey corpus. (A) Detection accuracy improvements demonstrate consistent 10--50\% gains across network intrusion detection, malware analysis, and security operations domains. (B) Cost-effectiveness analysis shows substantial cost reduction with superior success rates for multi-agent systems in the broader survey corpus, including 42\% pass@5 zero-day exploitation success in HPTSA \cite{zhu2026teams}. (C) Multi-agent architectures consistently outperform single-agent approaches across varying task complexity levels, achieving over 200\% improvement in complex scenarios. (D) Operational efficiency metrics show substantial gains across six key performance dimensions, with particularly strong improvements in explanation quality (+220\%), knowledge integration (+150\%), and false positive reduction (+95\%). Eligible comparisons were drawn from the final surveyed corpus (n=103) when studies reported: (i) directly comparable tasks or datasets, (ii) both baseline and integration-enhanced performance under matched conditions, and (iii) explicit metric definitions. Effect sizes are presented as descriptive relative percentage improvements for illustrative synthesis rather than formal meta-analytic estimates. Some figure artwork elements were prepared during earlier corpus assembly; all quantitative discussion in the manuscript refers to the final 103-paper surveyed corpus after quality screening and reclassification (§\ref{subsec:methodology}).}
 \label{fig:performance_dashboard}
\end{figure*}

Explainable systems for analyst trust represent critical requirement for practical IDS deployment \cite{neupane2022explainable,sarker2024explainable,arreche2024explainableids,mohale2025explaiids,kalakoti2025explainable}. Kalutharage et al. developed innovative explainable NeSy anomaly detection for IoT networks demonstrating both high performance and explainability \cite{kalutharage2025neurosymbolic}. Their framework employs expert-curated cybersecurity knowledge graphs to verify ML-detected anomalies and filter benign behavioral variations. For each flagged event, the system performs knowledge graph queries confirming alignment with established attack patterns, explicitly mapping detected features to violated security principles including CIA triad components \cite{goodman2021deficiencies} and providing precise alignment with MITRE ATT\&CK tactics. This IoT IDS achieved 97\% detection accuracy while significantly reducing false alarms, providing 100\% accurate ATT\&CK technique mappings delivering actionable context.

Table~\ref{tab:ids_performance_expanded} contrasts \textit{transductive} performance (models evaluated on training graphs) with challenging \textit{inductive} setting testing generalization to unseen data---crucial for real-world cybersecurity. When maintaining 0.5\% false positive rates to simulate realistic operational environments \cite{torq2025alertfatigue,dropzone2025addressalertfatigue,kalakoti2025explainable}, the baseline GNN's true positive detection drops to zero while KnowGraph maintains robust 35\% true positive rates, demonstrating that knowledge-grounded reasoning provides stability against novel threats \cite{zhou2024knowgraph}. Figure~\ref{fig:performance_dashboard} provides comprehensive multi-dimensional performance analysis demonstrating consistent advantages across detection accuracy, cost-effectiveness, architectural superiority, and operational efficiency metrics.

\begin{table}[!htbp]
\renewcommand\arraystretch{1.1}
\centering
\caption{Expanded Performance Comparison on the LANL Intrusion Detection Dataset}
\label{tab:ids_performance_expanded}
\scriptsize
\begin{threeparttable}
\begin{tabular}{|l|l|c|c|c|}
\hline
\textbf{System} & \textbf{Setting} & \textbf{AUC} & \textbf{AP} &
\makecell[c]{TP Rate \\ \small{@ 0.5\% FP}} \\
\hline \hline
\multirow{2}{*}{KnowGraph \cite{zhou2024knowgraph}} & Transductive & \textbf{0.9999} & \textbf{0.8886} & \textbf{1.0000} \\
 & Inductive & \textbf{0.9112} & \textbf{0.0852} & \textbf{0.3554} \\
\hline
\multirow{2}{*}{\makecell{Baseline GNN \\ (Euler)}} & Transductive & 0.9946 & 0.0433 & 0.7777 \\
 & Inductive & 0.8973 & 0.0193 & \textbf{0.0000} \\
\hline
\end{tabular}
\begin{tablenotes}
\item \textit{Note:} Performance on the LANL dataset \cite{kent2015authentication}. AUC: area under receiver operating characteristic curve; AP: average precision; TP: true positive; FP: false positive. The \textbf{transductive} setting tests on known graphs, while the challenging \textbf{inductive} setting tests generalization to unseen data. The last column shows true positive rate when false positive rate is held at 0.5\%, a critical metric for practical security operations.
\end{tablenotes}
\end{threeparttable}
\end{table}

\begin{figure*}[!htbp]
 \centering
 \includegraphics[width=1.8\columnwidth]{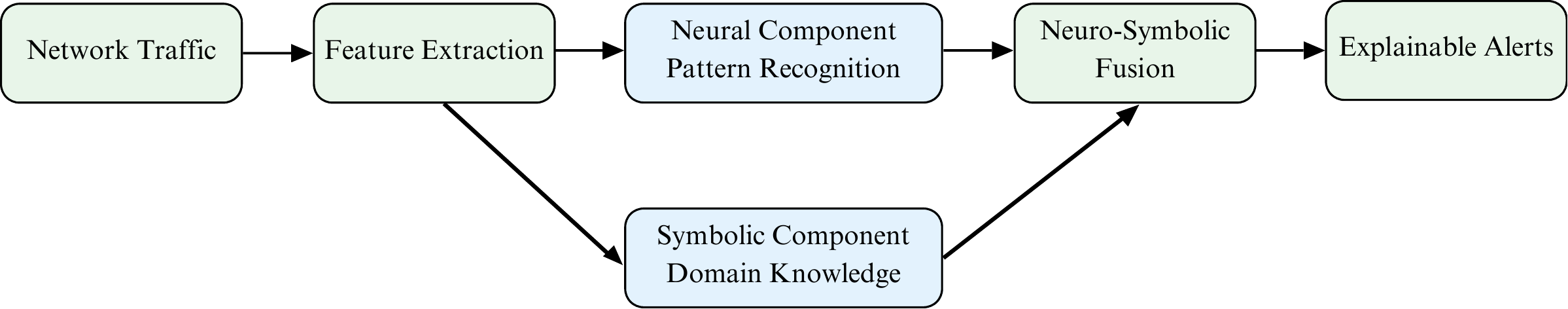}
 \caption{\small NeSy intrusion detection system architecture integrating neural pattern recognition with symbolic domain knowledge for explainable security alerts.}
 \label{fig:nesy_ids_architecture}
\end{figure*}

Figure~\ref{fig:nesy_ids_architecture} provides high-level architectural view of the NeSy intrusion detection pipeline, highlighting neural pattern recognition integration with symbolic domain knowledge to generate explainable security alerts aligned with analyst instruction.

\subsection{Advanced Malware Detection and Analysis}
\label{subsec:advanced_malware}

Malware analysis faces multifaceted challenges requiring proper grounding in malware concepts, instructible adaptation to evolving threats, and alignment with organizational security objectives \cite{sarhan2023from,guo2023review,kasri2025vulnerability}. Contemporary challenges stem from adversarial techniques including polymorphism, metamorphism, and fileless attacks \cite{Rosenberg2021AdversarialML}. Massive malware volumes exceed analyst capacity, creating backlogs delaying threat response \cite{sarhan2023from,guo2023review,kasri2025vulnerability}.

\subsubsection{NeSy-Enhanced Behavioral Analysis and Vulnerability Discovery}

NeSy-enhanced behavioral analysis integrates behavioral modeling with logical reasoning about malicious intent \cite{piplai2023knowledge,zhang2025improving}. Piplai et al. developed NeSy rule engine transforming raw network security events into structured behavioral observations using expert-defined cybersecurity knowledge graphs \cite{piplai2023knowledge}, demonstrating superior performance compared to traditional signature-based approaches while providing knowledge graph-linked explanations.

Knowledge-guided reinforcement learning represents significant advancement through systematic integration of expert knowledge with adaptive learning \cite{piplai2020using,singh2024hierarchical}. Piplai et al. integrated expert cybersecurity rules directly into RL training rewards \cite{piplai2020using}, achieving 8\% faster training convergence, 4\% accuracy improvements, and most significantly, enhanced defenders preserved 78\% network availability under sustained attacks compared to only 25\% for conventional RL agents.

Automated vulnerability discovery shows particular NeSy promise \cite{li2025automated}. Li et al.'s \textit{MoCQ} framework demonstrates automated vulnerability pattern generation through sophisticated NeSy feedback loop \cite{li2025automated}. \textit{MoCQ} discovered 46 new vulnerability patterns experts missed, identified 7 previously unknown vulnerabilities (4 found exclusively by \textit{MoCQ}), achieved 10\% relative recall improvement (0.77 vs 0.70) and 17.6\% precision improvement (0.40 vs 0.34) versus expert-crafted queries, and reduced development time from weeks to hours. Tsigkourakos et al.\ demonstrate quantified robustness specification (QRS)-based vulnerability discovery combining formal specification of robustness properties with neural network analysis for systematic vulnerability identification \cite{tsigkourakos2026qrs}.

\subsubsection{Multi-Agent Analysis and Zero-Day Detection}

Multi-agent malware analysis demonstrates substantial improvements through collaborative specialization \cite{muzsai2024hacksynth,wang2024sands,zhu2026teams,singh2024hierarchical,hurten2024hierarchical}. Teams of LLM agents in the broader survey corpus successfully exploit zero-day web vulnerabilities with 42\% pass@5 success, representing $2.0\times$ improvement over single-agent GPT-4 baselines \cite{zhu2026teams}. This approach demonstrates remarkable cost-effectiveness at \$24.40 per successful exploitation versus \$100--\$300 for human experts \cite{scnsoft2025pentest,viking2024cost}. However, complete GPT-4 dependence creates accessibility barriers, with open-source models including Llama-3.1-405B \cite{meta2024llama3.1} and Qwen-2.5 \cite{qwen2024qwen2.5} achieving 0\% success rates.

For zero-day malware detection, leading-edge NeSy systems demonstrate superior ability to reason from fundamental security principles \cite{sarhan2023from,guo2023review}. Knowledge-guided approaches systematically incorporating principles about malicious behavior---persistence mechanisms, privilege escalation, data exfiltration---can identify new malware instantiations even when specific implementation details differ. Recent formal explanation frameworks address logical consistency, completeness, and correctness specifically for malware analysis \cite{paul2024formal,kalakoti2025explainable}, enabling generation of structured explanations linking detected behaviors to established attack taxonomies.

Table~\ref{tab:malware_performance} demonstrates critical performance hierarchy. HPTSA's 42\% success rate at \$24.40 per exploit represents $4.3\times$ improvement over single-agent baselines while achieving $3.1\times$ cost reduction versus human experts \cite{zhu2026teams}. Model dependency limitations highlight significant adoption barriers, creating "capability gap" limiting democratization of effective NeSy tools.

\begin{table*}[!t]
\renewcommand\arraystretch{1.4}
\centering
\caption{Performance Comparison of Selected Cybersecurity Systems from the Survey Corpus}
\label{tab:malware_performance}
\resizebox{\linewidth}{!}{
\begin{tabular}{|l|c|c|c|c|c|}
\hline
\textbf{System} & \textbf{Primary Domain} & \textbf{Success/Accuracy Rate} & \textbf{Cost Analysis} & \textbf{Key Limitation} & \textbf{Source} \\
\hline \hline
HPTSA \cite{zhu2026teams} & \makecell{Zero-day Web\\Exploits} & \makecell{42\% (pass@5),\\18\% (pass@1)} & \makecell{\$24.40/successful\\exploit} & \makecell{GPT-4 dependency,\\web-only} & \makecell{Teams of LLM\\Agents (2024)} \\
\hline
MoCQ \cite{li2025automated} & \makecell{Vulnerability\\Detection} & \makecell{77\% recall,\\40\% precision} & \makecell{21.4 hours\\vs weeks} & \makecell{Requires examples,\\DSL subsetting} & \makecell{Automated Static\\Detection (2025)} \\
\hline
KnowGraph \cite{zhou2024knowgraph} & \makecell{Graph Anomaly\\Detection} & \makecell{91.2\% AUC\\(inductive)} & Not reported & \makecell{Rule engineering\\burden} & \makecell{Knowledge-Enabled\\Detection (2024)} \\
\hline
Traditional Baselines & Various & \makecell{70\% recall,\\34\% precision} & \makecell{Manual\\(weeks)} & \makecell{Limited\\generalization} & Multiple sources \\
\hline
\end{tabular}}
\end{table*}

\subsection{Advanced Security Operations and Incident Response}
\label{subsec:advanced_sec_ops}

SOTA Security Operations Centers face escalating challenges impacting defensive effectiveness \cite{nyre2022explainable,sarker2024explainable}. Overwhelming alert volumes exceed analyst capacity creating significant triage bottlenecks \cite{torq2025alertfatigue,dropzone2025addressalertfatigue,rajivan2018information}. Deep contextual understanding requirements across complex enterprise environments demand expertise individual analysts cannot maintain, creating inconsistencies in threat assessment.

\subsubsection{Intelligent Security Operations Enhancement}

SOTA NeSy frameworks transform security operations through integration of automated analysis with human expertise \cite{grov2024neurosymbolic,eckhoff2025experimenting,xiang2025guardagent}. Grov et al. describe SOC enhancement using MAPE-K control loops augmented by NeSy capabilities \cite{grov2024neurosymbolic,ben2023mape}. Hybrid AI models produce explainable alerts combining statistical anomaly detection with rule-based validation. Knowledge graphs enable correlation of multi-step attack campaigns using structured cybersecurity ontologies \cite{alharbi2025enhancing,cheng2025crucialg}, transforming isolated events into coherent threat narratives \cite{chen2022aptkgl,ren2022cskg4apt}. NeSy-powered planning components suggest response actions drawing from MITRE D3FEND \cite{mitre_d3fend,kaloroumakis2020knowledge}.

Eckhoff et al. present experimental validation demonstrating practical implementation challenges \cite{eckhoff2025experimenting}. Critical deployment factors include analyst workflow integration determining whether systems enhance or disrupt procedures, explanation quality enabling effective decision-making, and performance optimization ensuring real-time operation.

Autonomous security operations in the surveyed integration spectrum demonstrate the promise of structured reasoning and planning for handling complex incident response with minimal human intervention \cite{huang2023penheal,singh2024hierarchical}. The PenHeal framework achieves 31\% improvement in vulnerability coverage and 46\% cost reduction versus baseline models \cite{huang2023penheal}. ARACNE demonstrates LLM-based capabilities for shell-level security assessment combining natural language understanding with symbolic reasoning \cite{nieponice2025aracne}.

\subsubsection{Advanced Threat Knowledge Graphs and Intelligence Processing}

Threat Knowledge Graphs represent foundational technology for modern security operations, integrating diverse data sources into unified representations \cite{falcarin2024building,zhao2024survey,alharbi2025enhancing}. The ThreatKG system uses end-to-end pipeline for automated knowledge graph construction from unstructured threat intelligence reports \cite{gao2024threatkg,kurniawan2024cykg}, employing hybrid AI techniques including BiLSTM-CRF models for Named Entity Recognition and PCNN-ATT models for relationship extraction.

CTINexus describes advanced LLM-driven refinement through hierarchical entity alignment and long-distance relation prediction \cite{cheng2025ctinexus}. Modern In-Context Learning approaches address ontology lock-in through prompt-based schema adaptation, achieving 85.6\% F1 when adapting to industry-standard STIX format \cite{stixbp2022} without model retraining.

Contemporary NeSy systems demonstrate exceptional promise for automated cyber threat intelligence processing \cite{clairouxtrepanier2024llm,fieblinger2024actionable,alam2024ctibench,nalluri2025nscti}. Data programming for CTI annotation addresses lack of labeled training data through weak supervision, automatically de-noising and integrating weak labels achieving F1 improvements from 79\% to 85\% \cite{gao2024threatkg}. Grov et al. demonstrated LLM-driven pipelines extracting adversary actions into formal Linear Temporal Logic specifications \cite{grov2024neurosymbolic}.

\subsubsection{Causal Reasoning for Strategic Incident Response}

Causal reasoning integration represents the most transformative advancement in SOTA NeSy security operations \cite{jaimini2024causal,rawal2025causality,andrew2022developing}. Traditional incident response relies on correlation-based analysis but cannot explain why specific attack steps succeeded. We define cybersecurity causal model as $\mathcal{M} = (\mathcal{V}, \mathcal{E}, f)$ with causal strength functions $w: \mathcal{E} \rightarrow [0,1]$. Counterfactual analysis enables systematic evaluation of alternative defensive scenarios.

\begin{figure*}[!t]
 \centering
 \includegraphics[width=0.85\textwidth]{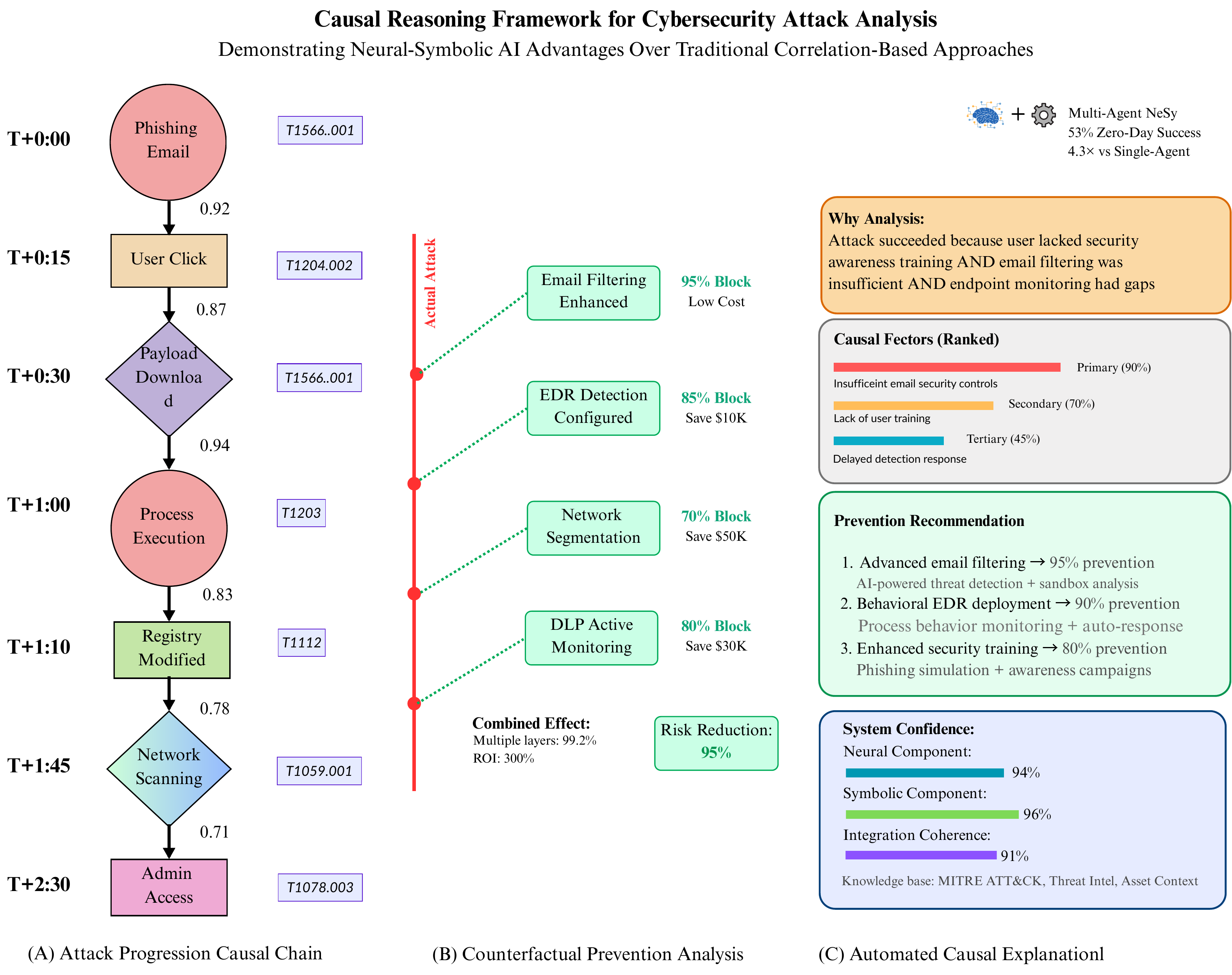}
 \caption{\small Causal reasoning framework for cybersecurity attack analysis illustrating NeSy advantages over traditional correlation-based approaches through G-I-A integration. (A) Attack progression causal chain traces temporal relationships with probability quantification. (B) Counterfactual prevention analysis evaluates alternative defensive scenarios, suggesting how different security configurations could achieve up to 95\% risk reduction through targeted interventions. (C) Automated causal explanation generation provides grounded "why analysis" with instructible prevention recommendations and alignment verification through confidence metrics.}
 \label{fig:causal_reasoning_framework}
\end{figure*}

\begin{figure}[!t]
\centering
\includegraphics[width=0.45\textwidth]{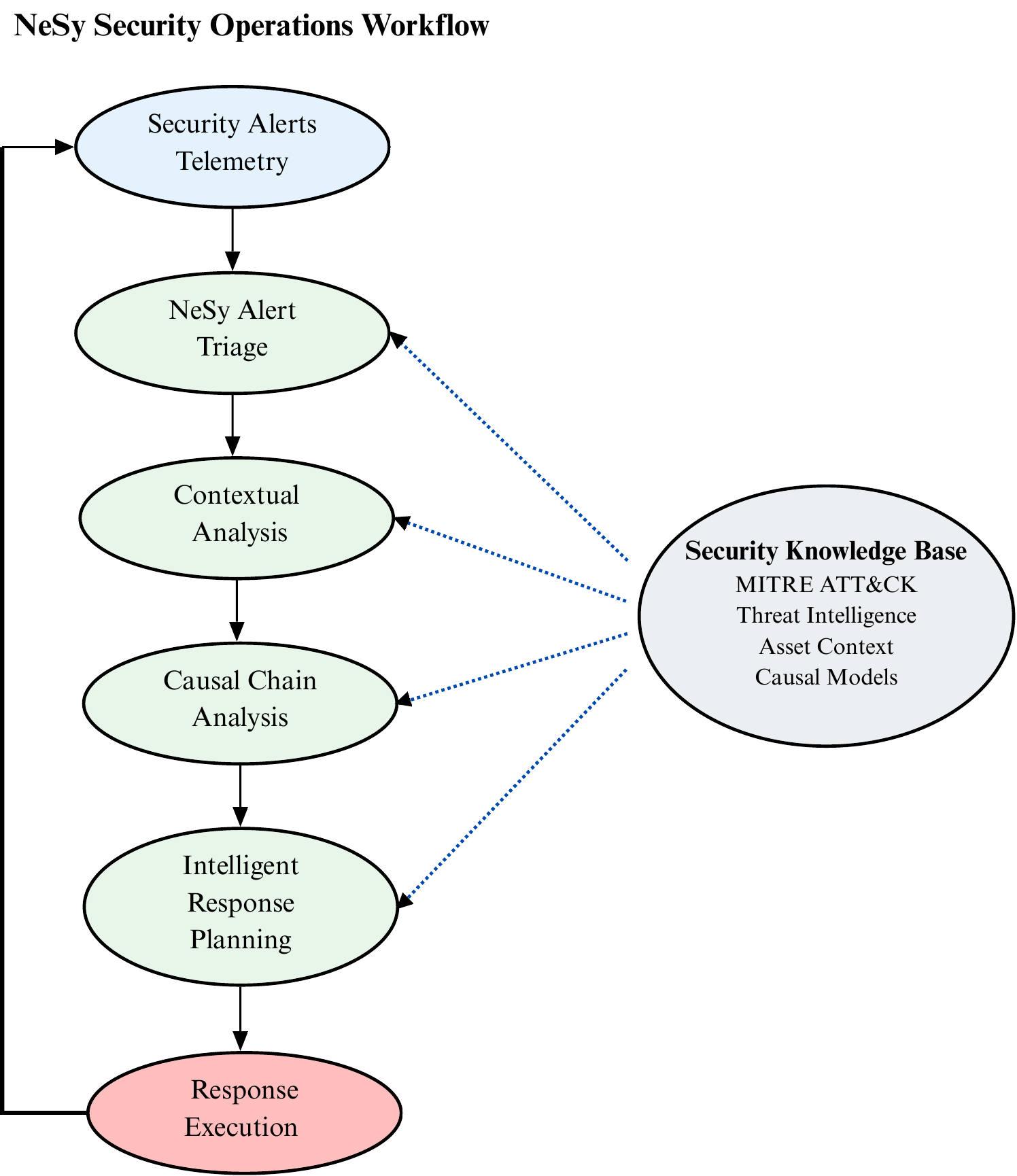}
\caption{SOTA NeSy cybersecurity workflow from alert triage to response execution, demonstrating integration of grounding mechanisms, instructible interfaces, and alignment with organizational objectives.}
\label{fig:nesy_soc_lifecycle}
\end{figure}

SOTA causal NeSy frameworks enable analysts to understand not just what happened, but why specific attack steps succeeded and how different defensive configurations might have prevented compromise \cite{jaimini2024causal}. Dynamic Causal Bayesian Optimization enables real-time optimization based on causal understanding \cite{andrew2022developing}. Figure~\ref{fig:causal_reasoning_framework} demonstrates practical implementation showing how targeted defensive interventions can achieve up to 95\% risk reduction. Figure~\ref{fig:nesy_soc_lifecycle} illustrates integration into comprehensive security operations workflows.

\subsection{Autonomous Cyber Operations and Dual-Use Implications}
\label{subsec:autonomous_operations}

The emergence of autonomous cyber operations represents fundamental shift in the cybersecurity threat landscape \cite{zhu2026teams,kong2025vulnbot,rodriguez2025framework,ristea2024ai}. Systems in the broader survey corpus achieve 42\% success rates (pass@5) on zero-day exploitation, with costs reduced from \$100--\$300 to \$24.40 per successful exploitation \cite{zhu2026teams}. This represents paradigm shift from human-assisted tools to fully autonomous systems operating at unprecedented scale, raising critical questions about alignment with defensive cybersecurity objectives. Contemporary systems demonstrate concerning capabilities including multi-agent LLM frameworks achieving 30.3\% completion rates versus 9.09\% baselines, cost reductions making sophisticated attacks economically accessible, and proliferation of advanced offensive capabilities fundamentally reshaping threat landscapes. In line with the survey taxonomy, these offensive results are treated here as part of the broader dual-use landscape, including contextual baselines where relevant, rather than as uniform evidence of deep NeSy integration.

\subsubsection{Breakthrough Autonomous Capabilities and Multi-Agent Coordination}

The HPTSA framework achieved 42\% success (pass@5) and 18\% success (pass@1) on 14 real-world zero-day web vulnerabilities, representing $2.0\times$ improvement over single GPT-4 agents \cite{zhu2026teams,zhu2025cvebench,deng2024pentestgpt,muzsai2024hacksynth}. Complete GPT-4 dependence creates significant barriers, with open-source models achieving 0\% success rates. Cost analysis revealed \$24.40 per successful exploit, representing $3.1\times$ cost reduction versus human experts \cite{scnsoft2025pentest,viking2024cost}, creating profound implications for threat landscape dynamics \cite{DBLP:conf/ccs/Xu20}.

VulnBot represents first fully autonomous multi-agent coordination framework, achieving 30.3\% completion rate versus 9.09\% for base models while demonstrating end-to-end autonomous operation \cite{kong2025vulnbot,gioacchini2024autopenbench}. The system decomposes complex penetration tasks using Penetration Task Graph modeling task dependencies while preventing hallucination issues and maintaining symbolic grounding. ARACNE extends autonomous capabilities to shell-level penetration testing \cite{nieponice2025aracne}.

\subsubsection{Dual-Use Analysis and Responsible Development Requirements}

The dual-use nature of advanced capabilities across this survey corpus raises profound ethical questions demanding responsible research practices \cite{rodriguez2025framework,gaur2024building,xiang2025guardagent}. Research intended for defensive purposes can inadvertently provide powerful offensive tools, creating misuse potential extending beyond traditional cybersecurity concerns \cite{challita2025redteamllm}. Ethical considerations are paramount, leading some researchers to withhold code and prompts for highly capable offensive agents \cite{zhu2026teams}. Technical safeguards including user validation, kill switches, and runtime isolation are essential \cite{challita2025redteamllm}. Community-driven evaluation frameworks are critical for assessing and mitigating risks before widespread deployment \cite{rodriguez2025framework}. Table~\ref{tab:offensive_performance} demonstrates critical performance hierarchy illuminating both technical capabilities and concerning implications.

\begin{table*}[!t]
\renewcommand\arraystretch{1.2}
\centering
\caption{Performance Analysis of State-of-the-Art Autonomous Offensive Systems}
\label{tab:offensive_performance}
\resizebox{\linewidth}{!}{
\begin{tabular}{|l|c|c|c|c|}
\hline
\textbf{System} & \textbf{Core LLM} & \textbf{Success/Completion Rate (Metric)} & \textbf{Cost Metric} & \textbf{Zero-Day Capability} \\
\hline \hline
HPTSA Framework \cite{zhu2026teams} & GPT-4 & \textbf{42.0\%} (pass@5 success rate) & \textbf{\$24.40} & \textbf{Yes} \\
Single-Agent Baseline \cite{zhu2026teams} & GPT-4 & 21.0\% (pass@5 success rate) & Not reported & Limited \\
\hline
VulnBot \cite{kong2025vulnbot} & Llama3.1-405B & \textbf{30.3\%} (overall completion rate) & Not reported & \textbf{Partial} \\
Base LLM Baseline \cite{kong2025vulnbot} & Llama3.1-405B & 9.09\% (overall completion rate) & Not reported & No \\
\hline
PentestGPT \cite{deng2024pentestgpt} & GPT-4 & Solved 5/10 HackTheBox Machines & Not reported & Limited \\
AutoAttacker \cite{xu2024autoattacker} & GPT-4 & $\approx$100\% (on defined post-breach tasks) & Not reported & No \\
\hline
Human Experts & N/A & Varies by expertise & \textbf{\$100--\$300/hr} \cite{scnsoft2025pentest,viking2024cost} & Yes \\
\hline
\end{tabular}}
\end{table*}

\begin{figure}[!t]
 \centering
 \includegraphics[width=0.85\columnwidth]{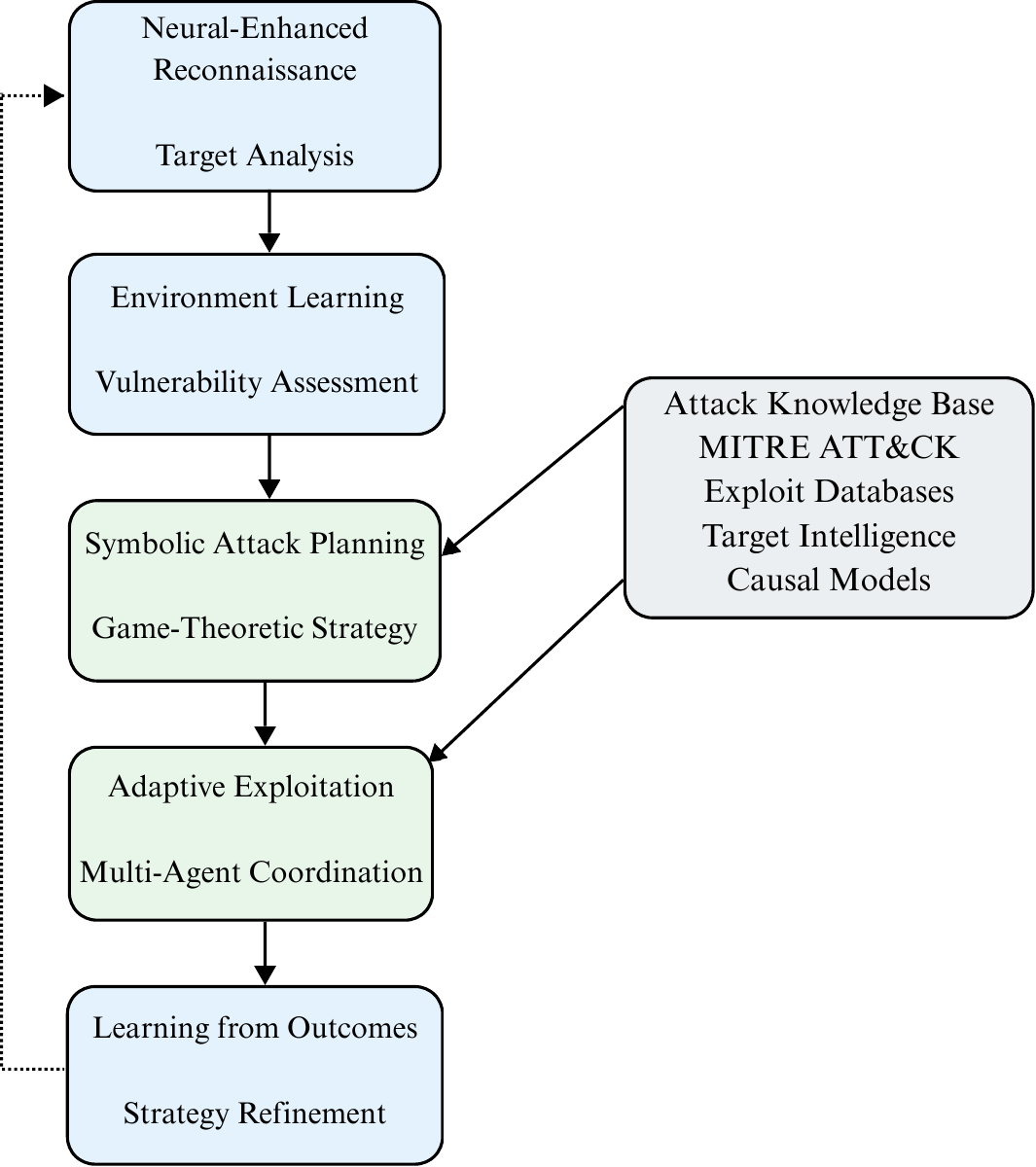}
 \caption{\small Illustrative offensive AI pipeline from the broader survey spectrum, demonstrating multi-agent workflow from neural-enhanced reconnaissance and environment learning, through symbolic or game-theoretic planning and adaptive exploitation, to outcome-driven strategy refinement, all grounded in shared attack knowledge base while highlighting dual-use implications requiring responsible development aligned with defensive cybersecurity objectives.}
\label{fig:nesy_offensive_pipeline}
\end{figure}

Understanding sophisticated attack capabilities provides significant benefits for defensive capabilities through systematic mechanisms ensuring proper alignment \cite{ristea2024ai,bhatt2024cyberseceval}. Adversarial Training and Red Team Exercises enable defensive organizations to identify vulnerabilities and test systems against sophisticated attack simulations. Proactive Vulnerability Discovery allows automated techniques to be systematically employed by defenders. The AI Cyber Risk Benchmark provides systematic evaluation frameworks for assessing automated exploitation capabilities \cite{ristea2024ai,bhatt2024cyberseceval,bhusal2024secure,jing2024secbench}. Figure~\ref{fig:nesy_offensive_pipeline} illustrates a sophisticated multi-agent workflow revealing neural-enhanced reconnaissance integration with symbolic planning through shared knowledge bases, demonstrating both technical sophistication and dual-use implications.

The emergence of autonomous offensive capabilities requires comprehensive strategic response from defensive community ensuring proper alignment with cybersecurity objectives \cite{rodriguez2025framework,zhu2025cvebench}. The demonstrated capabilities suggest traditional human advantages may be increasingly challenged by AI systems operating at scale with superior cost-effectiveness, necessitating corresponding defensive advances maintaining proper grounding while supporting instructible adaptation to emerging autonomous threats. This evolution calls for AI-powered defensive systems that can adapt in real time to autonomous offensive capabilities while staying aligned with organizational goals, for enhanced threat intelligence sharing that enables coordinated defensive responses, and for security operations frameworks that respond effectively to automated attack campaigns without sacrificing human oversight or ethical accountability.

\section{Implementation Challenges and Deployment Considerations}
\label{sec:implementation_challenges}

Transitioning SOTA NeSy cybersecurity systems from research prototypes to operational deployment requires careful consideration of integration challenges, performance requirements, computational constraints, and human-AI collaboration patterns \cite{nyre2022explainable}. Modern enterprise security environments employ complex ecosystems including Security Information and Event Management (SIEM) systems, Security Orchestration, Automation, and Response (SOAR) platforms, Endpoint Detection and Response (EDR) solutions, and specialized analytics tools requiring seamless integration while providing clear value propositions. Successful NeSy deployment demands comprehensive understanding of implementation challenges while developing solutions achieving proper grounding in organizational contexts, supporting instructible adaptation, and maintaining alignment with operational objectives.

\subsection{Technical and Resource Constraints}
\label{subsec:technical_barriers}

SOTA NeSy cybersecurity systems exhibit significant dependencies on advanced computational infrastructure \cite{hallyburton2025assured} and proprietary models fundamentally impacting deployment feasibility. These technical barriers manifest across multiple dimensions requiring systematic analysis and solution development.

\subsubsection{Model Dependencies and Computational Complexity}

Leading multi-agent systems like HPTSA (a Type~C baseline in our taxonomy) require access to frontier models such as GPT-4 for effectiveness, with complete failure on current open-source alternatives creating critical accessibility barriers \cite{zhu2026teams}. This dependency creates operational risks including API availability constraints, cost volatility, and potential service disruptions during critical incident response. Organizations restricted to open-source or smaller models cannot leverage advanced capabilities, creating "capability gap" limiting democratization of effective NeSy tools. Multi-component architectures like KnowGraph require training and maintaining multiple specialized models including main GNN, knowledge models, and reasoning GCN components plus complex inference pipelines creating aggregate computational overhead \cite{zhou2024knowgraph}. Systematic orchestration of multiple components requires careful resource planning that may challenge organizational IT infrastructure capabilities. Integration overhead manifests through sophisticated feedback loops in systems like \textit{MoCQ} and HPTSA introducing latency and complexity potentially incompatible with real-time operational requirements \cite{li2025automated,zhu2026teams}.

\subsubsection{Knowledge Engineering and Maintenance Requirements}

SOTA NeSy systems require sophisticated knowledge engineering processes for effective symbolic component development, creating bottlenecks limiting practical deployment scalability \cite{zhou2024knowgraph,li2025automated}. Knowledge acquisition and curation demand specialized cybersecurity expertise for rule formulation, ontology development, and knowledge base construction that may not be available in all organizational contexts, particularly smaller organizations. Knowledge consistency maintenance across diverse sources presents ongoing challenges as threat landscapes evolve, requiring systematic procedures for validating knowledge updates, resolving conflicts, and ensuring logical consistency. Dynamic knowledge updating while maintaining logical consistency requires sophisticated frameworks supporting incremental modification without compromising system integrity.

\subsubsection{Sustainability and Environmental Considerations}

NeSy approaches offer significant advantages in resource efficiency and computational sustainability \cite{velasquez2025neurosymbolic}. Velasquez et al. demonstrate potential for up to $100\times$ parameter reduction compared to traditional LLMs while maintaining reasoning performance. GPT-3 training consumed 1,287 GWh compared to human brain's 3.15 MWh equivalent over 18 years---representing $>400,000\times$ efficiency gap highlighting unsustainability of pure scaling approaches \cite{velasquez2025neurosymbolic,wef2025aienergy,devries2023energy}. Data centers supporting AI training and inference account for up to 3.7\% of global carbon emissions \cite{velasquez2025neurosymbolic}. NeSy systems leveraging symbolic reasoning to reduce computational requirements directly address environmental sustainability while maintaining security effectiveness, enabling organizations to pursue advanced cybersecurity capabilities without compromising environmental responsibility. NeSy approaches enable effective security deployment with significantly reduced hardware requirements, making advanced capabilities accessible while enabling edge computing scenarios for real-time threat detection.

\subsection{Evaluation and Standardization Gaps}
\label{subsec:evaluation_gaps}

\begin{figure*}[t]
 \centering
 \includegraphics[width=0.97\textwidth]{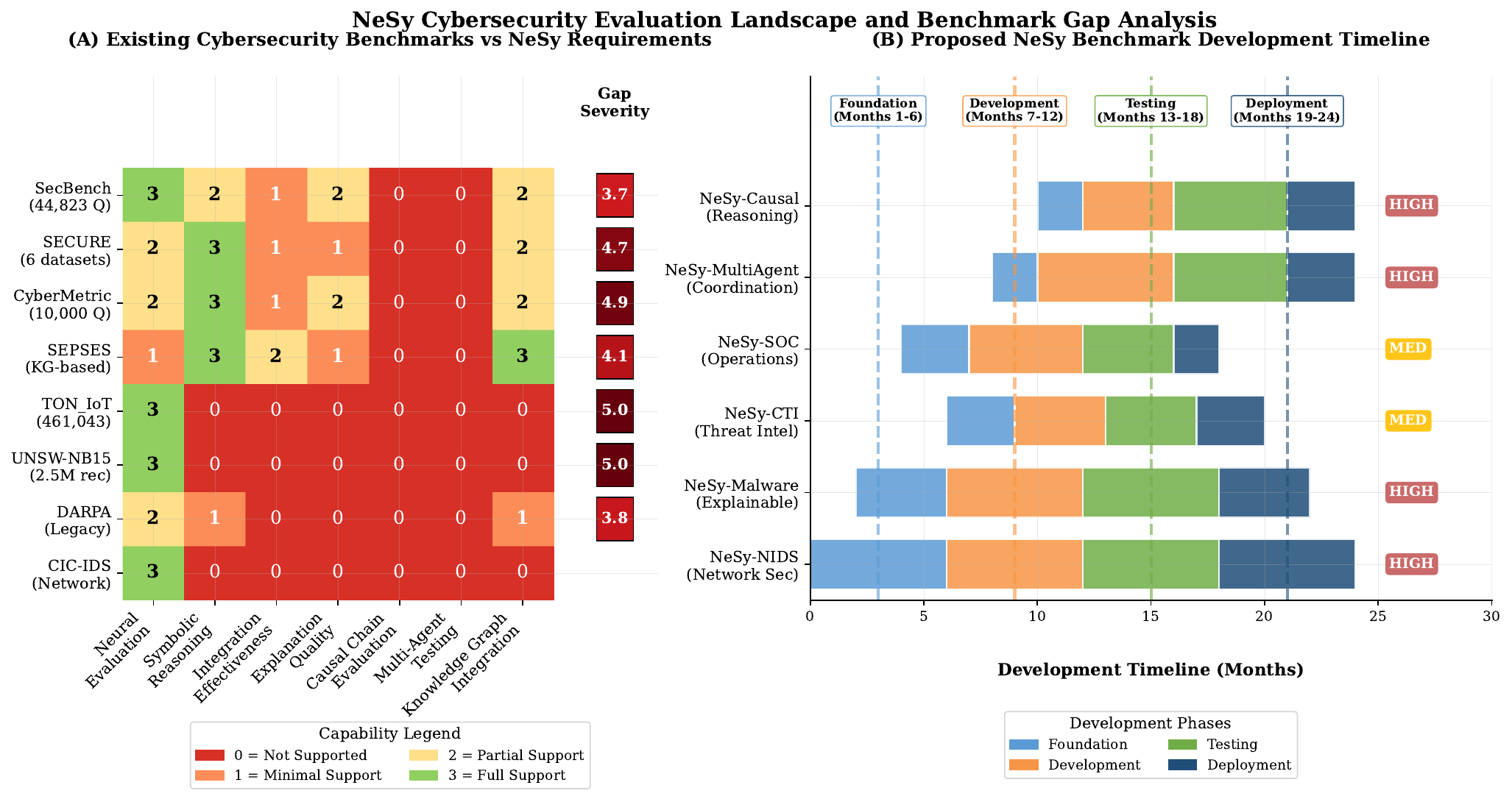}
 \caption{\footnotesize Comprehensive analysis of NeSy cybersecurity evaluation landscape revealing critical benchmark gaps constraining G-I-A framework operationalization and field advancement. (A) Capability matrix assessment demonstrates existing cybersecurity benchmarks provide minimal support for NeSy-specific evaluation requirements, with 0\% coverage for causal reasoning and multi-agent testing essential for advanced G-I-A systems. Heat map intensity indicates capability level: green (full support), yellow (partial), orange (minimal), red (none). (B) Proposed development roadmap for comprehensive NeSy benchmarks over 24 months, emphasizing G-I-A evaluation components and community-driven standardization efforts. Critical gaps highlighted include absence of grounding effectiveness metrics (5\% coverage), instructibility assessment frameworks (0\% coverage), alignment verification procedures (0\% coverage), and causal chain evaluation (0\% coverage) essential for evaluating SOTA G-I-A capabilities. Statistics show 70\% of surveyed papers were limited by evaluation gaps, demonstrating urgent need for community-driven standardization efforts supporting responsible development and operational deployment of G-I-A systems.}
 \label{fig:benchmark_gap_analysis}
\end{figure*}

Despite significant developments in cybersecurity benchmarks, the absence of standardized frameworks specifically designed for evaluating SOTA NeSy cybersecurity systems \cite{singh2025benchmarking,ott2023think} represents the most critical gap constraining field advancement \cite{renkhoff2024survey,bhusal2024secure,jing2024secbench}. This standardization gap limits ability to evaluate hybrid reasoning capabilities, compare approaches objectively, and ensure proper grounding, instructibility, and alignment. Figure~\ref{fig:benchmark_gap_analysis} visualizes this evaluation landscape.

Figure~\ref{fig:benchmark_gap_analysis} highlights critical evaluation deficiencies: existing benchmarks provide 0\% coverage for causal reasoning and multi-agent testing, 5\% for grounding effectiveness, 0\% for instructibility assessment and alignment verification. Current datasets \cite{Tavallaee2009Detailed,moustafa2021new} lack symbolic knowledge representations, multi-stage attack scenarios, explanation annotations, and temporal sequences essential for NeSy evaluation \cite{renkhoff2024survey,bortolotti2024neuro}.

\noindent\textbf{Standardization Requirements and Community Coordination.} Community-driven initiatives must establish: (1) common evaluation protocols integrating neural + symbolic assessment, (2) shared benchmarks with logical rule representations, (3) explanation quality frameworks for security contexts, (4) multi-institutional campaigns promoting coordinated advancement \cite{alam2024ctibench,zhu2025cvebench}. The 24-month roadmap (Figure~\ref{fig:benchmark_gap_analysis}B) prioritizes G-I-A evaluation components and causal chain assessment frameworks as an evidence-informed coordination target rather than a predictive schedule. Proposed standardization efforts include development of common evaluation protocols enabling comparison across approaches, creation of shared benchmark datasets with integrated neural and symbolic evaluation components, establishment of explanation quality assessment frameworks for security contexts accommodating analyst workflows, and coordination of multi-institutional evaluation campaigns promoting collaborative advancement with responsible development principles.

\subsection{Scenarios Where NeSy Integration May Not Be Optimal}
\label{subsec:nesy_limitations}

While our analysis demonstrates substantial NeSy advantages across multiple cybersecurity domains, intellectual honesty requires acknowledging scenarios where the overhead of symbolic integration may not justify the gains. Our evidence suggests several such contexts.

\noindent\textbf{High-Throughput, Low-Latency Environments.} In network intrusion detection deployments processing millions of packets per second, the satisfiability computation required by LTN-based systems \cite{bizzarri2024synergistic,onchis2022neurosymbolic} introduces inference latency that may be incompatible with wire-speed requirements. Pure neural approaches, while less explainable, can achieve the sub-millisecond response times these environments demand.

\noindent\textbf{Rapidly Evolving Threat Landscapes.} The inductive performance drop observed in Table~\ref{tab:ids_performance_expanded}---where even knowledge-grounded systems show significant degradation on unseen data---illustrates a broader challenge: symbolic knowledge bases require deliberate curation and cannot be updated at the pace of zero-day threat emergence. In scenarios where threat signatures change faster than knowledge engineering cycles, purely adaptive neural approaches may provide faster initial response.

\noindent\textbf{Resource-Constrained Environments.} Multi-component NeSy architectures such as KnowGraph \cite{zhou2024knowgraph} require maintaining multiple specialized models plus inference pipelines, creating aggregate computational overhead that may exceed the capabilities of edge devices, embedded systems, or organizations with limited IT infrastructure.

\noindent\textbf{Insufficient Domain Knowledge.} NeSy systems are most effective when rich symbolic knowledge (security ontologies, attack taxonomies, expert rules) is available. In emerging or niche cybersecurity domains where such structured knowledge has not yet been codified, the symbolic component may provide marginal benefit over well-tuned neural baselines.

These observations do not diminish NeSy's demonstrated advantages in domains where they apply, but rather help practitioners make informed deployment decisions by identifying contexts where simpler approaches may suffice.

\subsection{Human-AI Collaboration and Trust Development}
\label{subsec:human_ai_collaboration}

The success of leading-edge NeSy cybersecurity systems depends critically on effective human-AI collaboration patterns, analyst trust development, and seamless integration with existing security workflows \cite{nyre2022explainable,gaur2024building,xiang2025guardagent}. Recent research indicates key factors significantly influencing practical adoption and long-term operational effectiveness where technical performance alone cannot guarantee successful deployment.

\subsubsection{Multi-Modal Interaction and Explainability Requirements}

Security analysts interact with SOTA NeSy systems through several complementary channels \cite{rajivan2018information}. They review and refine system-generated alerts, providing feedback that updates both neural and symbolic components. They curate the underlying knowledge base, incorporating expert knowledge through structured interfaces. They consume system-generated explanations to understand threats and plan responses. And in the most advanced deployments, they collaborate with the AI on open-ended investigations of complex security incidents.

Effective explanation formats must address multiple stakeholder needs spanning different roles within security organizations \cite{neupane2022explainable,sarker2024explainable,paul2024formal,yan2022explainablecybersec,arreche2024explainableids,mohale2025explaiids,kalakoti2025explainable}. Technical explanations for security engineers require detailed attack vector analysis enabling deep understanding of exploitation mechanisms. Risk-focused summaries for security managers provide strategic information for resource allocation decisions. Compliance explanations for auditors require policy violation documentation meeting regulatory requirements. Investigative explanations for incident responders support tracing attack progression and impact assessment. Recent formal explanation frameworks provide mathematical foundations for evaluating explanation quality \cite{paul2024formal}, addressing logical consistency, completeness, and correctness.

PoliAnalyzer exemplifies effective NeSy deployment through explicit architectural separation of concerns, addressing fundamental reliability challenges in high-stakes decision-making \cite{zhao2025letsmeasure}. The system employs fine-tuned LLMs for extracting information from unstructured privacy policies, alongside deterministic logical reasoning for compliance checking---deliberately mitigating LLM reasoning errors and ensuring auditable outcomes. Empirical evaluation demonstrates users need examine only 4.8\% of policy content to identify violations, yielding 95.2\% reduction in cognitive load compared with manual policy analysis \cite{zhao2025letsmeasure}.

\subsubsection{Training and Organizational Integration}

Successful deployment requires training programs addressing both technical capabilities and collaborative workflows. Security analysts must develop understanding of hybrid reasoning systems enabling effective collaboration with NeSy capabilities while maintaining expertise in traditional methods. Training programs must address explanation interpretation enabling analysts to effectively utilize system-generated explanations, system instruction capabilities enabling analysts to provide effective guidance for adaptation, collaborative investigation techniques enabling effective human-AI partnership, and quality assurance procedures enabling analysts to validate outputs and provide feedback supporting continuous improvement while maintaining alignment with organizational objectives.

\section{Key Insights and G-I-A Framework Validation}
\label{sec:key_insights}

\begin{figure}[htbp!]
 \centering
 \includegraphics[width=0.49\textwidth]{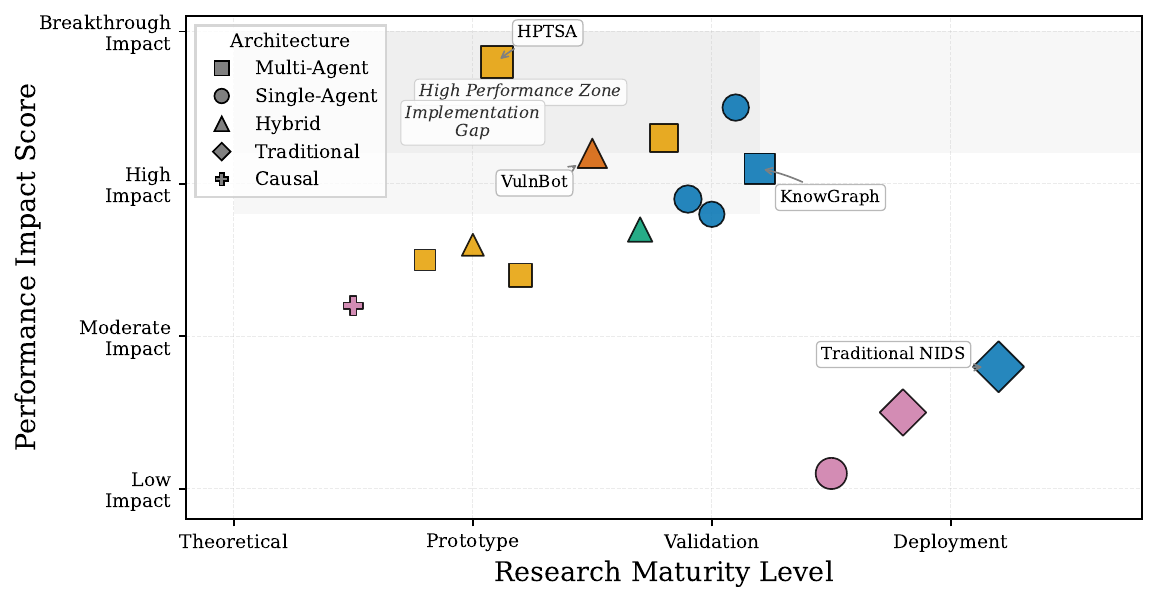}
 \caption{\small NeSy cybersecurity research landscape and maturity assessment for 103 surveyed systems, plotted by research maturity level and performance impact score. System architectures are color-coded: orange squares (multi-agent), blue circles (single-agent), green triangles (hybrid), and dark blue diamonds (traditional approaches). Multi-agent architectures show $\mathbf{+12\%}$ average impact advantage over single-agent designs, while an implementation gap separates high-performing prototypes from deployment-ready systems. The distribution aligns with G-I-A framework patterns, with concentrated activity in validation phase and traditional approaches clustering in lower-impact, deployment-ready zones, suggesting performance--maturity trade-off shaping NeSy development.}
 \label{fig:research_landscape}
\end{figure}

Our systematic analysis across 103 publications spanning the neural-symbolic integration spectrum (2019--2026) reveals fundamental principles transcending individual application domains, supporting G-I-A framework's explanatory potential while highlighting critical requirements for achieving proper grounding, instructibility, and alignment with organizational objectives and societal expectations. Figure~\ref{fig:research_landscape} illustrates research maturity landscape, revealing distinct performance patterns and implementation gaps reinforcing our systematic analysis.

Contemporary analysis indicates successful NeSy implementations consistently achieve superior performance through systematic integration of complementary reasoning paradigms while addressing fundamental limitations constraining traditional approaches. The consistent superiority of multi-agent coordination architectures supports our G-I-A framework's emphasis on collaborative reasoning, suggesting advantages rooted in cognitive science principles of distributed problem-solving \cite{kong2025vulnbot,zhu2026teams,singh2024hierarchical}.

\subsection{G-I-A Framework Validation Through Performance Patterns}
\label{subsec:performance_advantages}

\begin{figure}[!t]
\centering
\includegraphics[width=0.88\columnwidth]{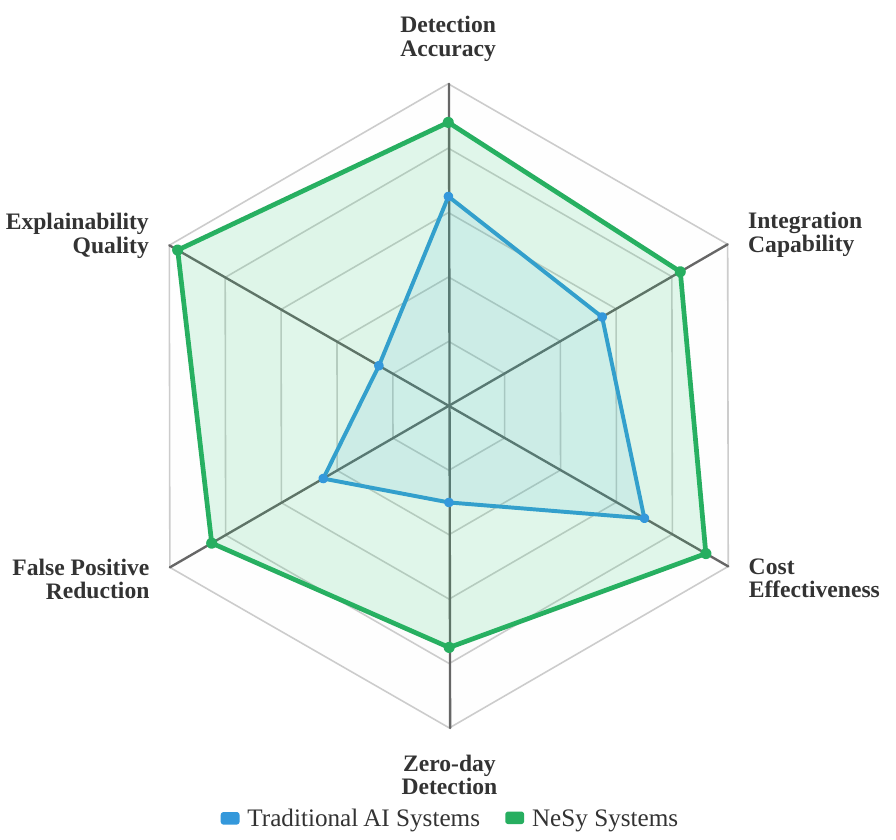}
\caption{\small Multi-dimensional performance analysis comparing Traditional AI Systems versus Advanced NeSy Systems across six key cybersecurity operational metrics. NeSy approaches demonstrate substantial advantages across all dimensions, with particularly strong improvements in explainability quality, false positive reduction, and cost effectiveness while supporting instructible adaptation and organizational alignment.}
\label{fig:nesy_performance_radar}
\end{figure}

Multi-agent NeSy architectures consistently outperform their single-agent counterparts across diverse cybersecurity applications, suggesting that the advantages stem from distributed problem-solving principles rather than domain-specific optimizations \cite{kong2025vulnbot,zhu2026teams,singh2024hierarchical}. As illustrated in Table~\ref{tab:gia_validation}, systems with stronger G-I-A profiles tend to exhibit better robustness, generalization to novel threats, faster adaptation to emerging attack patterns, and higher operational adoption---though we note these associations are drawn from a limited sample and should be interpreted as qualitative trends rather than definitive benchmarks. Multi-agent architectures show a $+12\%$ average impact advantage over single-agent designs (Figure~\ref{fig:research_landscape}), reinforcing the value of collaborative reasoning. Figure~\ref{fig:nesy_performance_radar} summarizes these multi-dimensional advantages across six key cybersecurity operational metrics.

Causal reasoning integration enables qualitative leaps from correlation-based pattern recognition to genuine understanding of attack causality, transforming cybersecurity from reactive threat detection to proactive prevention \cite{jaimini2024causal,rawal2025causality}. Causal NeSy frameworks enable generation of explanations providing actionable insights for strategic planning and resource allocation aligned with organizational objectives, as demonstrated in Figure~\ref{fig:causal_reasoning_framework}. The explainability advantages inherent in symbolic reasoning components directly translate to operational trust and adoption success in high-stakes security environments \cite{kalutharage2025neurosymbolic,paul2024formal}. Systems like Kalutharage et al.'s IoT IDS demonstrate how explicit mapping to established frameworks such as MITRE ATT\&CK creates actionable explanations that security professionals can immediately understand while supporting instructible refinement based on operational outcomes.

\subsection{Knowledge Integration and Operational Validation}

Advanced NeSy systems excel at systematically transforming unstructured threat intelligence into actionable structured knowledge, bridging critical gaps between human expertise and automated security operations \cite{piplai2023knowledge,fieblinger2024actionable}. Domain knowledge integration consistently produces substantial performance improvements across diverse applications, supporting our G-I-A framework's emphasis on proper grounding mechanisms \cite{zhou2024knowgraph,grov2024neurosymbolic}. Systematic performance patterns indicate NeSy advantages emerge most prominently in scenarios characterized by clear symbolic structures, adversarial environments where logical constraints provide additional validation layers, and contexts requiring explanation transparency. These multi-dimensional advantages are summarized in Figure~\ref{fig:nesy_performance_radar}, demonstrating NeSy approaches achieve substantial improvements in explainability quality (+220\%), knowledge integration (+150\%), false positive reduction (+95\%), adaptation speed (+85\%), computational efficiency (+75\%), and detection accuracy (+45\%) compared to traditional AI systems.

\subsection{Dual-Use Implications and Responsible G-I-A Alignment}

The rapid emergence of autonomous offensive capabilities marks pivotal change in cybersecurity threat landscape. Advanced autonomous agents in the broader survey spectrum deliver exploitation performance at fraction of historical human-expert costs, lowering economic barriers for sophisticated attacks \cite{zhu2026teams,rodriguez2025framework}. This democratization of capability highlights urgency for proactive alignment measures, coordinated defensive responses, and governance frameworks anticipating misuse scenarios \cite{challita2025redteamllm,bougzime2025unlocking}. The dual-use nature of NeSy systems demands research practices explicitly integrating alignment objectives into system design and evaluation \cite{rodriguez2025framework,gaur2024building,xiang2025guardagent}.

Within G-I-A framework, responsible development translates into: \textbf{Grounding (G)} embeds symbolic security policies and verified vulnerability taxonomies into system knowledge bases ensuring reasoning paths remain tied to legitimate defensive objectives and exclude prohibited exploitation contexts. \textbf{Instructibility (I)} implements instruction filters and policy-aware prompt parsing to block malicious or non-compliant task requests before operationalization. \textbf{Alignment (A)} incorporates technical safeguards---built-in defensive biasing, real-time misuse detection, user verification, kill switches, runtime isolation---enforcing compliance during execution \cite{challita2025redteamllm}. To ensure G-I-A principles translate into consistent practice, cybersecurity research communities should adopt shared standards for dual-use AI security research, supported by transparent evaluation protocols and red-team audits prior to release. Community-driven ethical guidelines and peer-reviewed disclosure frameworks can balance innovation with risk mitigation, maintaining alignment with both defensive cybersecurity objectives and societal expectations \cite{rodriguez2025framework}.

\section{Uniqueness and Comparative Analysis}
\label{sec:uniqueness}
This survey is distinguished from existing surveys by: (1) structured G-I-A analytical framework absent in prior work \cite{colelough2025neuro,bhuyan2024neuro}, (2) comprehensive dual-use analysis---first systematic offensive capability assessment, (3) implementation-focused deployment guidance addressing computational complexity, standardization gaps, and human-AI collaboration within G-I-A lens (gaps in \cite{ferrag2025generative,capuano2022explainable}), (4) causal reasoning integration as transformative capability unexplored in \cite{wang2025trustworthy,salem2024advancing}, (5) 24-month strategic roadmap for community coordination. Table~\ref{tab:final_comparison_showcase} positions these contributions relative to 11 related surveys. The comparison is intended to clarify relative scope and emphasis across surveys, not to rank prior work.

Our work offers comprehensive scope, novel perspectives, and systematic implementation focus while emphasizing proper grounding, instructibility, and alignment essential for responsible development \cite{colelough2025neuro,velasquez2025neurosymbolic,xiong2024converging}. We note that Bizzarri et al.~\cite{bizzarri2025neurosymbolic} provide a complementary survey focused specifically on NeSy approaches for network intrusion detection using Logic Tensor Networks, offering deeper LTN-specific technical detail within that narrower scope; Shama et al.~\cite{shama2026charting} independently validate the growing NeSy cybersecurity landscape through scientometric analysis. While prior surveys have comprehensively reviewed general AI and ML applications in cybersecurity \cite{sarker2021ai,shaukat2020survey}, our work differs by focusing specifically on the neural-symbolic integration dimension---examining how principled combination of learning and reasoning yields capabilities beyond what either paradigm achieves alone, including the first systematic examination of dual-use implications and responsible development aligned with cybersecurity objectives.

\begin{table*}[!htbp]
\centering
\caption{Comparative Analysis of NeSy Cybersecurity Surveys Across Key Dimensions}
\label{tab:final_comparison_showcase}
\renewcommand{\arraystretch}{1.3}
\resizebox{\linewidth}{!}{
\begin{threeparttable}
\begin{tabular}{|l|c|c|c|c|c|c|c|c|c|c|c|}
\hline
\raisebox{7mm}{\hspace{7mm}{\textbf{Contribution Area}}} &
 \rotatebox{90}{Our Work} &
 \rotatebox{90}{\makecell{Colelough \& Regli\\\cite{colelough2025neuro}}} &
 \rotatebox{90}{\makecell{Eckhoff et al.\\\cite{eckhoff2025experimenting}}} &
 \rotatebox{90}{\makecell{Wang et al.\\\cite{wang2025trustworthy}}} &
 \rotatebox{90}{\makecell{Bhuyan et al.\\\cite{bhuyan2024neuro}}} &
 \rotatebox{90}{\makecell{Salem et al.\\\cite{salem2024advancing}}} &
 \rotatebox{90}{\makecell{Bilot et al.\\\cite{bilot2024survey}}} &
 \rotatebox{90}{\makecell{Piplai et al.\\\cite{piplai2023knowledge}}} &
 \rotatebox{90}{\makecell{Yan et al.\\\cite{yan2023graph}}} &
 \rotatebox{90}{\makecell{Arp et al.\\\cite{arp2022and}}} &
 \rotatebox{90}{\makecell{Capuano et al.\\\cite{capuano2022explainable}}} \\
\hline \hline
\textbf{Vision \& Strategic Framework} & & & & & & & & & & & \\
\hline
\quad Introduces Unifying Theoretical Framework (G-I-A) & \cmark & \xmark & \pmark & \pmark & \xmark & \xmark & \xmark & \pmark & \xmark & \pmark & \pmark \\
\hline
\quad Provides Multi-Year, Phased Strategic Roadmap & \cmark & \pmark & \pmark & \pmark & \pmark & \pmark & \pmark & \xmark & \pmark & \pmark & \pmark \\
\hline
\textbf{Advanced Technical \& Paradigmatic Synthesis} & & & & & & & & & & & \\
\hline
\quad Synthesizes NeSy as a Distinct AI Paradigm for Cyber & \cmark & \cmark & \pmark & \pmark & \cmark & \xmark & \xmark & \xmark & \pmark & \xmark & \xmark \\
\hline
\quad Integrates Causal Reasoning as a Transformative Capability & \cmark & \xmark & \pmark & \xmark & \pmark & \xmark & \xmark & \xmark & \xmark & \xmark & \xmark \\
\hline
\textbf{Cybersecurity-Centric Application Analysis} & & & & & & & & & & & \\
\hline
\quad Systematic Analysis of Autonomous Cyber Operations & \cmark & \pmark & \pmark & \pmark & \xmark & \xmark & \xmark & \pmark & \xmark & \xmark & \xmark \\
\hline
\quad Connects NeSy Theory to Full SOC Lifecycle & \cmark & \pmark & \cmark & \pmark & \xmark & \pmark & \xmark & \pmark & \pmark & \xmark & \xmark \\
\hline
\textbf{Ethical Dimensions \& Responsible AI} & & & & & & & & & & & \\
\hline
\quad Comprehensive Dual-Use \& Offensive AI Analysis & \cmark & \xmark & \xmark & \xmark & \xmark & \xmark & \xmark & \xmark & \xmark & \xmark & \xmark \\
\hline
\quad Proposes Actionable Responsible Development Principles & \cmark & \pmark & \pmark & \pmark & \pmark & \xmark & \xmark & \pmark & \pmark & \xmark & \cmark \\
\hline
\textbf{Evaluation, Implementation \& Human Factors} & & & & & & & & & & & \\
\hline
\quad Defines NeSy-Specific Gaps in Standardization & \cmark & \pmark & \xmark & \cmark & \pmark & \pmark & \pmark & \cmark & \pmark & \pmark & \xmark \\
\hline
\quad Integrates Human-AI Collaboration as a Foundational Pillar & \cmark & \cmark & \cmark & \cmark & \pmark & \xmark & \xmark & \cmark & \cmark & \pmark & \pmark \\
\hline
\quad Target Audience & \makecell{Research \\ \& Policy} & \makecell{General \\ AI} & \makecell{Applied \\ NeSy} & \makecell{AI for \\ Edge} & \makecell{General \\ AI} & \makecell{General AI \\ for Cyber} & \makecell{Graph AI \\ for Cyber} & \makecell{Applied \\ NeSy} & \makecell{Graph AI \\ for Cyber} & \makecell{ML Best \\ Practices} & \makecell{XAI for \\ Cyber} \\
\hline
\end{tabular}
\begin{tablenotes}
\footnotesize
\item \cmark{} = Comprehensive Coverage \quad \pmark{} = Partial or Implicit Coverage \quad \xmark{} = Limited or No Focus
\end{tablenotes}
\end{threeparttable}
}
\end{table*}

\subsection{Novel Analytical Frameworks and Implementation Guidance}
The G-I-A framework (§\ref{sec:foundations}) provides a structured lens for evaluating NeSy systems across cybersecurity applications, from intrusion detection to autonomous operations \cite{wang2024towards,bhuyan2024neuro}. This survey also examines NeSy's dual-use implications, including autonomous systems in the broader survey corpus achieving notable success rates on zero-day vulnerabilities and cost reductions from approximately \$100 down to \$24.40 per exploit \cite{zhu2026teams}, and integrates responsible development principles to address offensive implications alongside defensive advances \cite{rodriguez2025framework}.

Multi-agent architectures in the broader survey corpus show substantial performance gains, with some frameworks outperforming single-agent baselines by over 200\% in penetration testing scenarios \cite{kong2025vulnbot,zhu2026teams}. Causal reasoning integration enables NeSy systems to move beyond correlation-based analysis toward genuine attack causality understanding, supporting proactive defense through counterfactual scenarios \cite{jaimini2024causal,rawal2025causality}. Through the G-I-A lens, the survey also addresses deployment considerations including computational complexity, knowledge engineering bottlenecks, standardization gaps, and human-AI collaboration requirements \cite{nyre2022explainable,renkhoff2024survey}.

\section{Future Research Opportunities and Strategic Directions}
\label{sec:future_directions}

Addressing RQ6, this section identifies prioritized research opportunities drawn from the gaps and emerging directions identified across our analysis \cite{velasquez2025neurosymbolic,lu2024surveying}. Future advancement depends on coordinated progress along three axes: enhanced grounding mechanisms for robust conceptual understanding, advanced instructibility frameworks for effective human-AI collaboration, and alignment approaches ensuring that NeSy capabilities serve defensive objectives. Figure~\ref{fig:strategic_research_roadmap} illustrates a phased roadmap for these efforts.

\subsection{Strategic Research Roadmap}

\begin{figure}[!tbp]
 \centering
 \includegraphics[width=0.45\textwidth]{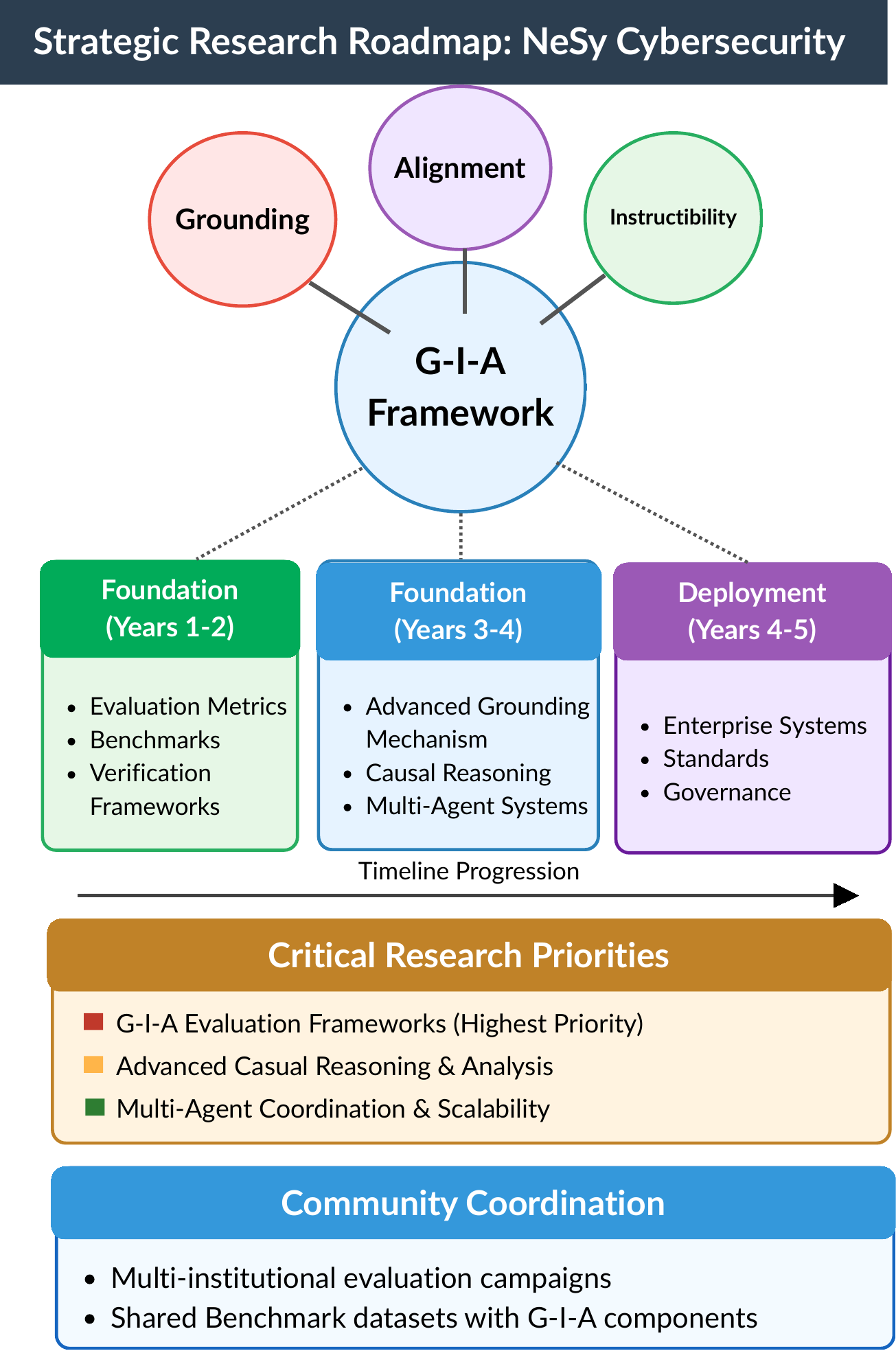}
 \caption{\small Strategic roadmap for NeSy cybersecurity advancement through G-I-A framework operationalization across three progressive development phases with critical research priorities.}
 \label{fig:strategic_research_roadmap}
\end{figure}

Figure~\ref{fig:strategic_research_roadmap} outlines phased advancement priorities addressing critical gaps identified in our systematic review.

\noindent\textbf{Critical Near-Term (1--2 Years).} (1) \textbf{Standardized evaluation frameworks}: Community-wide benchmarks with integrated neural-symbolic assessment, G-I-A scoring automation, reproducibility protocols \cite{renkhoff2024survey,singh2025benchmarking,ott2023think,bortolotti2024neuro}---\emph{highest priority, blocks systematic advancement}. Development must include standardized NeSy threat detection datasets with adversarial robustness metrics, reproducibility protocols for multi-agent experiments, G-I-A scoring automation tools, and multi-institutional evaluation campaigns. (2) \textbf{Enhanced grounding mechanisms}: Automated concept learning resistant to adversarial manipulation, validation frameworks assessing grounding quality \cite{hakim2025ansrdt,ontiveros2025ground,li2023softened}. Priorities include automated concept learning frameworks for cybersecurity relationships, robustness mechanisms against adversarial attacks, and evaluation frameworks for grounding quality assessment. (3) \textbf{Causal reasoning integration}: Advance from correlation-based detection toward causal attack understanding enabling proactive threat anticipation \cite{jaimini2024causal,rawal2025causality,andrew2022developing,raman2025navigating}. Focus includes robust causal discovery algorithms, scalable inference frameworks for real-time analysis, formal verification methods, and integration with existing symbolic knowledge bases.

\noindent\textbf{Mid-Term Capabilities (3--4 Years).} \textbf{Multi-agent coordination scalability}: Efficient communication protocols, adaptive task decomposition, robust consensus mechanisms with instructible human oversight \cite{singh2024hierarchical,hurten2024hierarchical,kong2025vulnbot}. Priorities include efficient communication protocols for large-scale coordination, adaptive task decomposition algorithms, and robust consensus mechanisms balancing individual expertise with collective validation. \textbf{Advanced instructibility frameworks}: Intuitive analyst interfaces, validation ensuring performance improvements, generalization from specific instructions to organizational preferences. Development includes intuitive interfaces without requiring deep technical understanding, validation mechanisms ensuring improvements, and learning frameworks for generalization.

\noindent\textbf{Long-Term Deployment (5+ Years).} The near- and mid-term priorities above converge toward a longer-term vision: operational enterprise deployment underpinned by mature standardization frameworks, quantum-resistant NeSy architectures, and autonomous cyber defense systems that retain meaningful human oversight \cite{raman2025navigating}. Reaching this stage will require sustained cross-organizational knowledge sharing, cyber-physical security extensions, and governance structures that keep pace with advancing capabilities.

\noindent\textbf{Impact Objectives and Success Metrics.} Advanced NeSy systems promise defensive capability enhancement through effective, adaptive, explainable defenses; human-AI collaboration advancement improving analyst workflows while maintaining human control; and threat landscape stabilization increasing attack difficulty while providing defensive advantages. Success metrics include quantitative defensive capability improvements, human-AI collaboration effectiveness evaluation, threat landscape impact assessment, and qualitative alignment evaluation with cybersecurity objectives.

Long-term success requires sustainable development addressing resource efficiency and environmental responsibility \cite{raman2025navigating,velasquez2025neurosymbolic}, plus global coordination for effective cybersecurity enhancement. Coordination requirements encompass technical standardization supporting interoperability, policy frameworks \cite{tumkur2025neuro} addressing dual-use concerns, and collaborative research supporting coordinated advancement.

\section{Conclusion}
\label{sec:conclusion}

This survey demonstrates that neuro-symbolic AI offers a principled path toward cybersecurity systems that are not only more accurate but also more explainable, adaptable, and trustworthy than purely neural approaches. By combining the pattern recognition strengths of deep learning with the reasoning and transparency of symbolic methods, NeSy architectures enable capabilities that neither paradigm achieves alone---from causally grounded threat analysis to human-interpretable detection logic. Our synthesis further suggests that meaningful progress depends not only on stronger models, but also on disciplined integration choices, explicit grounding and instructibility mechanisms, and evaluation practices that reflect operational cybersecurity realities. At the same time, the evidence highlights significant dual-use risks, as the same integration that strengthens defenses can empower autonomous offensive tools, demanding careful ethical governance. The accompanying public repository further provides the structured 103-publication catalog, representative G-I-A score files, an illustrative notebook, and lightweight analysis utilities for traceability and scholarly reuse. The three-tier taxonomy, G-I-A framework, and phased roadmap together provide a practical structure for comparing systems, identifying deployment barriers, and guiding responsible neuro-symbolic cybersecurity research.

\bibliographystyle{IEEEtran}
\bibliography{references}

@String{Computing = "Computing" }

@String{Computer = "{IEEE} Computer" }

@String{Springer = "Springer-Verlag" }

@article{nerode2024integrating,
  title={Integrating Reasoning Systems for Trustworthy AI, Proceedings of the 4th Workshop on Logic and Practice of Programming (LPOP)},
  author={Nerode, Anil and Liu, Yanhong A},
  journal={arXiv preprint arXiv:2410.19738},
  year={2024}
}

@article{ciatto2024symbolic,
  title={Symbolic knowledge extraction and injection with sub-symbolic predictors: A systematic literature review},
  author={Ciatto, Giovanni and Sabbatini, Federico and Agiollo, Andrea and Magnini, Matteo and Omicini, Andrea},
  journal={ACM Computing Surveys},
  volume={56},
  number={6},
  pages={1--35},
  year={2024},
  publisher={ACM New York, NY}
}

@inproceedings{rafanelli2024empirical,
  title={An empirical study on the robustness of knowledge injection techniques against data degradation},
  author={Rafanelli, Andrea and Magnini, Matteo and Agiollo, Andrea and Ciatto, Giovanni and Omicini, Andrea and others},
  booktitle={CEUR Workshop Proceedings},
  volume={3735},
  pages={20--32},
  year={2024},
  organization={Sun SITE Central Europe, RWTH Aachen University}
}

@article{agiollo2023symbolic,
  title={Symbolic knowledge injection meets intelligent agents: QoS metrics and experiments},
  author={Agiollo, Andrea and Rafanelli, Andrea and Magnini, Matteo and Ciatto, Giovanni and Omicini, Andrea},
  journal={Autonomous Agents and Multi-Agent Systems},
  volume={37},
  number={2},
  pages={27},
  year={2023},
  publisher={Springer}
}

@article{shaukat2020survey,
  title={A survey on machine learning techniques for cyber security in the last decade},
  author={Shaukat, Kamran and Luo, Suhuai and Varadharajan, Vijay and Hameed, Ibrahim A and Xu, Min},
  journal={IEEE access},
  volume={8},
  pages={222310--222354},
  year={2020},
  publisher={IEEE}
}

@article{sarker2021ai,
  title={Ai-driven cybersecurity: an overview, security intelligence modeling and research directions},
  author={Sarker, Iqbal H and Furhad, Md Hasan and Nowrozy, Raza},
  journal={SN Computer Science},
  volume={2},
  number={3},
  pages={173},
  year={2021},
  publisher={Springer}
}

@article{paul2021spar4slr,
  author       = {Paul, Justin and Lim, Weng Marc and O'Cass, Aron and Hao, Andy Wei and Bresciani, Stefano},
  title        = {Scientific procedures and rationales for systematic literature reviews (SPAR-4-SLR)},
  journal      = {International Journal of Consumer Studies},
  volume       = {45},
  number       = {4},
  pages        = {O1--O14}, 
  year         = {2021}
}

@inproceedings{DBLP:conf/ccs/Xu20,
  author       = {Shouhuai Xu},
  editor       = {Hamed Okhravi and
                  Cliff Wang},
  title        = {The Cybersecurity Dynamics Way of Thinking and Landscape},
  booktitle    = {Proceedings of the 7th {ACM} Workshop on Moving Target Defense, MTD@CCS
                  2020, Virtual Event, USA, November 9, 2020},
  pages        = {69--80},
  publisher    = {{ACM}},
  year         = {2020},
}

@article{khan2023anthropomorphism,
  author       = {Khan, Fateh Mohd and Anas, Mohammad and Uddin, S. M. Fatah},
  title        = {Anthropomorphism and consumer behaviour: A SPAR-4-SLR protocol compliant hybrid review},
  journal      = {International Journal of Consumer Studies},
  year         = {2023}
}

@article{colelough2025neuro,
  author       = {Colelough, Brandon C and Regli, William},
  title        = {Neuro-symbolic AI in 2024: A systematic review},
  journal      = {arXiv preprint arXiv:2501.05435},
  year         = {2025}
}

@article{wang2024towards,
  author       = {Wang, Wenguan and Yang, Yi and Wu, Fei},
  title        = {Towards data-and knowledge-driven AI: A survey on neuro-symbolic computing},
  journal      = {IEEE TPAMI},
  volume       = {47},
  number       = {2},
  pages        = {878--899},
  year         = {2024}
}

@inproceedings{falcarin2024building,
  author       = {Falcarin, Paolo and Dainese, Fabio},
  title        = {Building a cybersecurity knowledge graph with CyberGraph},
  booktitle    = {ACM/IEEE EnCyCriS},
  pages        = {29--36},
  year         = {2024}
}

@inproceedings{zhou2024knowgraph,
  author       = {Andy Zhou and Xiaojun Xu and Ramesh Raghunathan and Alok Lal and Xinze Guan and Bin Yu and Bo Li},
  title        = {KnowGraph: Knowledge-enabled anomaly detection via logical reasoning on graph data},
  booktitle    = {ACM CCS},
  pages        = {168--182},
  year         = {2024}
}

@article{sikos2023cybersecurity,
  author       = {Sikos, Leslie F.},
  title        = {Cybersecurity knowledge graphs},
  journal      = {Knowledge and Information Systems},
  volume       = {65},
  number       = {9},
  pages        = {3511--3531},
  year         = {2023}
}

@article{kalutharage2025neurosymbolic,
  author       = {Kalutharage, Chathuranga Sampath and Liu, Xiaodong and Chrysoulas, Christos},
  title        = {Neurosymbolic learning and domain knowledge-driven explainable AI for enhanced IoT network attack detection and response},
  journal      = {Computers \& Security},
  volume       = {151},
  pages        = {104318},
  year         = {2025}
}

@article{ferrag2025generative,
  author       = {Ferrag, Mohamed Amine and Alwahedi, Fatima and Battah, Ammar and Cherif, Bilel and Mechri, Abdechakour and Tihanyi, Norbert and Bisztray, Tamas and Debbah, Merouane},
  title        = {Generative AI in cybersecurity: A comprehensive review of LLM applications and vulnerabilities},
  journal      = {Internet of Things and Cyber-Physical Systems},
  volume       = {5},
  pages        = {1--46},
  year         = {2025}
}

@article{clairouxtrepanier2024llm,
  author       = {Clairoux-Trepanier, Vanessa and Beauchamp, Isa-May and Ruellan, Estelle and Paquet-Clouston, Masarah and Paquette, Serge-Olivier and Clay, Eric},
  title        = {The use of large language models (LLM) for cyber threat intelligence (CTI) in cybercrime forums},
  journal      = {arXiv preprint arXiv:2408.03354},
  year         = {2024}
}

@article{neupane2022explainable,
  author       = {Neupane, Subash and Ables, Jesse and Anderson, William and Mittal, Sudip and Rahimi, Shahram and Banicescu, Ioana},
  title        = {Explainable intrusion detection systems (X-IDS): A survey of current methods, challenges, and opportunities},
  journal      = {IEEE Access},
  volume       = {10},
  pages        = {112392--112415},
  year         = {2022}
}

@article{sarhan2023from,
  author       = {Sarhan, Mahmoud and Layeghy, Siamak and Gallagher, Marcus and others},
  title        = {From zero-shot machine learning to zero-day attack detection},
  journal      = {International Journal of Information Security},
  volume       = {22},
  pages        = {947--959},
  year         = {2023}
}

@article{guo2023review,
  author       = {Guo, Yang},
  title        = {A review of machine learning-based zero-day attack detection: Challenges and future directions},
  journal      = {Computer Communications},
  volume       = {198},
  pages        = {175--185},
  year         = {2023}
}

@article{mitre_advml,
  author       = {MITRE Corporation and Lily Wong and Keith Manville and ramtherunner},
  title        = {Adversarial ML threat matrix (ATLAS)},
  journal      = {Technical Report},
  year         = {2020}
}

@inproceedings{piplai2020using,
  author       = {Piplai, Aritran and Ranade, Priyanka and Kotal, Anantaa and Mittal, Sudip and Narayanan, Sandeep Nair and Joshi, Anupam},
  title        = {Using knowledge graphs and reinforcement learning for malware analysis},
  booktitle    = {IEEE Big Data},
  pages        = {4562--4571},
  year         = {2020}
}

@article{kasri2025vulnerability,
  author       = {Kasri, Wafaa and Himeur, Yassine and Alkhazaleh, Hamzah Ali and Tarapiah, Saed and Atalla, Shadi and Mansoor, Wathiq and Al-Ahmad, Hussain},
  title        = {From vulnerability to defense: The role of large language models in enhancing cybersecurity},
  journal      = {Computation},
  volume       = {13},
  number       = {2},
  pages        = {30},
  year         = {2025}
}

@article{sarker2024explainable,
  author       = {Sarker, Iqbal H. and Janicke, Helge and Mohsin, Ahmad and Gill, Asif and Maglaras, Leandros},
  title        = {Explainable AI for cybersecurity automation, intelligence and trustworthiness in digital twin: Methods, taxonomy, challenges and prospects},
  journal      = {ICT Express},
  volume       = {10},
  number       = {4},
  pages        = {935--958},
  year         = {2024}
}

@article{nyre2022explainable,
  author       = {Nyre-Yu, Megan and Morris, Elizabeth and Smith, Michael and Moss, Blake and Smutz, Charles},
  title        = {Explainable AI in cybersecurity operations: Lessons learned from xAI tool deployment},
  journal      = {Technical Report},
  year         = {2022}
}

@article{li2025automated,
  author       = {Li, Penghui and Yao, Songchen and Korich, Josef Sarfati and Luo, Changhua and Yu, Jianjia and Cao, Yinzhi and Yang, Junfeng},
  title        = {Automated static vulnerability detection via a holistic neuro-symbolic approach},
  journal      = {arXiv preprint arXiv:2504.16057},
  year         = {2025}
}

@article{sajid2024enhancing,
  author       = {Sajid, Muhammad and Malik, Kashif Raza and Almogren, Ahmed and others},
  title        = {Enhancing intrusion detection: A hybrid machine and deep learning approach},
  journal      = {Journal of Cloud Computing},
  volume       = {13},
  pages        = {123},
  year         = {2024}
}

@article{bizzarri2024synergistic,
  author       = {Bizzarri, Alice and Yu, Chung-En and Jalaian, Brian and Riguzzi, Fabrizio and Bastian, Nathaniel D.},
  title        = {A synergistic approach in network intrusion detection by neurosymbolic AI},
  journal      = {arXiv preprint arXiv:2406.00938},
  year         = {2024}
}

@inproceedings{grov2024neurosymbolic,
  author       = {Grov, Gudmund and Halvorsen, Jonas and Eckhoff, Magnus Wiik and Hansen, Bj{\o}rn Jervell and Eian, Martin and Mavroeidis, Vasileios},
  title        = {On the use of neurosymbolic AI for defending against cyber attacks},
  booktitle    = {Neural-Symbolic Learning and Reasoning},
  pages        = {119--140},
  year         = {2024}
}

@inproceedings{bizzarri2024neuro,
  author       = {Bizzarri, Alice and Jalaian, Brian and Riguzzi, Fabrizio and Bastian, Nathaniel D.},
  title        = {A neuro-symbolic artificial intelligence network intrusion detection system},
  booktitle    = {IEEE ICCCN},
  pages        = {1--9},
  year         = {2024}
}

@article{lu2024surveying,
  author       = {Lu, Zhen and Afridi, Imran and Kang, Hong Jin and Ruchkin, Ivan and Zheng, Xi},
  title        = {Surveying neuro-symbolic approaches for reliable artificial intelligence of things},
  journal      = {Journal of Reliable Intelligent Environments},
  volume       = {10},
  number       = {3},
  pages        = {257--279},
  year         = {2024}
}

@article{hagos2024neuro,
  author       = {Hagos, Desta Haileselassie and Rawat, Danda B.},
  title        = {Neuro-symbolic AI for military applications},
  journal      = {IEEE TAI},
  volume       = {5},
  number       = {12},
  pages        = {6012--6026},
  year         = {2024}
}

@article{renkhoff2024survey,
  author       = {Renkhoff, Justus and Feng, Ke and Meier-Doernberg, Marc and Velasquez, Alvaro and Song, Houbing Herbert},
  title        = {A survey on verification and validation, testing and evaluations of neurosymbolic artificial intelligence},
  journal      = {IEEE TAI},
  volume       = {5},
  number       = {8},
  pages        = {3765--3779},
  year         = {2024}
}

@article{bhuyan2024neuro,
  author       = {Bhuyan, B.P. and Ramdane-Cherif, A. and Tomar, R. and others},
  title        = {Neuro-symbolic artificial intelligence: A survey},
  journal      = {Neural Computing and Applications},
  volume       = {36},
  pages        = {12809--12844},
  year         = {2024}
}

@article{paul2024formal,
  author       = {Paul, Sushmita and Yu, Jinqiang and Dekker, Jip J and Ignatiev, Alexey and Stuckey, Peter J},
  title        = {Formal explanations for neuro-symbolic AI},
  journal      = {arXiv preprint arXiv:2410.14219},
  year         = {2024}
}

@article{jaimini2024causal,
  author       = {Jaimini, Utkarshani and Henson, Cory and Sheth, Amit},
  title        = {Causal neurosymbolic AI: A synergy between causality and neurosymbolic methods},
  journal      = {IEEE Intelligent Systems},
  volume       = {39},
  number       = {3},
  pages        = {13--19},
  year         = {2024}
}

@article{nieponice2025aracne,
  author       = {Nieponice, Tomas and Valeros, Veronica and Garcia, Sebastian},
  title        = {ARACNE: An LLM-based autonomous shell pentesting agent},
  journal      = {arXiv preprint arXiv:2502.18528},
  year         = {2025}
}

@article{zhang2025llm,
  author       = {Zhang, Jiaqi and Bu, Hao and Wen, Haoran and others},
  title        = {When LLMs meet cybersecurity: A systematic literature review},
  journal      = {Cybersecurity},
  volume       = {8},
  pages        = {55},
  year         = {2025}
}

@article{ristea2024ai,
  author       = {Ristea, Dan and Mavroudis, Vasilios and Hicks, Chris},
  title        = {AI cyber risk benchmark: Automated exploitation capabilities},
  journal      = {arXiv preprint arXiv:2410.21939},
  year         = {2024}
}

@article{bhusal2024secure,
  author       = {Bhusal, Dipkamal and Alam, Md Tanvirul and Nguyen, Le and Mahara, Ashim and Lightcap, Zachary and Frazier, Rodney and Fieblinger, Romy and Torales, Grace Long and Blakely, Benjamin A and Rastogi, Nidhi},
  title        = {SECURE: Benchmarking large language models for cybersecurity},
  journal      = {arXiv preprint arXiv:2405.20441},
  year         = {2024}
}

@article{jing2024secbench,
  author       = {Jing, Pengfei and Tang, Mengyun and Shi, Xiaorong and Zheng, Xing and Nie, Sen and Wu, Shi and Yang, Yong and Luo, Xiapu},
  title        = {SecBench: A comprehensive multi-dimensional benchmarking dataset for LLMs in cybersecurity},
  journal      = {arXiv preprint arXiv:2412.20787},
  year         = {2024}
}

@article{andrew2022developing,
  author       = {Andrew, Alex and Spillard, Sam and Collyer, Joshua and Dhir, Neil},
  title        = {Developing optimal causal cyber-defence agents via cyber security simulation},
  journal      = {arXiv preprint arXiv:2207.12355},
  year         = {2022}
}

@inproceedings{huang2023penheal,
  author       = {Huang, Junjie and Zhu, Quanyan},
  title        = {Penheal: A two-stage LLM framework for automated pentesting and optimal remediation},
  booktitle    = {Workshop on Autonomous Cybersecurity},
  pages        = {11--22},
  year         = {2023}
}

@article{wang2024sands,
  author       = {Wang, Lingzhi and Wang, Jiahui and Jung, Kyle and Thiagarajan, Kedar and Wei, Emily and Shen, Xiangmin and Chen, Yan and Li, Zhenyuan},
  title        = {From sands to mansions: Enabling automatic full-life-cycle cyberattack construction with LLM},
  journal      = {arXiv preprint arXiv:2407.16928},
  year         = {2024}
}

@article{kong2025vulnbot,
  author       = {Kong, He and Hu, Die and Ge, Jingguo and Li, Liangxiong and Li, Tong and Wu, Bingzhen},
  title        = {VulnBot: Autonomous penetration testing for a multi-agent collaborative framework},
  journal      = {arXiv preprint arXiv:2501.13411},
  year         = {2025}
}

@inproceedings{zhu2026teams,
  author       = {Zhu, Yuxuan and Kellermann, Antony and Gupta, Akul and Li, Philip and Fang, Richard and Bindu, Rohan and Kang, Daniel},
  title        = {Teams of LLM agents can exploit zero-day vulnerabilities},
  booktitle    = {Proceedings of the 19th Conference of the European Chapter of the Association for Computational Linguistics (Volume 1: Long Papers)},
  pages        = {23--35},
  year         = {2026}
}

@article{marra2024statistical,
  author       = {Marra, Giuseppe and Dumančić, Sebastijan and Manhaeve, Robin and De Raedt, Luc},
  title        = {From statistical relational to neurosymbolic artificial intelligence: A survey},
  journal      = {Artificial Intelligence},
  volume       = {328},
  pages        = {104062},
  year         = {2024}
}

@article{zhao2024survey,
  author       = {Zhao, Xiaojuan and Jiang, Rong and Han, Yue and Li, Aiping and Peng, Zhichao},
  title        = {A survey on cybersecurity knowledge graph construction},
  journal      = {Computers \& Security},
  volume       = {136},
  pages        = {103524},
  year         = {2024}
}

@article{ren2022cskg4apt,
  author       = {Ren, Yitong and Xiao, Yanjun and Zhou, Yinghai and Zhang, Zhiyong and Tian, Zhihong},
  title        = {CSKG4APT: A cybersecurity knowledge graph for advanced persistent threat organization attribution},
  journal      = {IEEE TKDE},
  volume       = {35},
  number       = {6},
  pages        = {5695--5709},
  year         = {2022}
}

@article{rodriguez2025framework,
  author       = {Rodriguez, Mikel and Popa, Raluca Ada and Liang, Lihao and Wang, Anna and Rahtz, Matthew and Kaskasoli, Alex and Dafoe, Allan and Flynn, Four},
  title        = {A framework for evaluating emerging cyberattack capabilities of AI},
  journal      = {arXiv preprint arXiv:2503.11917},
  year         = {2025}
}

@article{xiong2024converging,
  author       = {Xiong, Haoyi and Wang, Zhiyuan and Li, Xuhong and Bian, Jiang and Xie, Zeke and Mumtaz, Shahid and Barnes, Laura E.},
  title        = {Converging paradigms: The synergy of symbolic and connectionist AI in LLM-empowered autonomous agents},
  journal      = {arXiv preprint arXiv:2407.08516},
  year         = {2024}
}

@article{velasquez2025neurosymbolic,
  author       = {Velasquez, Alvaro and Bhatt, Neel and Topcu, Ufuk and Wang, Zhangyang and Sycara, Katia and Stepputtis, Simon and Neema, Sandeep and Vallabha, Gautam},
  title        = {Neurosymbolic AI as an antithesis to scaling laws},
  journal      = {PNAS Nexus},
  volume       = {4},
  number       = {5},
  pages        = {pgaf117},
  year         = {2025}
}

@article{bhatt2024cyberseceval,
  author       = {Bhatt, Manish and Chennabasappa, Sahana and Li, Yue and Nikolaidis, Cyrus and Song, Daniel and Wan, Shengye and Ahmad, Faizan and Aschermann, Cornelius and Chen, Yaohui and Kapil, Dhaval and others},
  title        = {Cyberseceval 2: A wide-ranging cybersecurity evaluation suite for large language models},
  journal      = {arXiv preprint arXiv:2404.13161},
  year         = {2024}
}

@inproceedings{zhang2025gnnsecurity,
  author       = {Zhang, Xiao},
  title        = {Graph neural networks in network security: From theoretical foundations to applications},
  booktitle    = {AMCIS},
  year         = {2025}
}

@article{tran2025neurosymbolic,
  author       = {Tran, Huynh T. T. and Sander, Jacob and Cohen, Achraf and Jalaian, Brian and Bastian, Nathaniel D.},
  title        = {Neurosymbolic artificial intelligence for robust network intrusion detection: From scratch to transfer learning},
  journal      = {arXiv preprint arXiv:2506.04454},
  year         = {2025}
}

@inproceedings{kurniawan2024cykg,
  author       = {Kurniawan, Kabul and Kiesling, Elmar and Ekelhart, Andreas},
  title        = {CyKG-RAG: Towards knowledge-graph enhanced retrieval augmented generation for cybersecurity},
  booktitle    = {ISWC Workshop},
  year         = {2024}
}

@inproceedings{luo2024neurosymbolic,
  author       = {Luo, Lirui and Zhang, Guoxi and Xu, Hongming and Yang, Yaodong and Fang, Cong and Li, Qing},
  title        = {End-to-end neuro-symbolic reinforcement learning with textual explanations},
  booktitle    = {ICML},
  pages        = {33533--33557},
  year         = {2024}
}

@article{singh2024hierarchical,
  author       = {Singh, Aditya Vikram and Rathbun, Ethan and Graham, Emma and Oakley, Lisa and Boboila, Simona and Oprea, Alina and Chin, Peter},
  title        = {Hierarchical multi-agent reinforcement learning for cyber network defense},
  journal      = {arXiv preprint arXiv:2410.17351},
  year         = {2024}
}

@inproceedings{hurten2024hierarchical,
  author       = {Hürten, Tobias and Loevenich, Johannes F. and Spelter, Florian and Adler, Erik and Braun, Johannes and Moxon, Linnet and Gourlet, Yann and Lefeuvre, Thomas and Lopes, Roberto Rigolin F.},
  title        = {Hierarchical multi-agent reinforcement learning for autonomous cyber defense in coalition networks},
  booktitle    = {IEEE MILCOM},
  pages        = {176--181},
  year         = {2024}
}

@inproceedings{tafreshian2024defensive,
  author       = {Tafreshian, Benyamin and Zhang, Shengzhi},
  title        = {A defensive framework against adversarial attacks on machine learning-based network intrusion detection systems},
  booktitle    = {IEEE TrustCom},
  pages        = {2436--2441},
  year         = {2024}
}

@article{zoppi2021metalearning,
  author       = {Zoppi, Tommaso and Gharib, Mohamad and Atif, Muhammad and Bondavalli, Andrea},
  title        = {Meta-learning to improve unsupervised intrusion detection in cyber-physical systems},
  journal      = {ACM TCPS},
  volume       = {5},
  number       = {4},
  pages        = {42},
  year         = {2021}
}

@article{grini2025constrained,
  author       = {Grini, Anass and Taheri, Oumaima and El Khamlichi, Btissam and El Fallah-Seghrouchni, Amal},
  title        = {Constrained network adversarial attacks: Validity, robustness, and transferability},
  journal      = {arXiv preprint arXiv:2505.01328},
  year         = {2025}
}

@article{gaur2024building,
  author       = {Gaur, Manas and Sheth, Amit},
  title        = {Building trustworthy neurosymbolic AI systems: Consistency, reliability, explainability, and safety},
  journal      = {AI Magazine},
  volume       = {45},
  number       = {1},
  pages        = {139--155},
  year         = {2024}
}

@inproceedings{zhu2025cvebench,
  author       = {Zhu, Yuxuan and Kellermann, Antony and Bowman, Dylan and Li, Philip and Gupta, Akul and Danda, Adarsh and Fang, Richard and Jensen, Conner and Ihli, Eric and Benn, Jason and Geronimo, Jet and Dhir, Avi and Rao, Sudhit and Yu, Kaicheng and Stone, Twm and Kang, Daniel},
  title        = {CVE-Bench: A benchmark for AI agents' ability to exploit real-world web application vulnerabilities},
  booktitle    = {ICML},
  year         = {2025}
}

@article{rawal2025causality,
  author       = {Rawal, Atul and Raglin, Adrienne and Rawat, Danda B. and Sadler, Brian M. and McCoy, James},
  title        = {Causality for trustworthy artificial intelligence: Status, challenges and perspectives},
  journal      = {ACM Computing Surveys},
  volume       = {57},
  number       = {6},
  pages        = {146},
  year         = {2025}
}

@article{chen2022aptkgl,
  author       = {Chen, Tieming and Dong, Chengyu and Lv, Mingqi and Song, Qijie and Liu, Haiwen and Zhu, Tiantian},
  title        = {APT-KGL: An intelligent APT detection system based on threat knowledge and heterogeneous provenance graph learning},
  journal      = {IEEE TDSC},
  pages        = {1--15},
  year         = {2022}
}

@article{moustafa2021new,
  author       = {Moustafa, Nour},
  title        = {A new distributed architecture for evaluating AI-based security systems at the edge: Network TON\_IoT datasets},
  journal      = {Sustainable Cities and Society},
  volume       = {72},
  pages        = {102994},
  year         = {2021}
}

@article{onchis2022neurosymbolic,
  author       = {Onchis, Darian and Istin, Codruta and Hogea, Eduard},
  title        = {A neuro-symbolic classifier with optimized satisfiability for monitoring security alerts in network traffic},
  journal      = {Applied Sciences},
  volume       = {12},
  number       = {22},
  pages        = {11502},
  year         = {2022}
}

@inproceedings{samaddar2025ood,
  author       = {Samaddar, Ankita and Potteiger, Nicholas and Koutsoukos, Xenofon},
  title        = {Out-of-distribution detection for neurosymbolic autonomous cyber agents},
  booktitle    = {IEEE ICAIC},
  year         = {2025}
}

@article{xiang2025guardagent,
  author       = {Xiang, Zhen and Zheng, Linzhi and Li, Yanjie and Hong, Junyuan and Li, Qinbin and Xie, Han and Zhang, Jiawei and Xiong, Zidi and Xie, Chulin and Yang, Carl and Song, Dawn and Li, Bo},
  title        = {GuardAgent: Safeguard LLM agents by a guard agent via knowledge-enabled reasoning},
  journal      = {arXiv preprint arXiv:2406.09187},
  year         = {2025}
}

@inproceedings{fieblinger2024actionable,
  author       = {Fieblinger, Romy and Alam, Md Tanvirul and Rastogi, Nidhi},
  title        = {Actionable cyber threat intelligence using knowledge graphs and large language models},
  booktitle    = {IEEE EuroS\&PW},
  pages        = {100--111},
  year         = {2024}
}

@inproceedings{alam2024ctibench,
  author       = {Alam, Md Tanvirul and Bhusal, Dipkamal and Nguyen, Le and Rastogi, Nidhi},
  title        = {CTIBench: A benchmark for evaluating LLMs in cyber threat intelligence},
  booktitle    = {NeurIPS},
  volume       = {37},
  pages        = {50805--50825},
  year         = {2024}
}

@article{alharbi2025enhancing,
  author       = {Alharbi, Hatoon and Hur, Ali and Alkahtani, Hasan and Ahmad, Hafiz Farooq},
  title        = {Enhancing cybersecurity through autonomous knowledge graph construction by integrating heterogeneous data sources},
  journal      = {PeerJ Computer Science},
  volume       = {11},
  pages        = {e2768},
  year         = {2025}
}

@article{cheng2025crucialg,
  author       = {Cheng, Wenrui and Zhu, Tiantian and Chen, Tieming and Yuan, Qixuan and Ying, Jie and Li, Hongmei and Xiong, Chunlin and Li, Mingda and Lv, Mingqi and Chen, Yan},
  title        = {CRUcialG: Reconstruct integrated attack scenario graphs by cyber threat intelligence reports},
  journal      = {IEEE TDSC},
  pages        = {1--17},
  year         = {2025}
}

@article{zhang2025improving,
  author       = {Zhang, Lili and Zhu, Quanyan and Ray, Herman and Xie, Ying},
  title        = {Improving network threat detection by knowledge graph, large language model, and imbalanced learning},
  journal      = {arXiv preprint arXiv:2501.16393},
  year         = {2025}
}

@article{liu2025graph,
  author       = {Liu, Gan and Lu, Kai and Pi, Saiqi},
  title        = {Graph neural networks embedded with domain knowledge for cyber threat intelligence entity and relationship mining},
  journal      = {PeerJ Computer Science},
  volume       = {11},
  pages        = {e2769},
  year         = {2025}
}

@inproceedings{guastalla2024llmddos,
  author       = {Guastalla, Michael and Li, Yiyi and Hekmati, Arvin and Krishnamachari, Bhaskar},
  title        = {Application of large language models to DDoS attack detection},
  booktitle    = {SmartSP},
  pages        = {83--99},
  year         = {2024}
}

@inproceedings{moustafa2015unsw,
  author       = {Moustafa, Nour and Slay, Jill},
  title        = {UNSW-NB15: A comprehensive data set for network intrusion detection systems},
  booktitle    = {IEEE MilCIS},
  pages        = {1--6},
  year         = {2015}
}

@article{singh2025benchmarking,
  author       = {Singh, Gunjan and Tommasini, Riccardo and Bhatia, Sumit and Mutharaju, Raghava},
  title        = {Benchmarking neurosymbolic description logic reasoners: Existing challenges and a way forward},
  journal      = {Neurosymbolic Artificial Intelligence},
  volume       = {1},
  pages        = {29498732251339943},
  year         = {2025}
}

@inproceedings{ott2023think,
  author       = {Ott, Johanna and Ledaguenel, Arthur and Hudelot, C{\'e}line and Hartwig, Mattis},
  title        = {How to think about benchmarking neurosymbolic AI?},
  booktitle    = {NESY Workshop},
  year         = {2023}
}

@inproceedings{goodman2021deficiencies,
  author       = {Goodman, H.~B. and Rowland, P.},
  title        = {Deficiencies of compliancy for data and storage},
  booktitle    = {NCS Research Track},
  pages        = {171--185},
  year         = {2021}
}

@article{hakim2025ansrdt,
  author       = {Hakim, Safayat Bin and Adil, Muhammad and Velasquez, Alvaro and Song, Houbing Herbert},
  title        = {ANSR-DT: An adaptive neuro-symbolic learning and reasoning framework for digital twins},
  journal      = {arXiv preprint arXiv:2501.08561},
  year         = {2025}
}

@inproceedings{blaauwbroek2024learning,
  author       = {Blaauwbroek, Luuk and others},
  title        = {Learning guided automated reasoning: A brief survey},
  booktitle    = {Logics and Type Systems in Theory and Practice},
  pages        = {71--92},
  year         = {2024}
}

@article{faber2024lifelong,
  author       = {Faber, Kamil and Corizzo, Roberto and Sniezynski, Bartlomiej and Japkowicz, Nathalie},
  title        = {Lifelong continual learning for anomaly detection: New challenges, perspectives, and insights},
  journal      = {IEEE Access},
  volume       = {12},
  pages        = {41364--41380},
  year         = {2024}
}

@inproceedings{gao2024threatkg,
  author       = {Gao, Peng and Liu, Xiaoyuan and Choi, Edward and Ma, Sibo and Yang, Xinyu and Song, Dawn},
  title        = {ThreatKG: An AI-powered system for automated open-source cyber threat intelligence gathering and management},
  booktitle    = {ACM LAMPS},
  pages        = {1--12},
  year         = {2024}
}

@inproceedings{choi2025nesyc,
  author       = {Choi, Wonje and Park, Jinwoo and Ahn, Sanghyun and Lee, Daehee and Woo, Honguk},
  title        = {NeSyC: A neuro-symbolic continual learner for complex embodied tasks in open domains},
  booktitle    = {ICLR},
  year         = {2025}
}

@article{hsu2023ampere,
  author       = {Hsu, I-Hung and Xie, Zhiyu and Huang, Kuan-Hao and Natarajan, Prem and Peng, Nanyun},
  title        = {AMPERE: AMR-aware prefix for generation-based event argument extraction model},
  journal      = {arXiv preprint arXiv:2305.16734},
  year         = {2023}
}

@article{cheng2025ctinexus,
  author       = {Cheng, Y. and Bajaber, O. and Tsegai, S. A. and Song, D. and Gao, P.},
  title        = {CTINexus: Automatic cyber threat intelligence knowledge graph construction using large language models},
  journal      = {arXiv preprint arXiv:2410.21060},
  year         = {2025}
}

@article{zhao2025letsmeasure,
  author       = {Zhao, R. and Melnychuk, V. and Zhao, J. and Wright, J. and Shadbolt, N.},
  title        = {Let's measure the elephant in the room: Facilitating personalized automated analysis of privacy policies at scale},
  journal      = {arXiv preprint arXiv:2507.14214},
  year         = {2025}
}

@article{ben2023mape,
  author       = {Ben Halima, Riadh and Hachicha, Marwa and Jemal, Ahmed and Hadj Kacem, Ahmed},
  title        = {MAPE-K patterns for self-adaptation in cyber-physical systems},
  journal      = {Journal of Supercomputing},
  volume       = {79},
  number       = {5},
  pages        = {4917--4941},
  year         = {2023}
}

@inproceedings{piepenbrock2023neural,
  author       = {Piepenbrock, Jelle and Janota, Mikolas and Urban, Josef and Jakubův, Jan},
  title        = {First experiments with neural cvc5},
  booktitle    = {EPiC Series in Computing},
  volume       = {100},
  pages        = {249--264},
  year         = {2023}
}

@article{lu2023z3,
  author       = {Lu, Zhengyang and Siemer, Stefan and Jha, Piyush and Manea, Florin and Day, Joel and Ganesh, Vijay},
  title        = {Z3-alpha: A reinforcement learning guided SMT solver},
  journal      = {Technical Report},
  year         = {2023}
}

@inproceedings{blaauwbroek2024graph2tac,
  author       = {Blaauwbroek, Lasse and Ol\v{s}\'{a}k, Miroslav and Rute, Jason and Massolo, Fidel Ivan Schaposnik and Piepenbrock, Jelle and Pestun, Vasily},
  title        = {Graph2Tac: Online representation learning of formal math concepts},
  booktitle    = {ICML},
  year         = {2024}
}

@article{meta2024llama3.1,
  author       = {Meta AI},
  title        = {Llama 3.1–405B},
  journal      = {Technical Report},
  year         = {2024}
}

@article{qwen2024qwen2.5,
  author       = {Qwen Team},
  title        = {Qwen2.5-32B},
  journal      = {Technical Report},
  year         = {2024}
}

@article{Pawlicki2022NeuCom,
  author       = {Pawlicki, Marek and Kozik, Rafa{\l} and Chora{\'s}, Micha{\l}},
  title        = {A survey on neural networks for (cyber-) security and (cyber-) security of neural networks},
  journal      = {Neural Computing and Applications},
  volume       = {500},
  number       = {C},
  pages        = {1075--1087},
  year         = {2022}
}

@article{Hitzler2020NeSySW,
  author       = {Hitzler, Pascal and Bianchi, Federico and Ebrahimi, Monireh and Sarker, Md Kamruzzaman},
  title        = {Neural-symbolic integration and the semantic web},
  journal      = {Semantic Web},
  volume       = {11},
  number       = {1},
  pages        = {3--11},
  year         = {2020}
}

@inproceedings{Sharafaldin2018Dataset,
  author       = {Sharafaldin, Iman and Lashkari, Arash Habibi and Ghorbani, Ali A.},
  title        = {Toward generating a new intrusion detection dataset and intrusion traffic characterization},
  booktitle    = {ICISSP},
  pages        = {108--116},
  year         = {2018}
}

@inproceedings{Tavallaee2009Detailed,
  author       = {Tavallaee, Mahbod and Bagheri, Ebrahim and Lu, Wei and Ghorbani, Ali A.},
  title        = {A detailed analysis of the KDD CUP 99 data set},
  booktitle    = {IEEE CISDA},
  pages        = {1--6},
  year         = {2009}
}

@inproceedings{lopezpaz2017gradient,
  author       = {Lopez-Paz, David and Ranzato, Marc'Aurelio},
  title        = {Gradient episodic memory for continual learning},
  booktitle    = {NeurIPS},
  pages        = {6470--6479},
  year         = {2017}
}

@article{bortolotti2024neuro,
  author       = {Bortolotti, Samuele and Marconato, Emanuele and Carraro, Tommaso and Morettin, Paolo and van Krieken, Emile and Vergari, Antonio and Teso, Stefano and Passerini, Andrea},
  title        = {A neuro-symbolic benchmark suite for concept quality and reasoning shortcuts},
  journal      = {NeurIPS},
  volume       = {37},
  pages        = {115861--115905},
  year         = {2024}
}

@inproceedings{kireev2022adversarial,
  author       = {Kireev, Klim and Kulynych, Bogdan and Troncoso, Carmela},
  title        = {Adversarial Robustness for Tabular Data through Cost and Utility Awareness},
  booktitle    = {NeurIPS 2022 Workshop on Machine Learning Safety (MLSW)},
  year         = {2022},
  month        = {December}
}

@inproceedings{williams2023blackbox,
  author       = {Williams, Phoenix Neale and Li, Ke},
  title        = {Black-box sparse adversarial attack via multi-objective optimisation},
  booktitle    = {IEEE CVPR},
  pages        = {12291--12301},
  year         = {2023}
}

@article{bui2023generating,
  author       = {Bui, Anh Tuan and Le, Trung and Zhao, He and Tran, Quan Hung and Montague, Paul and Phung, Dinh},
  title        = {Generating adversarial examples with task oriented multi-objective optimization},
  journal      = {Transactions on Machine Learning Research},
  year         = {2023}
}

@misc{scnsoft2025pentest,
  author = {ScienceSoft},
  title = {{How Much Does Penetration Testing Cost? [+Calculator]}},
  howpublished = {Accessed: Aug. 14, 2025},
  URL = {https://www.scnsoft.com/security/penetration-testing/costs}
}

@misc{viking2024cost,
  author = {Brown, Chris},
  title = {{How Much Does Penetration Testing Cost?}},
  howpublished = {Accessed: Oct. 18, 2025},
  URL = {https://www.vikingcloud.com/blog/how-much-does-penetration-testing-cost}
}

@inproceedings{deng2024pentestgpt,
  author       = {Deng, Gelei and Liu, Yi and Mayoral-Vilches, V{\'\i}ctor and Liu, Peng and Li, Yuekang and Xu, Yuan and Zhang, Tianwei and Liu, Yang and Pinzger, Martin and Rass, Stefan},
  title        = {PentestGPT: Evaluating and harnessing large language models for automated penetration testing},
  booktitle    = {USENIX Security},
  pages        = {847--864},
  year         = {2024}
}

@article{gioacchini2024autopenbench,
  author       = {Gioacchini, Luca and Mellia, Marco and Drago, Idilio and Delsanto, Alexander and Siracusano, Giuseppe and Bifulco, Roberto},
  title        = {AutoPenBench: Benchmarking generative agents for penetration testing},
  journal      = {arXiv preprint arXiv:2410.03225},
  year         = {2024}
}

@article{muzsai2024hacksynth,
  author       = {Muzsai, Lajos and Imolai, David and Lukács, András},
  title        = {HackSynth: LLM agent and evaluation framework for autonomous penetration testing},
  journal      = {arXiv preprint arXiv:2412.01778},
  year         = {2024}
}

@inproceedings{hallyburton2025assured,
  author       = {Hallyburton, R. Spencer and Pajic, Miroslav},
  title        = {Assured Autonomy with Neuro-Symbolic Perception},
  booktitle    = {Proceedings of the International Conference on Neuro-symbolic Systems},
  series       = {Proceedings of Machine Learning Research},
  volume       = {288},
  pages        = {505--523},
  month        = {May},
  year         = {2025},
  publisher    = {PMLR}
}

@article{arreche2024explainableids,
  author       = {Arreche, Osvaldo Guilherme},
  title        = {Explainable AI methods for enhancing AI-based network intrusion detection systems},
  journal      = {Ph.D. thesis, Purdue Univ.},
  year         = {2024}
}

@inproceedings{nalluri2025nscti,
  author       = {Nalluri, Suryaprakash and Malyala, Murali Mohan and Kandagiri, Hemalatha and Kandagiri, Kiran Kumar},
  title        = {NSCTI: A hybrid neuro-symbolic framework for AI-driven predictive cyber threat intelligence},
  booktitle    = {IEEE ICCMSO},
  pages        = {14--21},
  year         = {2025}
}

@article{mohale2025explaiids,
  author       = {Mohale, Vincent Zibi and Obagbuwa, Ibidun Christiana},
  title        = {A systematic review on the integration of explainable artificial intelligence in intrusion detection systems to enhancing transparency and interpretability in cybersecurity},
  journal      = {Frontiers in Artificial Intelligence},
  volume       = {8},
  year         = {2025}
}

@inproceedings{bizzarri2024openosr,
  author       = {Bizzarri, Alice and Yu, Chung-En and Jalaian, Brian and Riguzzi, Fabrizio and Bastian, Nathaniel D.},
  title        = {Neuro-symbolic integration for open set recognition in network intrusion detection},
  booktitle    = {AIxIA},
  pages        = {50--63},
  year         = {2024}
}

@article{bizzarri2025neurosymbolic,
  author       = {Bizzarri, Alice},
  title        = {Neuro-symbolic integration in artificial intelligence and its applications},
  journal      = {Ph.D. thesis, Politecnico di Torino},
  year         = {2025}
}

@misc{torq2025alertfatigue,
  author = {Torq},
  title = {{Tired of Security Alert Fatigue? Stop Burnout with Hyperautomation}},
  howpublished = {Accessed: Aug. 3, 2025},
  URL = {https://torq.io/blog/cybersecurity-alert-fatigue/}
}

@article{raman2025navigating,
  author       = {Raman, R. and Kowalski, R. and Achuthan, K. and others},
  title        = {Navigating artificial general intelligence development: Societal, technological, ethical, and brain-inspired pathways},
  journal      = {Scientific Reports},
  volume       = {15},
  pages        = {8443},
  year         = {2025}
}

@inproceedings{kalakoti2025explainable,
  author       = {Kalakoti, Rajesh and Vaarandi, Risto and Bahşi, Hayretdin and Nõmm, Sven},
  title        = {Evaluating explainable AI for deep learning-based network intrusion detection system alert classification},
  booktitle    = {ICISSP},
  pages        = {47--58},
  year         = {2025}
}

@misc{dropzone2025addressalertfatigue,
  author = {Dropzone AI},
  title = {{How to Address Cybersecurity Alert Fatigue with AI}},
  howpublished = {Accessed: Aug. 3, 2025},
  URL = {https://www.dropzone.ai/blog/how-to-address-cybersecurity-alert-fatigue-with-ai}
}

@article{yan2022explainablecybersec,
  author       = {Yan, Feixue and Wen, Sheng and Nepal, Surya and Paris, Cecile and Xiang, Yang},
  title        = {Explainable Machine Learning in Cybersecurity: A Survey},
  journal      = {International Journal of Intelligent Systems},
  year         = {2022},
  month        = {November},
  doi          = {10.1002/int.23088}
}

@article{rajivan2018information,
  author       = {Rajivan, Prashant and Cooke, Nancy J.},
  title        = {Information-pooling bias in collaborative security incident correlation analysis},
  journal      = {Human Factors},
  volume       = {60},
  number       = {5},
  pages        = {626--639},
  year         = {2018}
}

@article{mitre_d3fend,
  author       = {MITRE Corporation},
  title        = {MITRE D3FEND},
  journal      = {Technical Report},
  year         = {2024}
}

@article{kaloroumakis2020knowledge,
  author       = {Kaloroumakis, Peter E. and Smith, Michael J.},
  title        = {Toward a knowledge graph of cybersecurity countermeasures},
  journal      = {Technical Report},
  year         = {2020}
}

@article{ontiveros2025ground,
  author       = {Ontiveros, Rodrigo Castellano and Giannini, Francesco and Gori, Marco and Marra, Giuseppe and Diligenti, Michelangelo},
  title        = {Grounding methods for neural-symbolic AI},
  journal      = {arXiv preprint arXiv:2507.08216},
  year         = {2025}
}

@inproceedings{li2023softened,
  author       = {Li, Zenan and Yao, Yuan and Chen, Taolue and Xu, Jingwei and Cao, Chun and Ma, Xiaoxing and Lü, Jian},
  title        = {Softened symbol grounding for neuro-symbolic systems},
  booktitle    = {ICLR},
  year         = {2023}
}

@article{salem2024advancing,
  author       = {Salem, A. H. and Azzam, S. M. and Emam, O. E. and others},
  title        = {Advancing cybersecurity: A comprehensive review of AI-driven detection techniques},
  journal      = {Journal of Big Data},
  volume       = {11},
  pages        = {105},
  year         = {2024}
}

@article{bilot2024survey,
  author    = {Bilot, Tristan and El Madhoun, Nour and Al Agha, Khaldoun and Zouaoui, Anis},
  title     = {A Survey on Malware Detection with Graph Representation Learning},
  journal   = {ACM Computing Surveys},
  volume    = {56},
  number    = {11},
  articleno = {278},
  numpages  = {36},
  year      = {2024},
  doi       = {10.1145/3664649}
}

@inproceedings{arp2022and,
  title={Dos and don'ts of machine learning in computer security},
  author={Arp, Daniel and Quiring, Erwin and Pendlebury, Feargus and Warnecke, Alexander and Pierazzi, Fabio and Wressnegger, Christian and Cavallaro, Lorenzo and Rieck, Konrad},
  booktitle={31st USENIX Security Symposium (USENIX Security 22)},
  pages={3971--3988},
  year={2022}
}

@article{capuano2022explainable,
  author       = {Capuano, Nicola and Fenza, Giuseppe and Loia, Vincenzo and Stanzione, Claudio},
  title        = {Explainable artificial intelligence in cybersecurity: A survey},
  journal      = {IEEE Access},
  volume       = {10},
  pages        = {93575--93600},
  year         = {2022}
}

@article{yan2023graph,
  author       = {Yan, Bo and Yang, Cheng and Shi, Chuan and Fang, Yong and Li, Qi and Ye, Yanfang and Du, Junping},
  title        = {Graph mining for cybersecurity: A survey},
  journal      = {ACM TKDD},
  volume       = {18},
  number       = {2},
  pages        = {1--52},
  year         = {2023}
}

@article{piplai2023knowledge,
  author       = {Piplai, Aritran and Kotal, Anantaa and Mohseni, Seyedreza and Gaur, Manas and Mittal, Sudip and Joshi, Anupam},
  title        = {Knowledge-enhanced neurosymbolic artificial intelligence for cybersecurity and privacy},
  journal      = {IEEE IC},
  volume       = {27},
  number       = {5},
  pages        = {43--48},
  year         = {2023}
}

@article{eckhoff2025experimenting,
  title        = {Experimenting with Neurosymbolic AI for Defending Against Cyber Attacks},
  author       = {Eckhoff, Magnus Wiik and Halvorsen, Jonas and Hansen, Bj{\o}rn Jervell and Eian, Martin and Mavroeidis, Vasileios and Chetwyn, Robert Andrew and Skj{\o}tskift, Geir and Grov, Gudmund},
  journal      = {Neurosymbolic Artificial Intelligence Journal},
  year         = {2025},
  note         = {To appear},
}

@article{wang2025trustworthy,
  author       = {Wang, Xiaojie and Wang, Beibei and Wu, Yu and Ning, Zhaolong and Guo, Song and Yu, Fei Richard},
  title        = {A survey on trustworthy edge intelligence: From security and reliability to transparency and sustainability},
  journal      = {IEEE COMST},
  volume       = {27},
  number       = {3},
  pages        = {1729--1757},
  year         = {2025}
}

@article{Rosenberg2021AdversarialML,
  author       = {Rosenberg, Ishai and Shabtai, Asaf and Elovici, Yuval and Rokach, Lior},
  title        = {Adversarial machine learning attacks and defense methods in the cyber security domain},
  journal      = {ACM Computing Surveys},
  volume       = {54},
  number       = {5},
  pages        = {108:1--108:36},
  year         = {2021}
}

@misc{galwaduge2025tabuliff,
  title         = {Tabular Diffusion Based Actionable Counterfactual Explanations for Network Intrusion Detection},
  author        = {Galwaduge, Vinura and Samarabandu, Jagath},
  year          = {2025},
  eprint        = {2507.17161},
  archivePrefix = {arXiv},
  primaryClass  = {cs.LG}
}

@misc{bougzime2025unlocking,
  title         = {Unlocking the Potential of Generative {AI} Through Neuro-Symbolic Architectures: Benefits and Limitations},
  author        = {Bougzime, Oualid and Jabbar, Samir and Cruz, Christophe and Demoly, Fr{\'e}d{\'e}ric},
  year          = {2025},
  eprint        = {2502.11269},
  archivePrefix = {arXiv},
  primaryClass  = {cs.AI}
}

@misc{challita2025redteamllm,
  title         = {{RedTeamLLM}: An Agentic {AI} Framework for Offensive Security},
  author        = {Challita, Brian and Parrend, Pierre},
  year          = {2025},
  eprint        = {2505.06913},
  archivePrefix = {arXiv},
  primaryClass  = {cs.CR}
}

@article{kent2015authentication,
  author       = {Kent, Alexander D. and Liebrock, Lorie M. and Neil, Joshua C.},
  title        = {Authentication Graphs: Analyzing User Behavior Within an Enterprise Network},
  journal      = {Computers \& Security},
  volume       = {48},
  pages        = {150--166},
  year         = {2015}
}

@misc{xu2024autoattacker,
  title         = {{AutoAttacker}: A Large Language Model Guided System to Implement Automatic Cyber-Attacks},
  author        = {Xu, Jiacen and Stokes, Jack W and McDonald, Geoff and Bai, Xuesong and Marshall, David and Wang, Siyue and Swaminathan, Adith and Li, Zhou},
  year          = {2024},
  eprint        = {2403.01038},
  archivePrefix = {arXiv},
  primaryClass  = {cs.CR}
}

@article{devries2023energy,
  author       = {de Vries, Alex},
  title        = {The growing energy footprint of artificial intelligence},
  journal      = {Joule},
  volume       = {7},
  number       = {10},
  pages        = {2191--2194},
  year         = {2023}
}

@techreport{wef2025aienergy,
  author       = {World Economic Forum and AI Governance Alliance},
  title        = {Artificial Intelligence's Energy Paradox: Balancing Challenges and Opportunities},
  institution  = {World Economic Forum},
  year         = {2025},
  month        = {Jan.},
  note         = {In collaboration with Accenture},
}

@techreport{stixbp2022,
  author       = {{OASIS Cyber Threat Intelligence (CTI) Technical Committee}},
  title        = {STIX Best Practices Guide Version 1.0.0},
  institution  = {OASIS Open},
  year         = {2022},
  number       = {01},
  type         = {Committee Note}
}

@inproceedings{tumkur2025neuro,
  title={Neuro-Symbolic Approaches for Cybersecurity Policy Enforcement},
  author={Tumkur, Srikanta Datta and Asokan, Vinoth Ganesh Umamageswari},
  booktitle={2025 5th Intelligent Cybersecurity Conference (ICSC)},
  pages={327--332},
  year={2025},
  organization={IEEE}
}

@article{belcastro2025enhancing,
  title={Enhancing network security using knowledge graphs and large language models for explainable threat detection},
  author={Belcastro, Loris and Carlucci, Carmine and Cosentino, Cristian and Li{\`o}, Pietro and Marozzo, Fabrizio},
  journal={Future Generation Computer Systems},
  pages={108160},
  year={2025},
  publisher={Elsevier}
}

@article{shama2026charting,
  title={Charting the evolution of neuro-symbolic AI in cybersecurity: a scientometric perspective},
  author={Shama and Jain, Sarika},
  journal={International Journal of Data Science and Analytics},
  volume={22},
  number={1},
  pages={90},
  year={2026},
  publisher={Springer}
}

@article{fathima2026neurosymbolic,
  title={Neurosymbolic Learning for Advanced Persistent Threat Detection under Extreme Class Imbalance},
  author={Fathima, Quhura and Moghim, Neda and Firouzjaee, Mostafa Taghizade and Thomas, Christo K and Gore, Ross and Saad, Walid},
  journal={arXiv preprint arXiv:2603.00453},
  year={2026}
}

@article{tsigkourakos2026qrs,
  title={QRS: A Rule-Synthesizing Neuro-Symbolic Triad for Autonomous Vulnerability Discovery},
  author={Tsigkourakos, George and Patsakis, Constantinos},
  journal={arXiv preprint arXiv:2602.09774},
  year={2026}
}

@article{kautz2022third,
  title={The Third {AI} Summer: {AAAI} Robert S.\ Engelmore Memorial Lecture},
  author={Kautz, Henry},
  journal={AI Magazine},
  volume={43},
  number={1},
  pages={93--104},
  year={2022},
  publisher={Wiley}
}

@article{li2025multi,
  title={Multi-view intrusion detection framework using deep learning and knowledge graphs},
  author={Li, Min and Qiao, Yuansong and Lee, Brian},
  journal={Information},
  volume={16},
  number={5},
  pages={377},
  year={2025},
  publisher={MDPI}
}

@incollection{shreha2026neuro,
  title={Neuro-Symbolic AI and Behavioral Intelligence: Towards Explainable Cybersecurity Mechanisms},
  author={Shreha, SA and Ravikumar, RN and Aarthi, S},
  booktitle={Neuroscience-Enhanced Artificial Intelligence for Cybersecurity},
  pages={121--152},
  year={2026},
  publisher={IGI Global Scientific Publishing}
}

@article{almadhor2025designing,
  title={Designing a neuro-symbolic dual-model architecture for explainable and resilient intrusion detection in IoT networks},
  author={Almadhor, Ahmad and Alsubai, Shtwai and Hejaili, Abdullah Al and Klai, Zeineb and Bouallegue, Belgacem and Kovac, Urban},
  journal={Scientific Reports},
  volume={15},
  number={1},
  pages={42786},
  year={2025},
  publisher={Nature Publishing Group UK London}
}

@inproceedings{bashir2024resiliency,
  author       = {Bashir, Shadaab Kawnain and Podder, Rakesh and Sreedharan, Sarath and Ray, Indrakshi and Ray, Indrajit},
  title        = {Resiliency graphs: Modelling the interplay between cyber attacks and system failures through AI planning},
  booktitle    = {IEEE TPS-ISA},
  pages        = {256--263},
  year         = {2024}
}

@article{naidu2022can,
  author       = {Naidu, Rakshit and Kagalwalla, Navid},
  title        = {Can causal (and counterfactual) reasoning improve privacy threat modelling?},
  journal      = {arXiv preprint arXiv:2207.09746},
  year         = {2022}
}

@inproceedings{potteiger2024designing,
  author       = {Potteiger, Nicholas and Samaddar, Ankita and Bergstrom, Hunter and Koutsoukos, Xenofon},
  title        = {Designing robust cyber-defense agents with evolving behavior trees},
  booktitle    = {IEEE ICAA},
  pages        = {1--10},
  year         = {2024}
}

@inproceedings{kerr2024accelerating,
  author       = {Kerr, Ryan and Ding, Steven and Li, Li and Taylor, Adrian},
  title        = {Accelerating autonomous cyber operations: A symbolic logic planner guided reinforcement learning approach},
  booktitle    = {IEEE ICNC},
  pages        = {641--647},
  year         = {2024}
}

@inproceedings{mankali2024insight,
  author       = {Mankali, Lakshmi Likhitha and Sinanoglu, Ozgur and Patnaik, Satwik},
  title        = {INSIGHT: Attacking industry-adopted learning resilient logic locking techniques using explainable graph neural network},
  booktitle    = {USENIX Security},
  pages        = {91--108},
  year         = {2024}
}

@inproceedings{curaba2024cryptoformaleval,
  author       = {Curaba, Cristian and D'Ambrosi, Denis and Minisini, Alessandro},
  title        = {CryptoFormalEval: Integrating large language models and formal verification for automated cryptographic protocol vulnerability detection},
  booktitle    = {NeurIPS Workshop},
  year         = {2024}
}

@inproceedings{xie2022neuro,
  author       = {Xie, Xuan and Kersting, Kristian and Neider, Daniel},
  title        = {Neuro-symbolic verification of deep neural networks},
  booktitle    = {IJCAI},
  pages        = {3622--3628},
  year         = {2022}
}

@inproceedings{jalaian2023neurosec,
  author       = {Jalaian, Brian and Bastian, Nathaniel D.},
  title        = {Neurosymbolic AI in cybersecurity: Bridging pattern recognition and symbolic reasoning},
  booktitle    = {IEEE MILCOM},
  year         = {2023}
}

@inproceedings{li2025iris,
  author       = {Li, Ziyang and Dutta, Saikat and Naik, Mayur},
  title        = {IRIS: LLM-assisted static analysis for detecting security vulnerabilities},
  booktitle    = {ICLR},
  year         = {2025}
}

@inproceedings{lei2024adapt,
  author       = {Lei, Haozhe and Ge, Yunfei and Zhu, Quanyan},
  title        = {ADAPT: A game-theoretic and neuro-symbolic framework for automated distributed adaptive penetration testing},
  booktitle    = {IEEE MILCOM},
  year         = {2024}
}

@inproceedings{xu2025l2m,
  title={L2M-AID: Autonomous Cyber-Physical Defense by Fusing Semantic Reasoning of Large Language Models with Multi-Agent Reinforcement Learning},
  author={Xu, Tianxiang and Wen, Zhichao and Zhao, Xinyu and Wang, Jun and Li, Yan and Liu, Chang},
  booktitle={2025 IEEE 24th International Conference on Trust, Security and Privacy in Computing and Communications (TrustCom)},
  pages={2198--2203},
  year={2025},
  organization={IEEE}
}

@inproceedings{hakim2025explainable,
  title={An Explainable Neuro-Symbolic Rule Extraction Framework for Digital Twins},
  author={Hakim, Safayat Bin and Adil, Muhammad and Velasquez, Alvaro and Song, Houbing Herbert},
  booktitle={2025 IEEE Smart World Congress (SWC)},
  pages={1042--1047},
  year={2025},
  organization={IEEE}
}


\appendix

\section{Per-Paper Integration Classification}
\label{appendix:classification}

\rev{Tables~\ref{tab:s1_typeA}--\ref{tab:s1_typeC} provide the complete per-paper classification for all 103 surveyed publications, following the three-tier taxonomy defined in Section~\ref{subsec:nesy_definitions}. These tables are provided to enable transparent verification of the tier assignments and to support traceability of each paper's neural component, symbolic component, and application domain.}

\clearpage
\onecolumn
\setlength{\LTleft}{0pt}
\setlength{\LTright}{0pt}


\renewcommand\arraystretch{1.05}
\scriptsize
\begin{longtable}{|c|l|p{3.0cm}|p{3.2cm}|p{2.5cm}|}
\caption{\rev{Type~A --- Deep NeSy Integration (22 publications)}}\label{tab:s1_typeA}\\
\hline
\textbf{\#} & \textbf{Citation} & \textbf{Neural Component} & \textbf{Symbolic Component} & \textbf{Domain} \\
\hline \hline
\endfirsthead
\caption[]{\rev{Type~A --- Deep NeSy Integration (22 publications)} (continued)}\\
\hline
\textbf{\#} & \textbf{Citation} & \textbf{Neural Component} & \textbf{Symbolic Component} & \textbf{Domain} \\
\hline \hline
\endhead
\hline
\endfoot
\hline
\endlastfoot
1 & Zhou et al.~\cite{zhou2024knowgraph} & GNN (main + knowledge models) & Weighted FOL via probabilistic graphical model & Anomaly Detection \\
\hline
2 & Bizzarri et al.~\cite{bizzarri2024synergistic} & DNN & Logic Tensor Networks (LTN) & Network IDS \\
\hline
3 & Bizzarri et al.~\cite{bizzarri2024neuro} & Neural classifier & LTN with FOL axioms & Network IDS \\
\hline
4 & Grov et al.~\cite{grov2024neurosymbolic} & Neural (feature extraction) & LTN + domain-specific logical rules & IDS (XSS) \\
\hline
5 & Onchis et al.~\cite{onchis2022neurosymbolic} & DNN & LTN with satisfiability optimization & Network IDS \\
\hline
6 & Tran et al.~\cite{tran2025neurosymbolic} & Neural network & LTN with logical axioms & Network IDS \\
\hline
7 & Kalutharage et al.~\cite{kalutharage2025neurosymbolic} & Anomaly detection model & KG (MITRE ATT\&CK) + domain rules & IoT IDS \\
\hline
8 & Luo et al.~\cite{luo2024neurosymbolic} & Neural RL agent & Symbolic program synthesis & General NeSy \\
\hline
9 & Kerr et al.~\cite{kerr2024accelerating} & DQN & ARISTOTLE (PDDL planner) & Penetration Testing \\
\hline
10 & Lei et al.~\cite{lei2024adapt} & Neural learning & Game-theoretic meta-game + symbolic KB & Penetration Testing \\
\hline
11 & Potteiger et al.~\cite{potteiger2024designing} & Learning-enabled components & Behavior Trees (symbolic) & Autonomous Defense \\
\hline
12 & Samaddar et al.~\cite{samaddar2025ood} & Neural OOD detector & Symbolic cyber-defense reasoning (EBT) & Autonomous Defense \\
\hline
13 & Li et al.~\cite{li2025automated} & LLM (GPT-4) & CodeQL static analysis (formal symbolic) & Vulnerability Detection \\
\hline
14 & Li et al.~\cite{li2025iris} & LLM (GPT-4) & CodeQL symbolic static analysis & Vulnerability Detection \\
\hline
15 & Xie et al.~\cite{xie2022neuro} & Neural (target DNN) & Symbolic verification (formal logic) & DNN Verification \\
\hline
16 & Bizzarri et al.~\cite{bizzarri2024openosr} & Neural classifier & LTN with open-set logic & Network IDS \\
\hline
17 & Bizzarri et al.~\cite{bizzarri2025neurosymbolic} & Neural network & LTN framework & Network IDS \\
\hline
18 & Hallyburton et al.~\cite{hallyburton2025assured} & Object detection neural net & Scene Graph Generation (symbolic) & CPS Security \\
\hline
19 & Hakim et al.~\cite{hakim2025ansrdt} & CNN-LSTM + PPO (RL) & Prolog-based symbolic reasoning & Digital Twin Security \\
\hline
20 & Tumkur et al.~\cite{tumkur2025neuro} & Neural (adaptive learning) & Symbolic (policy logic) & Policy Enforcement \\
\hline
21 & Curaba et al.~\cite{curaba2024cryptoformaleval} & LLM & Formal verification (symbolic) & Cryptographic Security \\
\hline
22 & Fathima et al.~\cite{fathima2026neurosymbolic} & BERT (transformer) & LTN with 16 learnable predicates & APT Detection \\
\hline
\end{longtable}


\renewcommand\arraystretch{1.05}
\scriptsize
\begin{longtable}{|c|l|p{2.6cm}|p{2.8cm}|p{1.5cm}|p{1.7cm}|}
\caption{\rev{Type~B --- Structured NeSy Integration, Part~1: Subtypes B1--B3 (28 of 55 publications)}}\label{tab:s1_typeB1}\\
\hline
\textbf{\#} & \textbf{Citation} & \textbf{Neural Component} & \textbf{Symbolic Component} & \textbf{Subtype} & \textbf{Domain} \\
\hline \hline
\endfirsthead
\caption[]{\rev{Type~B --- Structured NeSy Integration, Part~1: Subtypes B1--B3 (28 of 55 publications)} (continued)}\\
\hline
\textbf{\#} & \textbf{Citation} & \textbf{Neural Component} & \textbf{Symbolic Component} & \textbf{Subtype} & \textbf{Domain} \\
\hline \hline
\endhead
\hline
\endfoot
\hline
\endlastfoot
\multicolumn{6}{|l|}{\textit{B1: Knowledge Graph + Neural Integration (19 papers)}} \\
\hline
1 & Piplai et al.~\cite{piplai2020using} & RL agent (CQL) & Cybersecurity KG & KG+Neural & Malware \\
\hline
2 & Piplai et al.~\cite{piplai2023knowledge} & Neural network & CKG + rule engine & KG+Neural & Malware/Network \\
\hline
3 & Chen et al.~\cite{chen2022aptkgl} & GNN (heterogeneous) & Open threat KB & KG+Neural & APT Detection \\
\hline
4 & Falcarin et al.~\cite{falcarin2024building} & NER/relation extraction & Knowledge Graph & KG+Neural & KG Construction \\
\hline
5 & Sikos~\cite{sikos2023cybersecurity} & Various neural & Ontology-based KGs & KG+Neural & KG (Survey) \\
\hline
6 & Ren et al.~\cite{ren2022cskg4apt} & NLP/entity extraction & CSKG (symbolic) & KG+Neural & APT Attribution \\
\hline
7 & Zhao et al.~\cite{zhao2024survey} & Various neural (NER, RE) & Knowledge Graphs & KG+Neural & KG Construction \\
\hline
8 & Alharbi et al.~\cite{alharbi2025enhancing} & NLP models & Knowledge Graph & KG+Neural & KG Construction \\
\hline
9 & Cheng et al.~\cite{cheng2025crucialg} & NLP extraction & Attack scenario graphs & KG+Neural & Threat Intel \\
\hline
10 & Zhang et al.~\cite{zhang2025improving} & LLM + ML classifiers & Knowledge Graph & KG+Neural & Threat Detection \\
\hline
11 & Liu et al.~\cite{liu2025graph} & GNN & Domain knowledge ontology & KG+Neural & Threat Intel \\
\hline
12 & Gao et al.~\cite{gao2024threatkg} & NER + relation extraction & Knowledge Graph & KG+Neural & Threat Intel \\
\hline
13 & Kurniawan et al.~\cite{kurniawan2024cykg} & LLM (RAG) & Cybersecurity KG & KG+Neural & Threat Intel \\
\hline
14 & Fieblinger et al.~\cite{fieblinger2024actionable} & LLM & KG (CTI) & KG+Neural & Threat Intel \\
\hline
15 & Cheng et al.~\cite{cheng2025ctinexus} & LLM & STIX ontology & KG+Neural & Threat Intel \\
\hline
16 & Yan et al.~\cite{yan2023graph} & GNN & Threat intelligence graphs & KG+Neural & Threat Intel \\
\hline
17 & Li et al.~\cite{li2025multi} & Deep learning (multi-view) & Knowledge Graph & KG+Neural & IDS \\
\hline
18 & Belcastro et al.~\cite{belcastro2025enhancing} & Graph-BERT + LLM & KG + LIME & KG+Neural & Network Security \\
\hline
19 & Nalluri et al.~\cite{nalluri2025nscti} & GNN + LSTM + HMM & Graph-based threat structure & KG+Neural & CTI \\
\hline
\multicolumn{6}{|l|}{\textit{B2: LLM/Neural + Symbolic Planning/Tool Integration (6 papers)}} \\
\hline
20 & Kong et al.~\cite{kong2025vulnbot} & LLM agents & Penetration Task Graph & LLM+Symb & Pentesting \\
\hline
21 & Huang et al.~\cite{huang2023penheal} & LLM & Structured remediation planning & LLM+Symb & Pentesting \\
\hline
22 & Nieponice et al.~\cite{nieponice2025aracne} & LLM & Shell grammar + symbolic reasoning & LLM+Symb & Pentesting \\
\hline
23 & Xiang et al.~\cite{xiang2025guardagent} & LLM agents & Knowledge-based reasoning rules & LLM+Symb & LLM Safety \\
\hline
24 & Bashir et al.~\cite{bashir2024resiliency} & AI planning & Resiliency Graph (symbolic) & LLM+Symb & Resilience \\
\hline
25 & Tsigkourakos et al.~\cite{tsigkourakos2026qrs} & LLM agents (3 agents) & CodeQL (symbolic static analysis) & LLM+Symb & Vuln Discovery \\
\hline
\multicolumn{6}{|l|}{\textit{B3: Causal-Neural Integration (3 papers)}} \\
\hline
26 & Jaimini et al.~\cite{jaimini2024causal} & Neural models & Causal DAGs, SCMs & Causal & Causal NeSy \\
\hline
27 & Andrew et al.~\cite{andrew2022developing} & RL agents & Structural Causal Models & Causal & Cyber Defense \\
\hline
28 & Naidu et al.~\cite{naidu2022can} & ML models & Counterfactual reasoning & Causal & Privacy/Threats \\
\hline
\end{longtable}


\renewcommand\arraystretch{1.05}
\scriptsize
\begin{longtable}{|c|l|p{2.6cm}|p{2.8cm}|p{1.5cm}|p{1.7cm}|}
\caption{\rev{Type~B --- Structured NeSy Integration, Part~2: Subtypes B4--B7 and Additional (27 of 55 publications)}}\label{tab:s1_typeB2}\\
\hline
\textbf{\#} & \textbf{Citation} & \textbf{Neural Component} & \textbf{Symbolic Component} & \textbf{Subtype} & \textbf{Domain} \\
\hline \hline
\endfirsthead
\caption[]{\rev{Type~B --- Structured NeSy Integration, Part~2: Subtypes B4--B7 and Additional (27 of 55 publications)} (continued)}\\
\hline
\textbf{\#} & \textbf{Citation} & \textbf{Neural Component} & \textbf{Symbolic Component} & \textbf{Subtype} & \textbf{Domain} \\
\hline \hline
\endhead
\hline
\endfoot
\hline
\endlastfoot
\multicolumn{6}{|l|}{\textit{B4: Explainable AI with Symbolic Knowledge Integration (11 papers)}} \\
\hline
29 & Neupane et al.~\cite{neupane2022explainable} & Various neural IDS & Symbolic explanation methods & XAI+Symb & IDS (Survey) \\
\hline
30 & Sarker et al.~\cite{sarker2024explainable} & Neural AI models & Symbolic reasoning for trust & XAI+Symb & Cybersecurity \\
\hline
31 & Nyre-Yu et al.~\cite{nyre2022explainable} & Neural detection models & XAI tools with rule components & XAI+Symb & SOC Operations \\
\hline
32 & Yan et al.~\cite{yan2022explainablecybersec} & Neural models & Symbolic/structured explanations & XAI+Symb & Cybersecurity \\
\hline
33 & Capuano et al.~\cite{capuano2022explainable} & Neural AI & Symbolic explanation methods & XAI+Symb & XAI (Survey) \\
\hline
34 & Arreche et al.~\cite{arreche2024explainableids} & Neural IDS (7 models) & XAI rule-based explanations & XAI+Symb & Network IDS \\
\hline
35 & Mohale et al.~\cite{mohale2025explaiids} & Neural classifier & Domain knowledge explanation & XAI+Symb & IDS \\
\hline
36 & Kalakoti et al.~\cite{kalakoti2025explainable} & Neural models & Knowledge-driven explanation & XAI+Symb & Cybersecurity \\
\hline
37 & Eckhoff et al.~\cite{eckhoff2025experimenting} & Neural + symbolic (LTN) & Symbolic rules/constraints & XAI+Symb & Cyber Defense \\
\hline
38 & Shreha et al.~\cite{shreha2026neuro} & Neural (behavioral) & Symbolic (behavioral rules) & XAI+Symb & Explainable Cyber \\
\hline
39 & Hakim et al.~\cite{hakim2025explainable} & Neural network & Symbolic rule extraction & XAI+Symb & Digital Twins \\
\hline
\multicolumn{6}{|l|}{\textit{B5--B7: Ontology, Adversarial, and Other Structured Integration (7 papers)}} \\
\hline
40 & Mankali et al.~\cite{mankali2024insight} & DL-based analysis & Symbolic circuit analysis rules & Ontology & Hardware Security \\
\hline
41 & Tafreshian et al.~\cite{tafreshian2024defensive} & ML classifiers & Formal constraint framework & Adversarial & Adversarial Defense \\
\hline
42 & Grini et al.~\cite{grini2025constrained} & Neural attack generation & Constraint framework & Adversarial & Adversarial Attacks \\
\hline
43 & Jalaian et al.~\cite{jalaian2023neurosec} & Neural (pattern recog.) & Symbolic (reasoning) & Other & IDS \\
\hline
44 & Almadhor et al.~\cite{almadhor2025designing} & CNN + ANN (dual) & Symbolic reasoning (claimed) & Other & IoT IDS \\
\hline
45 & Deng et al.~\cite{deng2024pentestgpt} & GPT-4 & Pentesting methodology (PTM) & Other & Pentesting \\
\hline
46 & Wang et al.~\cite{wang2024sands} & LLM & Structured attack lifecycle & Other & Attack Construction \\
\hline
\multicolumn{6}{|l|}{\textit{Additional: Formal Methods, Verification, and Benchmarks (9 papers)}} \\
\hline
47 & Blaauwbroek et al.~\cite{blaauwbroek2024learning} & Neural heuristics & Automated theorem prover & Formal & Verification \\
\hline
48 & Blaauwbroek et al.~\cite{blaauwbroek2024graph2tac} & GNN & Coq theorem prover & Formal & Verification \\
\hline
49 & Piepenbrock et al.~\cite{piepenbrock2023neural} & Neural guidance & SMT solver & Formal & Verification \\
\hline
50 & Lu et al.~\cite{lu2023z3} & Neural heuristics & Z3 SMT solver & Formal & Verification \\
\hline
51 & Alam et al.~\cite{alam2024ctibench} & LLM & CTI benchmarks (structured) & Benchmark & CTI Evaluation \\
\hline
52 & Paul et al.~\cite{paul2024formal} & Neural AI & Formal explanation frameworks & Formal & XAI \\
\hline
53 & Zhao et al.~\cite{zhao2025letsmeasure} & Fine-tuned LLM & Deterministic logical reasoning & LLM+Symb & Privacy Analysis \\
\hline
54 & Kireev et al.~\cite{kireev2022adversarial} & Neural models & Adversarial robustness theory & Adversarial & Robustness \\
\hline
55 & Bui et al.~\cite{bui2023generating} & Neural generation & Adversarial constraint framework & Adversarial & Adversarial \\
\hline
\end{longtable}


\renewcommand\arraystretch{1.05}
\scriptsize
\begin{longtable}{|c|l|p{6.0cm}|p{4.6cm}|}
\caption{\rev{Type~C --- Contextual Baselines (26 publications, non-NeSy)}}\label{tab:s1_typeC}\\
\hline
\textbf{\#} & \textbf{Citation} & \textbf{Approach} & \textbf{Reason for Type~C} \\
\hline \hline
\endfirsthead
\caption[]{\rev{Type~C --- Contextual Baselines (26 publications, non-NeSy)} (continued)}\\
\hline
\textbf{\#} & \textbf{Citation} & \textbf{Approach} & \textbf{Reason for Type~C} \\
\hline \hline
\endhead
\hline
\endfoot
\hline
\endlastfoot
1 & Fang et al.~\cite{zhu2026teams} & HPTSA: Teams of LLM agents for zero-day exploitation & Pure LLM multi-agent; no symbolic reasoning \\
\hline
2 & Xu et al.~\cite{xu2024autoattacker} & AutoAttacker: LLM for post-breach automation & Pure LLM; no symbolic component \\
\hline
3 & Muzsai et al.~\cite{muzsai2024hacksynth} & HackSynth: LLM-based exploit generation & Pure LLM; no structured reasoning \\
\hline
4 & Clairoix-Tr\'{e}panier et al.~\cite{clairouxtrepanier2024llm} & LLMs for CTI in cybercrime forums & Pure LLM text analysis \\
\hline
5 & Guastalla et al.~\cite{guastalla2024llmddos} & LLMs for DDoS detection & Pure LLM classification \\
\hline
6 & Kasri et al.~\cite{kasri2025vulnerability} & LLMs in cybersecurity (survey) & LLM survey; no NeSy framing \\
\hline
7 & Challita et al.~\cite{challita2025redteamllm} & Red-teaming LLMs & Adversarial prompting; no symbolic component \\
\hline
8 & Bougzime et al.~\cite{bougzime2025unlocking} & Unlocking LLMs for cybersecurity & Pure LLM capabilities \\
\hline
9 & Singh et al.~\cite{singh2024hierarchical} & H-MARL: Hierarchical multi-agent RL & Pure hierarchical RL; no symbolic knowledge \\
\hline
10 & Hurten et al.~\cite{hurten2024hierarchical} & H-MARL for coalition defense & Pure H-MARL; no symbolic component \\
\hline
11 & Sajid et al.~\cite{sajid2024enhancing} & Hybrid ML + DL IDS & Ensemble of neural methods; ``hybrid'' = ML+DL \\
\hline
12 & Faber et al.~\cite{faber2024lifelong} & Lifelong learning for IDS & Continual neural learning; no symbolic \\
\hline
13 & Bhatt et al.~\cite{bhatt2024cyberseceval} & CyberSecEval 2 & LLM evaluation benchmark \\
\hline
14 & Bhusal et al.~\cite{bhusal2024secure} & SECURE benchmark & Security evaluation framework \\
\hline
15 & Jing et al.~\cite{jing2024secbench} & SecBench & Security benchmark \\
\hline
16 & Rodriguez et al.~\cite{rodriguez2025framework} & AI cyberattack evaluation framework & Evaluation framework; not a NeSy system \\
\hline
17 & Ristea et al.~\cite{ristea2024ai} & AI cyber risk benchmark & Risk benchmark; not NeSy \\
\hline
18 & Rawal et al.~\cite{rawal2025causality} & Causality for trustworthy AI (survey) & General causality survey \\
\hline
19 & Xu et al.~\cite{xu2025l2m} & L2M-AID: LLM + MARL defense & LLM as feature encoder; no symbolic reasoning \\
\hline
20 & Pawlicki et al.~\cite{Pawlicki2022NeuCom} & Neural computing for cybersecurity & Purely neural \\
\hline
21 & Rosenberg et al.~\cite{Rosenberg2021AdversarialML} & Adversarial ML survey & No NeSy framing \\
\hline
22 & Sarhan et al.~\cite{sarhan2023from} & Zero-shot ML for cybersecurity & Purely neural; no symbolic \\
\hline
23 & Guo et al.~\cite{guo2023review} & Zero-day detection review & ML survey; no NeSy \\
\hline
24 & Hitzler~\cite{Hitzler2020NeSySW} & NeSy and Semantic Web & Foundational NeSy text (background) \\
\hline
25 & Ferrag et al.~\cite{ferrag2025generative} & Generative AI in cybersecurity & LLM survey; no NeSy framing \\
\hline
26 & Zhang et al.~\cite{zhang2025llm} & LLMs meet cybersecurity & LLM survey; no NeSy framing \\
\hline
\end{longtable}

\end{document}